\documentclass[%
 reprint,
superscriptaddress,
 amsmath,amssymb,
 aps,
pra,
]{revtex4-2}

\usepackage{graphicx}% Include figure files
\usepackage{dcolumn}% Align table columns on decimal point
\usepackage{bm}% bold math
\usepackage{multirow}
\usepackage{xspace}
\usepackage[implicit=false]{hyperref}% add hypertext capabilities
\usepackage{xr}
\makeatletter
\newcommand*{\addFileDependency}[1]{% argument=file name and extension
  \typeout{(#1)}
  \@addtofilelist{#1}
  \IfFileExists{#1}{}{\typeout{No file #1.}}
}
\makeatother

\newcommand{\ba}{\bar{\alpha}}
\newcommand{\br}{\bm{r}}
\newcommand{\mo}[1]{\mathcal{O}[|\nabla n|^{#1}]}
\newcommand{\nup}{n_{\uparrow}}
\newcommand{\ndn}{n_{\downarrow}}
\newcommand{\sus}{_{\sigma}}
\newcommand{\rs}{r_{\mathrm{s}}}
\newcommand{\gn}{\nabla n}
\newcommand{\lan}{\nabla^2 n}
\newcommand{\tauu}{\tau_U}

\newcommand{\tss}[1]{\ensuremath{^{\text{#1}}}}
\newcommand{\rrscan}{r\tss{2}SCAN\xspace}
\newcommand{\rfscan}{r\tss{4}SCAN\xspace}
\newcommand{\smx}{_{\mr{x}}}
\newcommand{\smc}{_{\mr{c}}}

\usepackage{color}
\usepackage{booktabs}
\usepackage{longtable}

\renewcommand{\v}[1]	{\ensuremath{\bm{#1}}} % for vectors
\newcommand{\mr}[1]     {\ensuremath{\mathrm{#1}}}
\newcommand{\mol}[1]	{\ensuremath{_{\text{#1}}}}
\newcommand{\excite}[1]    {\ensuremath{^{\text{#1}}}}
\newcommand{\rppscan}{r++SCAN\xspace}
\newcommand{\muak}      {\ensuremath{\mu_{\mr{AK}}}}

\begin{document}

\preprint{N/A}

\title{Construction of meta-GGA functionals through restoration of exact constraint adherence to regularized SCAN functionals.}

\author{James W. Furness}
\email{jfurness@tulane.edu}
\affiliation{Department of Physics and Engineering Physics, Tulane University, New Orleans, LA 70118}

\author{Aaron D. Kaplan}
\affiliation{Department of Physics, Temple University, Philadelphia, PA 19122}

\author{Jinliang Ning}
\affiliation{Department of Physics and Engineering Physics, Tulane University, New Orleans, LA 70118}

\author{John P. Perdew}
\affiliation{Department of Physics, Temple University, Philadelphia, PA 19122}
\affiliation{Department of Chemistry, Temple University, Philadelphia, PA 19122}

 \author{Jianwei Sun}
 \email{jsun@tulane.edu}
 \affiliation{Department of Physics and Engineering Physics, Tulane University, New Orleans, LA 70118}

\date{\today}% It is always \today, today,

\begin{abstract}
The SCAN meta-GGA exchange-correlation functional [Phys. Rev. Lett. \textbf{115}, 036402 (2015)] is constructed as a chemical environment-determined interpolation between two separate energy densities: one describes single orbital electron densities accurately, and another describes slowly-varying densities accurately. To conserve constraints known for the exact exchange-correlation functional, the derivatives of this interpolation vanish in the slowly-varying limit. While theoretically convenient, this choice introduces numerical challenges that degrade the functional's efficiency. We have recently reported a modification to the SCAN functional, termed \rrscan [J. Phys. Chem. Lett. \textbf{11}, 8208 (2020)] that introduces two regularizations into SCAN which improve its numerical performance at the expense of not recovering the fourth order term of the slowly-varying density gradient expansion for exchange. Here we show the derivation of a progression of functionals (rSCAN, \rppscan, \rrscan, and \rfscan) with increasing adherence to exact conditions while maintaining a smooth interpolation. The greater smoothness of \rrscan seems to lead to better general accuracy than the additional exact constraint of SCAN or \rfscan does.
\end{abstract}

\maketitle

% This is the document to develop

% \tableofcontents

\section{\label{sec:intro} Introduction}

The importance of efficient computational modeling in chemistry and materials science cannot be understated, and for many applications, Kohn--Sham density functional theory presents the most appealing compromise between accuracy and efficiency. The favorable position of this compromise has been enabled by the steady progression of ever more accurate density functionals produced over the last 60 years. These functionals are commonly characterized by the Perdew--Schmidt hierarchy \cite{Perdew2001}, a progression of increasing non-locality where successive levels can be expected to give greater accuracy at the cost of increased computational complexity.

The meta-generalized gradient approximations (meta-GGAs) stand as an appealing level at which the highest accuracy can be expected from semi-local ingredients, including the electron density, its gradient, and the kinetic energy density. While hybrid functionals incorporating admixtures of non-local exact exchange have become most prominent for molecular applications, the prohibitive cost scaling of exact-exchange with number of electrons has limited their utility for extended systems.

A meta-GGA is commonly designed by either enforcing constraints on the exchange-correlation (XC) functional, or by fitting to reference data sets. Those that take the latter route, called empirical functionals, can be inaccurate for systems outside their respective fitting set, or can suffer from difficulties due to over-fitting. General purpose functionals that are accurate for diverse systems have tended to be of the former, so-called non-empirical, variety in which transferable accuracy is promoted by adherence to physical constraints that are necessarily true for all systems of electrons.
The first generation of meta-GGAs were non-empirical, and predate most GGAs.
Becke and Roussel \cite{Becke1983,Becke1989} derived generalized Taylor series of the exact-exchange hole by enforcing sum rule and non-negativity constraints on a hole model.
Perdew \cite{Perdew1985} derived a Laplacian-level meta-GGA for the exchange energy by enforcing the same set of constraints.

At the meta-GGA level, the strongly-constrained and appropriately-normed (SCAN) functional has incorporated all 17 known constraints on the exact XC energy appropriate to the semi-local level \cite{Sun2015}.
(These constraints are listed together in the Supplementary Material of Ref. \cite{Sun2015}, and the references for them are presented in the main text of Ref. \cite{Sun2015}.)
From this foundation, further works have proposed modifications to the SCAN energy densities to improve its accuracy in certain domains. revSCAN is a simple modification to the slowly-varying limit of SCAN's correlation energy that modifies the fourth-order term in its density-gradient expansion \cite{Mezei2018}. The TASK functional is a complete revision of SCAN, retaining only its fulfillment of exact constraints for the exchange energy \cite{Aschebrock2019}. TASK is designed to accurately predict band gaps, and has recently been itself extended for accuracy in 2D systems, in a modification termed ``mTASK'' \cite{Neupane2021}. Note TASK and mTASK use a correlation density functional at the local density approximation.

The SCAN functional has proved broadly transferable and has shown good accuracy for many systems normally challenging for DFT methods \cite{Kitchaev2016, Sun2016, Peng2017, Zhang2017, Chen2017, Furness2018, Lane2018, SaiGautam2018, Zhang2019, Zhang2020b, Pulkkinen2020, Zhang2020a, Ning2021}, though its numerical difficulties have hindered some applications such as pseudo-potential generation \cite{Bartok2019, Furness2019}. To address this, Bart\'ok and Yates proposed a regularized SCAN, termed ``rSCAN'', that aims to control numerical challenges while remaining as close to the original SCAN functional as possible. While the regularizations are effective in improving numerical performance, removing the grid sensitivity of SCAN that is problematic in some electronic structure codes, they break five of the exact conditions SCAN was designed to obey. Recent work by Mej\'ia-Rodr\'iguez and Trickey shows that some transferablity is lost in rSCAN, with atomization energies particularly degraded \cite{Mejia-Rodriguez2019, Bartok2019a}. We have recently proposed a restored-regularized-SCAN, called \rrscan, which maintains the regularizations of rSCAN while restoring exact constraint adherence \cite{Furness2020c}. Compared to SCAN, the resulting \rrscan functional has shown pronounced improvements in numerical efficiency, alongside small systematic improvements in accuracy \cite{Furness2020c, Mejia-Rodriguez2020f, Mejia-Rodriguez2020g, Ehlert2021b, Grimme2021,Ning2021}. On the extensive GMTKN55 test set \cite{Goerigk2017} for main-group chemistry, the overall error measure WTMAD-2 was \cite{Ehlert2021b} 8.6 kcal/mol for SCAN+D4 and 7.5 kcal/mol for \rrscan+D4, where D4 is a dispersion correction. Note that SCAN and \rrscan are not fitted to any bonded system, but are genuinely predictive for bonded systems. Applying SCAN+D4 to the unrestricted Hartree-Fock density \cite{Santra2021} instead of its own self-consistent density leads to a remarkably small WTMAD-2 of 5.079 kcal/mol, better than nearly all the (necessarily empirical) hybrid functionals tested thus far. This ``density correction'' \cite{Song2021} to SCAN also leads to a nearly-perfect many-body expansion and molecular dynamics for water \cite{Dasgupta2021}.

The present publication completes Ref \cite{Furness2020c}, providing the necessary detail of how each exact constraint was restored in \rrscan. We show how these restorations affect the numerical performance of the functional and how smoothness can be maintained for all constraints except the fourth order term of the slowly-varying density gradient expansion for exchange, which is less easy to enforce following the present interpolation-based model.
This work also provides context for the exact constraints enforced by SCAN, and demonstrates how a meta-GGA can be constructed to enforce those constraints.

This work builds upon Ref. \cite{Furness2020c} by expanding it to a progression of functionals (rSCAN, \rppscan, \rrscan, and \rfscan) that ultimately restore all the exact constraints obeyed by SCAN to a regularized form. Thus our presentation expands upon and completes the letter version of Ref. \cite{Furness2020c}, by filling in the details in the constructions of the last three functionals, and by presenting numerical results for \rppscan and \rfscan. We also present individual errors of these functionals on their appropriate norms (used to determine their parameters) in Table \ref{tab:norms}, and on the lattice constants of solids in Table \ref{tab:LC20_table}. Figures \ref{fig:fx_osc}, \ref{fig:setting_r4_params}, \ref{fig:Xe_derivatives}, and \ref{fig:G3_progression} will be familiar to readers of Ref \cite{Furness2020c}, but are expanded to incorporate results for the new meta-GGAs. We also include a more detailed analysis of the construction of \rrscan than was presented in Ref. \cite{Furness2020c}: in addition to a derivation of the \rrscan gradient expansion
in supplemental material A, %\ref{SM-AP:exc_deriv}
Fig. \ref{fig:d2_Set} shows how the damping factor $d_{p2}$ of \rrscan was determined. Variations of Figs. \ref{fig:alpha_comp} and \ref{fig:iefcomp} were presented in Ref. \cite{Furness2020c}; they are included here for completeness. The Tables in Appendices C--F report results for the same test sets considered in Ref. \cite{Furness2020c}, but including the novel meta-GGAs.

\section{\label{sec:theory} Constraint restoration}

\subsection{Coordinate scaling and uniform density limit \label{sec:theory_scaling}}

The SCAN functional is comprised of independent exchange and correlation functionals each constructed as an interpolation and extrapolation of two semi-local energy densities: one for single-orbital densities $\epsilon_\mr{x/c}^0$, and one for slowly-varying densities $\epsilon_\mr{x/c}^1$, where ``x/c'' is either exchange or correlation, respectively. Here, we will use $\epsilon$ to refer to the energy density, and $\varepsilon=\epsilon/n$ to refer to the energy per electron. The single-orbital and slowly-varying energy densities are joined by way of an interpolation function,
\begin{equation}
    \epsilon_\mr{x/c}^{\mr{SCAN}}(\v{r}) = \epsilon_\mr{x/c}^1(\v{r}) + f_\mr{x/c}(\alpha(\v{r}))\left[\epsilon_\mr{x/c}^0(\v{r}) - \epsilon_\mr{x/c}^1(\v{r})\right], \label{eq:scan_framework}
\end{equation}
which is controlled by the iso-orbital indicator variable,
\begin{equation}
    \alpha = \frac{\tau - \tau_W}{\tau_U}. \label{eq:alpha}
\end{equation}
$\alpha$ is built from three kinetic energy densities: the positive-definite conventional $\tau = 1/2\sum_i^{\mr{occ.}}|\nabla \phi_i|^2$ defined with the occupied Kohn-Sham orbitals \{$\phi_i$\}, von Weizs\"acker $\tau_W = |\nabla n|^2/(8n)$ that is the single-orbital limit of $\tau$ as a function of the electron density $n=\sum_i^{\mr{occ.}}|\phi_i|^2$, and $\tau_U = \frac{3}{10}k_{\mr{F}}^2\, n\, d_s(\zeta)$ the uniform electron gas limit of $\tau$. Here, the Fermi wavevector $k_{\mr{F}}=[3\pi^2 n]^{1/3}$, and $d_s(\zeta) = [(1 + \zeta)^{5/3} + (1 - \zeta)^{5/3}]/2$ is a function of the spin polarization $\zeta = (n_\uparrow - n_\downarrow)/(n_\uparrow + n_\downarrow)$. We refer to Refs. \cite{Becke1990, Ruzsinszky2012, Sun2012, Sun2013a, Sun2015, Furness2019, Furness2020a} for a detailed discussion of the properties of $\alpha$ and related quantities.

The first change made in rSCAN \cite{Bartok2019} is to regularize the iso-orbital indicator $\alpha$ to prevent divergence of the XC potential in the asymptotic regions of single orbital systems \cite{Furness2019}. In this region, the derivative $\partial\alpha/\partial\tau$ diverges faster than the decay of $\epsilon_{\mr{x}}^{\mr{LDA}}=-3k_{\mr{F}} n/(4\pi)$ (the local density approximation for exchange), resulting in a diverging exchange-correlation potential when $\partial\epsilon_\mr{xc}/\partial\alpha \neq 0$ \cite{Furness2019}, as is the case for SCAN. This diverging potential is problematic for pseudo-potential generation \cite{Bartok2019, Furness2019} and is avoided in rSCAN using a regularized $\alpha^\prime$,
\begin{align}
    \tilde\tau_U &= \left (\frac{3}{10}(3\pi^2)^{2/3}n^{5/3} + \tau_r \right)d_s(\zeta), \label{eq:regUEG}\\
    \tilde\alpha &= \frac{\tau - \tau_W}{\tilde\tau_U}, \label{eq:tildealpha}\\
    \alpha^\prime &= \frac{\tilde\alpha^3}{\tilde\alpha^2 + \alpha_r}.
\end{align}
The regularizing constants are $\tau_r = 10^{-4}$ and $\alpha_r = 10^{-3}$. Whilst successful in preventing the rSCAN potential from diverging, the $\alpha^\prime$ regularization breaks two exact constraints: the 1) uniform density limit, and 2) the uniform coordinate scaling of the exchange energy \cite{Levy1985}.

The exact uniform density limit is recovered in SCAN by recognizing,
\begin{equation}
    \lim_{|\nabla n|\to0} \tau = \lim_{|\nabla n|\to0}\tau_U,
\end{equation}
and,
\begin{equation}
    \lim_{|\nabla n|\to0} \tau_W = 0, \label{eq:tauwlim}
\end{equation}
hence,
\begin{equation}
    \lim_{|\nabla n|\to0} \alpha = 1.
\end{equation}
Then by construction,
\begin{equation}
    \lim_{|\nabla n|\to0} f_{\mr{x/c}}^{\mr{SCAN}}(\alpha) = 0,
\end{equation}
and Eq. \ref{eq:scan_framework} exclusively selects $\epsilon_{\mr{x}}^1$ and $\epsilon_{\mr{c}}^1$ in the uniform density limit. These energy densities satisfy the uniform density limit by design.

The uniform density limit is broken by the regularization parameters in $\alpha^\prime$ as,
\begin{align}
    &\lim_{|\nabla n|\to0}\tilde\tau_U \neq \lim_{|\nabla n|\to0} \tau , \\
    &\lim_{|\nabla n|\to0} \tilde\alpha = \frac{\lim_{|\nabla n|\to0}\tau_U}{\lim_{|\nabla n|\to0}\tilde\tau_U} \neq 1, \\
    &\lim_{|\nabla n|\to0} \alpha^\prime \neq 1,
\end{align}
hence,
\begin{equation}
    \lim_{|\nabla n|\to0} f_{\mr{x/c}}^{\mr{rSCAN}}(\alpha^\prime) \neq 0.
\end{equation}
The final inequality results in a slight scaling of $\epsilon_{\mr{x/c}}^1$ and a small inclusion of $\epsilon_\mr{x/c}^0$, which does not recover the uniform density limit. Hence the constraint is broken. The uniform density limit is important for metallic elements. For a uniform electron gas of density parameter $\rs=4$ (roughly characteristic of the valence electron density in solid sodium) $\alpha'\approx 0.719$, for example.

The regularized uniform electron gas kinetic energy density, $\tilde\tau_U$, also causes the exchange energy density to scale incorrectly under the uniform coordinate scaling transformations. To see this, we define a uniform coordinate scaling of the density $n$ and Kohn-Sham orbital $\phi_i$ as
\begin{align}
    n_\lambda(\bm{r}) &= \lambda^3 n(\lambda \bm{r}), \\
    \phi_{i,\lambda}(\bm{r}) &= \lambda^{3/2}\phi_i(\lambda \bm{r}),
\end{align}
with $\lambda \geq 0$, such that the standard meta-GGA variables scale as,
\begin{align}
    \tau_\lambda (\bm{r}) &= \lambda^5 \frac{1}{2} \sum_i^{\mr{occ.}}\left| \frac{\partial \phi_i(\lambda \bm{r})}{\partial (\lambda \bm{r})} \right|^2 = \lambda^5 \tau (\lambda \bm{r}) \\
    \tau_{W,\lambda}(\bm{r}) &= \lambda^5 \tau_W(\lambda \bm{r}) \\
    \tau_{U,\lambda}(x,y,z) &= \lambda^{5}\tau_U(\lambda \bm{r}).
\end{align}
Thus, while $\alpha(\bm{r}) \to \alpha(\lambda \bm{r})$ under uniform coordinate scaling, $\tilde \alpha$ does not have this correct behavior except in the limit $\lambda \to \infty$, because the regularization $\tau_r$ in $\tilde\tau_U$ doesn't vary with the coordinate scaling parameter $\lambda$. This clearly violates the uniform coordinate scaling behavior of the exchange energy \cite{Levy1985}
\begin{equation}
  E_\mr{x}[n_{\lambda}] = \lambda E_\mr{x}[n],
\end{equation}
as $\tilde \alpha$ does not scale correctly, and the exchange and correlation models here are highly nonlinear in $\tilde \alpha$.
The exact correlation energy evaluated on a uniformly scaled density tends to distinct limits \cite{Levy1991,Levy1993}
\begin{align}
    \lim_{\lambda \to \infty} E_\mr{c}[n_\lambda] &= \text{constant} \leq 0  \\
    \lim_{\lambda \to 0} E_\mr{c}[n_\lambda] &= \lambda D_\mr{c}[n],
\end{align}
where the constant and functional $D_\mr{c}$ are unknown. It can be seen that SCAN, and the functionals presented here satisfy both limits, but rSCAN satisfies only the $\lambda \to \infty$ limit.

It should be noted that under nonuniform coordinate scaling, rSCAN does not violate known exact constraints \cite{Levy1991,Gorling1992,Pollack2000} because of the robustness of the underlying SCAN model. It does, however, tend to distinct limits from SCAN, likely impacting performance for real systems. To see this, we define a non-uniform coordinate scaling of the density $n$ and Kohn--Sham orbital $\phi_i$ in one dimension as,
\begin{align}
    n_\lambda^{x}(x, y, z) &= \lambda n(\lambda x, y, z), \label{eq:nus_dens} \\
    \phi_{i,\lambda}^x(x, y, z) &= \lambda^{1/2}\phi_i(\lambda x, y, z), \label{eq:nus_ks_orb}
\end{align}
again with $\lambda \geq 0$.
Under this coordinate transformation, the exact exchange energy satisfies \cite{Levy1991,Gorling1992}
\begin{align}
    \lim_{\lambda \to 0}\frac{1}{N} E_\mr{x}[n_\lambda^x] &> -\infty\\
    \lim_{\lambda \to \infty} \frac{1}{N} E_\mr{x}[n_\lambda^x] &> -\infty, \label{eq:nus_ex_inf}
\end{align}
with $N$ the number of electrons.
Identical inequalities hold for the exact correlation energy \cite{Gorling1992,Pollack2000}.
It should be emphasized that these constraints imply that the exact exchange and correlation energies per electron tend to finite constants under either limit of non-uniform coordinate scaling.
The constant limit for Eq. \ref{eq:nus_ex_inf} is a non-zero negative constant \cite{Pollack2000}, the exchange energy per electron for a two-dimensional system.

To recover these constraints on an approximate exchange energy functional, the exchange enhancement factor $F_\mr{x} = \epsilon_\mr{x}/\epsilon_\mr{x}^\text{LDA}$ must satisfy \cite{Chiodo2012,Perdew2014}
\begin{equation}
    \lim_{p\to \infty} F_\mr{x} \propto p^{-1/4}.
\end{equation}
$p = [|\nabla n|/(2k_\mr{F} n)]^2$ is the square of a dimensionless gradient of the density on the appropriate length scale for the exchange energy.
We will discuss $p$ further in the ensuing section on gradient expansions, but it suffices here to note that $p$ scales as $\lambda^{-2/3}$ as $\lambda \to 0$, and as $\lambda^{4/3}$ as $\lambda \to \infty$, as shown in
supplemental material B. %\ref{SM-AP:non_unif_scl}.
Therefore, $p$ is always divergent under the extreme limits of non-uniform coordinate scaling.
In SCAN, rSCAN, and the functionals developed here, the set of coordinate scaling constraints for exchange are imposed through a function $g_\mr{x}(p)$
\begin{align}
    F_\mr{x}(n,|\nabla n|,\tau) &= \left\{h_{\mr{1x}} + f_\mr{x}(\alpha)\left[h_{0\mr{x}} - h_{\mr{1x}}\right ]\right\} g_\mr{x}(p) \\
    g_\mr{x}(p) &= 1 - \exp[-a_1p^{-1/4}].
\end{align}
Referring to Eq. (\ref{eq:scan_framework}), it can be seen that $h_\mr{jx}=\epsilon_\mr{x}^j/[\epsilon_\mr{x}^\text{LDA}g_\mr{x}]$, with $j=0,1$.
In the limit $p\to \infty$,
\begin{align}
    \lim_{p\to \infty} g_\mr{x}(p) &= a_1 p^{-1/4} + \mathcal{O}(p^{-1/2}) \\
    \lim_{p\to \infty} h_\mr{xj} &= \text{constant} > 0, \quad j = 0,1.
\end{align}
Thus $F_\mr{x}\sim p^{-1/4}$.
In SCAN, rSCAN, and this work, $h_\mr{x0}=1.174$ identically, and $h_\mr{x1}(p\to\infty)=1.065$.
The LDA, most GGAs, and most meta-GGAs do not recover the right asymptotic behavior for exchange.

Recovering the analogous set of non-uniform coordinate scaling constraints for correlation is more straightforward, and requires that, for $j=0,1$,
\begin{align}
    \lim_{p \to \infty} \epsilon_\mr{c}^j &= \text{constant} \leq 0 \\
    \lim_{\rs \to 0} \epsilon_\mr{c}^j &= \text{constant} \leq 0,
\end{align}
where $\rs = [3/(4\pi n)]^{1/3}$ is the Wigner-Seitz radius.
In SCAN, rSCAN, and the functionals developed here, both constants are chosen to be zero.
Many non-empirical GGAs and meta-GGAs for correlation satisfy the non-uniform coordinate scaling constraints.
LDA, which has a logarithmic divergence as $\rs \to 0$, does not.

We can now consider the iso-orbital indicators used in SCAN, rSCAN, and \rrscan. Under the non-uniform coordinate scaling defined in Eqs. \ref{eq:nus_dens} and \ref{eq:nus_ks_orb}, the standard meta-GGA variables scale as
\begin{align}
    \tau_\lambda^x (x, y, z) &= \frac{\lambda}{2} \sum_i^{\mr{occ.}}\left|\hat{\bm{x}}\lambda \frac{\partial \phi_i(\lambda x, y, z)}{\partial(\lambda x)} + \nabla_\perp \phi_i(\lambda x, y, z)\right|^2 \label{eq:nus_tau} \\
    \tau_{W,\lambda}^x(x,y,z) &= \lambda \frac{\left|\hat{\bm{x}}\lambda\frac{\partial n(\lambda x, y, z)}{\partial(\lambda x)} + \nabla_\perp n(\lambda x, y, z)\right|^2}{8n(\lambda x, y, z)} \\
    \tau_{U,\lambda}^{x}(x,y,z) &= \lambda^{5/3}\tau_U(\lambda x, y, z), \label{eq:nus_tauu}
\end{align}
with
\[
 \nabla_\perp = \hat{\bm{y}} \frac{\partial}{\partial y} + \hat{\bm{z}} \frac{\partial}{\partial z}.
\]
From these equations, we see that, when $\lambda \to 0$, $\tau$ and $\tau_W$ scale as $\lambda$, but $\tau_U$ scales as $\lambda^{5/3}$.
Thus, $\alpha$ scales with a leading order of $\lambda^{-2/3}$ in this limit.
When $\lambda \to \infty$, $\alpha$ can either scale as $\lambda^{4/3}$ or $\lambda^{-2/3}$. Examples and analysis of both scaling limits are given in
supplemental material B.%\ref{SM-AP:non_unif_scl}.}
%From these we see that $\tau$ and $\tau_W$ scale with $\lambda$ at the same rate, but $\tau_U$ scales more strongly as $\lambda^{5/3}$. Thus, $\alpha$ scales with a leading order of $\lambda^{-2/3}$.

Due to the $\tau_r$ constant in the denominator of $\tilde\alpha$ (Eq. \ref{eq:tildealpha}), the leading order behavior of the regularized $\tilde \alpha$ under non-uniform scaling, with $\lambda \to 0$, is $\lambda$. Then, whereas $\alpha$ tends to infinity in the $\lambda \to 0$ limit, $\tilde \alpha$ tends to zero. $\tilde \alpha$ has the correct leading-order behavior in the limit $\lambda \to \infty$, which can be a single-orbital limit where exact constraints, including the finite exchange and correlation energies under the nonuniform coordinate scaling, were built for SCAN.

In Ref. \cite{Furness2021a}, we proposed an alternative regularization of $\alpha$ to restore these constraints,
\begin{equation}
    \bar{\alpha} = \frac{\tau - \tau_W}{\tau_U + \eta\tau_W} = \frac{\alpha}{1 + \eta \frac{5}{3}p}, \label{eq:alphabar} \\
\end{equation}
where $\eta$ is a regularization parameter to be determined later.
Clearly, $\bar\alpha$ has the correct behavior, $\bar\alpha(\br)\to \bar\alpha(\lambda\br)$ under uniform coordinate scaling.
This regularization eliminates the asymptotic region ($|\br|\to\infty$) divergence
%This regularization controls the asymptotic region divergence through the slower decay of $\tau_W$
and, by Eq. (\ref{eq:tauwlim}), does not change the uniform density limit,
\begin{equation}
    \lim_{|\nabla n|\to 0} \bar\alpha = 1.
\end{equation}
The nonuniform coordinate scaling of $\bar\alpha$ is also maintained as $\lambda^{-2/3}$ to leading order in the $\lambda \to 0$ limit. But, for $\lambda \to \infty$, the leading-order term of $\bar\alpha$ is independent of $\lambda$, for non-homogeneous densities. This is demonstrated in supplemental material B.%\ref{SM-AP:non_unif_scl}.

\begin{figure}[t]
    \centering
    \includegraphics[width=0.45\textwidth]{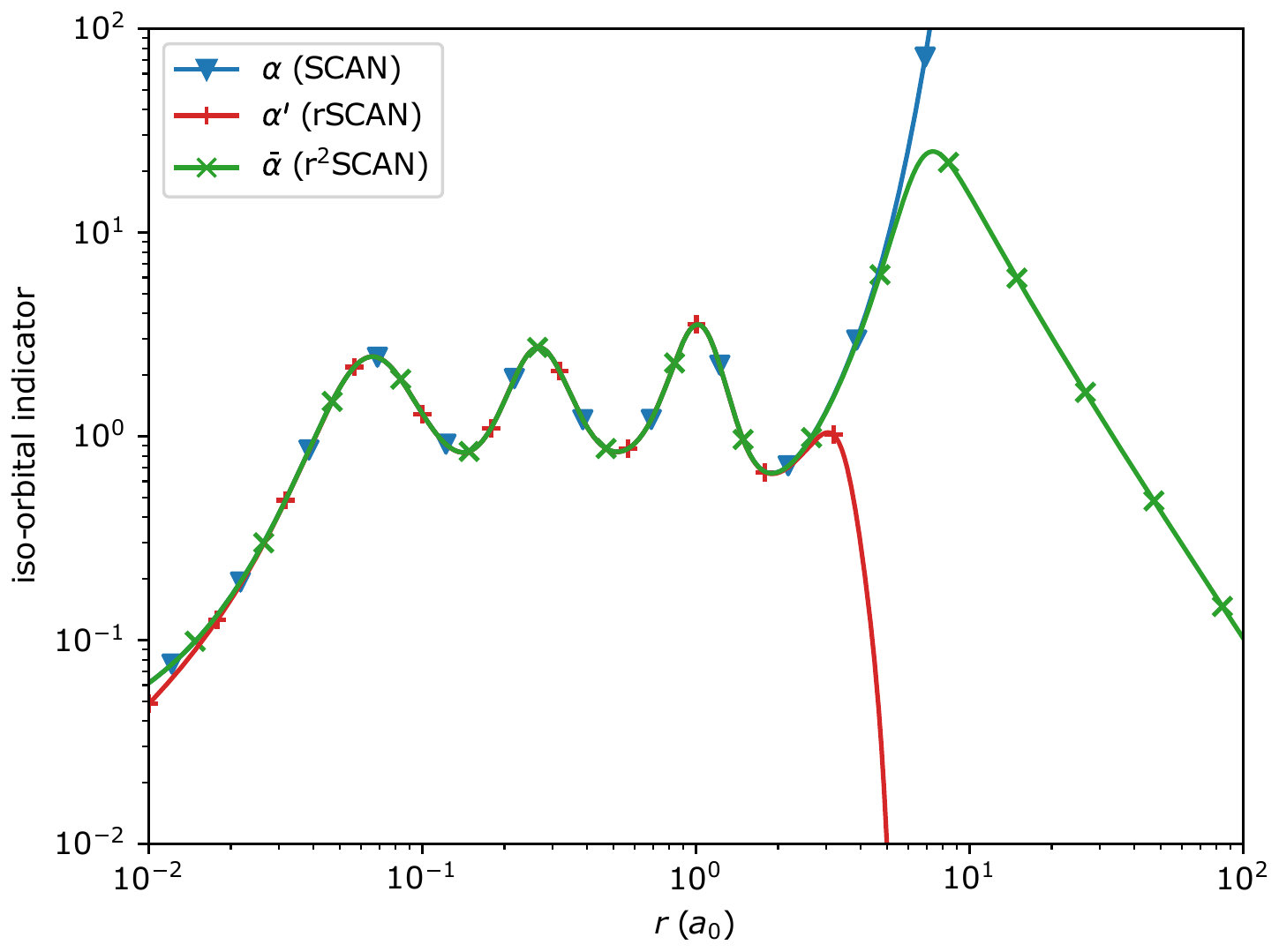}
    \caption{\label{fig:alpha_comp} Comparison between the conventional $\alpha$ (e.g. from SCAN), $\alpha^\prime$ (from rSCAN), and the new $\bar{\alpha}$ as a function of distance from the Kr nucleus (in Bohr radii) computed from accurate spherical Hartree--Fock Slater type orbitals \cite{Clementi1974, Furness2021a}. Regularization parameters are $\tau_r = 10^{-4}$ and $\alpha_r = 10^{-3}$ in $\alpha^\prime$, and $\eta = 10^{-3}$ in $\bar{\alpha}$.}
\end{figure}

Figure \ref{fig:alpha_comp} shows a comparison of $\alpha$, $\alpha^\prime$, and $\bar{\alpha}$ for the krypton atom. The divergence of the conventional $\alpha$ is apparent in the asymptotic region, while $\alpha^\prime$ and $\bar{\alpha}$ decay to 0. Close to the nucleus, the $\alpha_r$ regularization constant causes $\alpha^\prime$ to behave differently to $\alpha$ and $\bar{\alpha}$, but otherwise all three indicators behave similarly.

Substituting $\bar\alpha$ for $\alpha^\prime$ in rSCAN is sufficient to restore the uniform density limit and coordinate scaling behaviors, and we refer to rSCAN with this replacement as ``r++SCAN'' throughout.

\subsection{Gradient expansions\label{sec:gradexp}}

\begin{figure}
    \centering
    \includegraphics[width=0.45\textwidth]{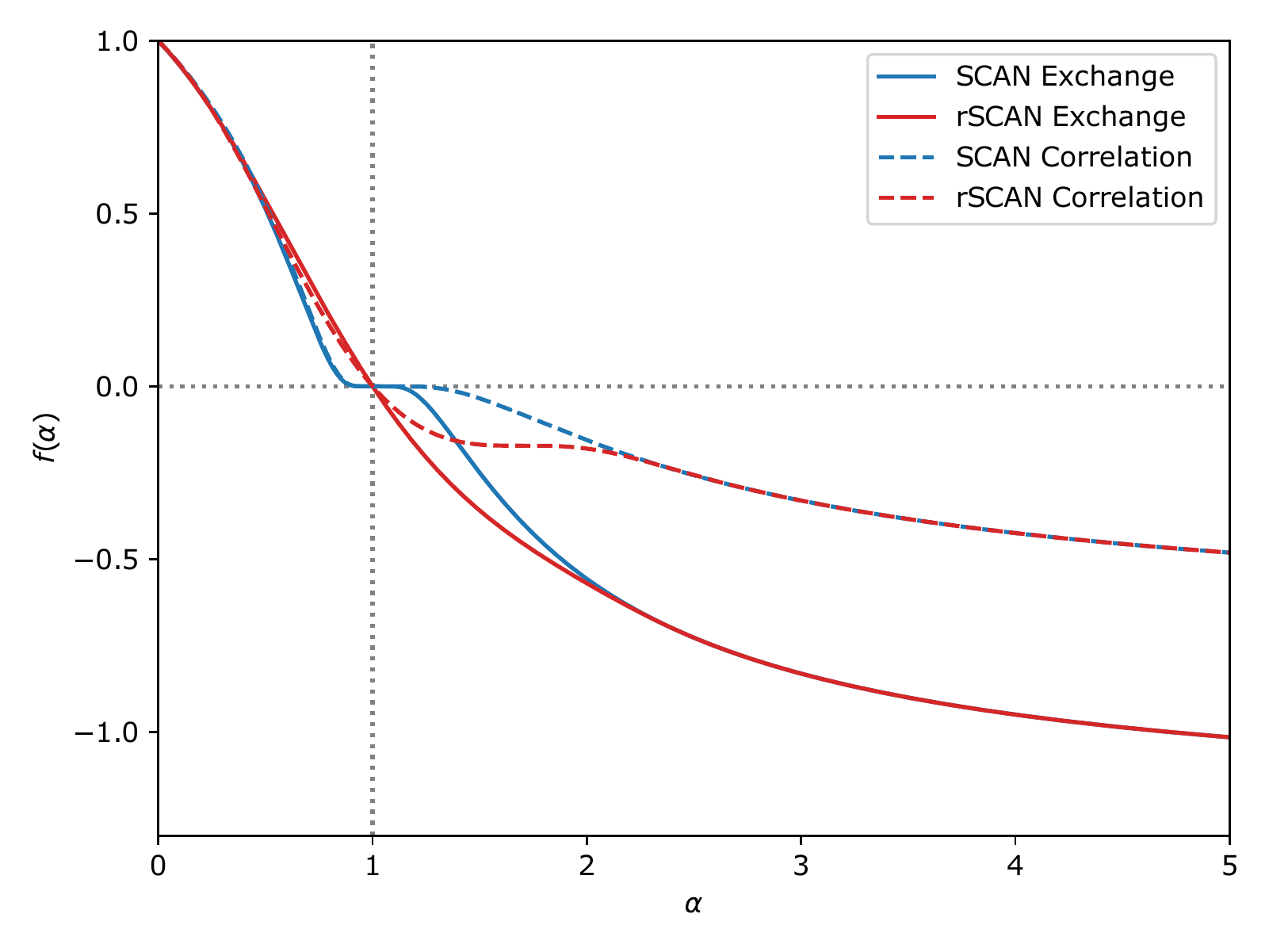}
    \caption{The exchange (solid) and correlation (dashed) interpolation functions for the SCAN (blue, Eq. \ref{eq:scanf}) and rSCAN (orange, Eq. \ref{eq:rscan_f}) functionals.}
    \label{fig:iefcomp}
\end{figure}

The interpolative design of SCAN allows construction of the single-orbital ($\epsilon^0$) and slowly-varying ($\epsilon^1$) energy densities that consider only the exact constraints relevant to their respective limits. In SCAN, the interpolation function is a piece-wise combination of two exponential terms,
\begin{equation}
    f_\mr{x/c}(\alpha) =
    \begin{cases}
        \exp[\frac{-c_{1\mr{x/c}}\alpha}{1 - \alpha}] & \alpha \leq 1 \\
        -d_\mr{x/c}\exp[\frac{c_{2\mr{x/c}}}{1 - \alpha}] & \alpha > 1,
    \end{cases}
    \label{eq:scanf}
\end{equation}
where $\{d_{\mr{x/c}}, c_\mr{1x/c}, c_\mr{2x/c}\}$ are separate parameters for exchange and correlation determined by fitting to appropriate norms \cite{Sun2015}. This function was chosen such that: 1) $f(\alpha = 0) = 1$ exclusively selects $\epsilon^0$ for single-orbital densities, 2) $f(\alpha = 1) = 0$ exclusively selects $\epsilon^1$ in slowly-varying densities, and 3)
\begin{equation}
    \left.\frac{df(\alpha)}{d\alpha}\right |_{\alpha \to 1} = \left.\frac{d^2f(\alpha)}{d\alpha^2}\right |_{\alpha \to 1} = 0,\label{eq:interp_const}
\end{equation}
which prevents $\epsilon^0$ contributing to the slowly-varying density gradient expansions to the 4th order in $|\nabla n|$. Note $\left.\frac{d^mf(\alpha)}{d\alpha^m}\right |_{\alpha \to 1} = 0$ with $m$ to be any integer in Eq. \ref{eq:scanf} by design. While theoretically convenient, Eq. \ref{eq:scanf} introduces a twist into the function around $\alpha = 1$, see Figure \ref{fig:iefcomp}. This twist destroys the overall smoothness of the functional, introduces oscillations into the XC potential \cite{Yang2016, Bartok2019}, and harms its performance on numerical integration grids.

The rSCAN functional uses a smooth polynomial interpolation function in place of the SCAN piece-wise exponential for the range $0 \leq \alpha^{\prime} < 2.5$,
\begin{equation}
    f(\alpha^\prime) = \begin{cases}
        \sum_{i=0}^7c_{i}\alpha^{\prime \, i} & 0 \le \alpha^\prime \leq 2.5 \\
        -d_\mr{x/c}\exp[\frac{c_{2\mr{x/c}}}{1 - \alpha^\prime}] & \alpha^\prime > 2.5
        \end{cases},\label{eq:rscan_f}
\end{equation}
where $\{c_i\}$ are polynomial coefficients determined to smoothly join $f^\mr{SCAN}(\alpha^\prime)$ at $\alpha^\prime = 0$ and $2.5$, see Ref. \citenum{Bartok2019}. A comparison of the two interpolation functions is shown in Figure \ref{fig:iefcomp}. While this replacement smooths the XC energy density and potential, it breaks the third constraint on the interpolation function, and hence $\epsilon^0$ makes spurious contributions to the slowly-varying gradient density expansion of $\epsilon^1$. Here, we restore the correct expansions by directly subtracting the extra terms that result from breaking the third condition (Eq. (\ref{eq:interp_const})) around $p \to 0$ and $\alpha \to 1$ where the expansion is relevant.

The exact gradient expansion for exchange around the slowly-varying density limit to the 2nd order (GE2X) and to the 4th order (GE4X) was derived in terms of the exchange enhancement factor $F_\mr{x}$ in Refs. \cite{Kirzhnits1957, Svendsen1996} as,
\begin{equation}
    \lim_{|\nabla n|\to 0} F_{\mr{x}}^{\mr{GE}} = 1 + \mu p + \frac{146}{2025}\tilde q^2 - \frac{73}{405}p\tilde q + \mo{6}, \label{eq:GE_x}
\end{equation}
where $\mu = 10/81$ and $\tilde q = (9/20)(\bar\alpha - 1) + (\eta 3/4 + 2/3)p$ recovers the reduced density Laplacian at $|\nabla n| \to 0$. The gradient expansion for correlation was derived to the 2nd order (GE2C) in Ref. \cite{Ma1968,Wang1991,Perdew1996,Sun2015} as,
\begin{equation}
    \varepsilon_{\mr{c}} = \varepsilon_{\mr{c}}^{\mr{LSDA}} + \beta(\rs)\phi(\zeta)^3 t(\rs, \zeta, p)^2 + \mo{4},
\end{equation}
where $\phi(\zeta) = [(1 + \zeta)^{2/3} + (1 - \zeta)^{2/3}]/2$ and $t(\rs, \zeta, p) = (3\pi^2/16)^{1/3} \sqrt{p/\rs}/\phi(\zeta)$. We will restore each expansion to r++SCAN in turn.

\subsubsection{Exchange}

The SCAN exchange enhancement factor for a spin-unpolarized system is,
\begin{align}
    F_\mr{x}^{\mr{SCAN}}&(p,\alpha) \\
    &= \left\{h_{\mr{1x}}(p, \alpha) + f_\mr{x}(\alpha)\left[h_{0\mr{x}} - h_{\mr{1x}}(p,\alpha)\right ]\right\} g_\mr{x}(p),\nonumber \\
    g_\mr{x}(p) &= 1 - \exp[-a_1p^{-1/4}], \\
    h_{\mr{0x}} &= 1 + \kappa_0 = 1.174, \\
    h_\mathrm{1x}(p,\alpha) &= 1 + \kappa_1 - \frac{\kappa_1}{1 + \frac{x(p,\alpha)}{\kappa_1}},\\
    x(p, \alpha) &= \mu p\left[1 + \left(\frac{b_4p}{\mu}\right)\exp\left(\frac{-|b_4|p}{\mu}\right)\right] \nonumber\\
    &+ \left\{b_1 p + b_2(1 - \alpha)\exp\left[-b_3(1 - \alpha)^2\right]\right\}^2, \label{eq:scan_x(pa)}
\end{align}
where $\mu = 10/81$ and $\{b_1, b_2, b_3, b_4\}$ are chosen such that SCAN yields GE2X and GE4X, noting that the expansion of $g_\mr{x}(p)$ around $p=0$ has only zero-order contributions (see supplemental material A 3).%\ref{SM-AP:r4_x_deriv}).

In rSCAN, the interpolation function derivatives are non-zero at $\bar\alpha = 1$, so $h_\mr{0x}$ also contributes to the gradient expansion. This changes the functional's gradient expansion, spoiling the correct gradient expansion of the $x(p,\alpha)$ inherited from SCAN. Thus both rSCAN and r++SCAN fail to recover GE2X and GE4X.

We restore GE2X to give the \rrscan exchange functional by redesigning $x(p,\alpha)$ as,
\begin{equation}
    x(p) = \left(C_\eta C_{2\mr{x}} \exp[-p^2/d_{p2}^4] + \mu\right)p, \label{eq:x_change}
\end{equation}
where $C_\eta$ and $C_{2\mr{x}}$ are constants set to cancel spurious contributions from $h_{\mr{0x}}$ to the 2nd order in $\nabla n$. The restoring constants are multiplied by the damping function $\exp[-p^2/d_{p2}^4]$ to prevent them dominating as $p$ becomes large. The damping parameter $d_{p2}$ derives from scaling the reduced density gradient as $s \to s/d_{p2}$ with $d_{p2}$ fit to recover the appropriate norms in Section \ref{sec:settingParams}.

To find $C_\eta$ and $C_{2\mr{x}}$ we take the Taylor expansion of the rSCAN interpolation function (Eq. \ref{eq:rscan_f}) around $\bar\alpha = 1$, noting that $1-\bar\alpha$ is $\mo{2}$,
\begin{align}
    \lim_{|\nabla n|\to0} &f^{\mr{rSCAN}}(\bar\alpha)\label{eq:lim_f} \\
    &= -(1 - \bar{\alpha})\Delta f_2 + \frac{(1 - \bar{\alpha})^2}{2}\Delta f_4 + \mo{6}, \nonumber
\end{align}
where,
\begin{align}
    \Delta f_\mathrm{2} &= \sum_{i=1}^7 ic_{i}, \label{eq:del_f2} \\
	\Delta f_\mathrm{4} &= \sum_{i=2}^7i(i - 1)c_{i},
\end{align}
are determined by the first and second derivatives of the interpolation function with respect to $\bar{\alpha}$ respectively.

The $(1 - \bar\alpha)$ term of Eq. \ref{eq:lim_f} indicates a fixed slope for the $\bar\alpha$ dependence of the enhancement factor across the slowly-varying limit that is found to be numerically problematic, analyzed further in Section \ref{sec:numerics}. This can be avoided to second order by expressing $(1 - \bar\alpha)$ in terms of $p$ through an integration by parts on the exchange energy density \cite{Sun2015a},
\begin{equation}
    \lim_{|\nabla n|\to 0}(1 - \bar\alpha) = \left(\frac{20}{27} + \eta\frac{5}{3}\right)p + \mo{4}. \label{eq:oma_to_p}
\end{equation}
This substitution, derived and discussed in supplemental material A 2, %\ref{SM-AP:r2_x_deriv},
is used in r$^2$SCAN and identifies,
\begin{equation}
    C_\eta = \left(20/27 + \eta5/3\right). \label{eq:c_eta_expr}
\end{equation}

To second order the slowly-varying gradient expansion of r$^2$SCAN is then,
\begin{equation}
    \lim_{|\nabla n|\to0} F_\mr{x}^{\mr{r^2SCAN}} = \lim_{|\nabla n|\to0} h_\mr{1x} - C_\eta p \Delta f_2\left (h_\mr{0x} - \lim_{|\nabla n|\to0}h_\mr{1x}\right).
\end{equation}
Finding,
\begin{equation}
    \lim_{|\nabla n|\to0}h_\mr{1x} = 1 + \left(\mu + C_\eta C_{2\mr{x}}\right)p + \mo{4},
\end{equation}
and collecting terms gives,
\begin{align}
    \lim_{|\nabla n|\to0} &F_\mr{x}^{\mr{r^2SCAN}}\\
    &= 1 + \mu p + C_\eta \left[ C_{2\mr{x}} - \Delta f_2 h_{\mr{0x}} + \Delta f_2 \right]p + \mo{4}, \nonumber
\end{align}
equating this to GE2X (second order and below terms of Eq. \ref{eq:GE_x}) and solving for $C_{2\mr{x}}$ gives,
\begin{equation}
    C_{2\mr{x}} = -\Delta f_2(1 - h_\mr{0x}) \approx -0.162742, \label{eq:c_2x_expr}
\end{equation}
as shown in supplemental material A1.%\ref{SM-AP:r2_x_deriv}.

GE4X can be restored to give the ``r$^4$SCAN'' functional by including a further correcting term in the exchange enhancement factor outside the interpolation. This introduces three more constants, derived in supplemental material A 3, %\ref{SM-AP:r4_x_deriv}
for all terms in Eq. \ref{eq:GE_x}:
  \begin{align}
    F_\mr{x}^\mr{r^4SCAN}(p, \ba) &= \left\{h_\mr{1x}(p) + f_\mr{x}(\bar{\alpha})\left [h_\mr{0x} - h_\mr{1x}(p)\right ] \right. \nonumber \\
    & \left. +\Delta F_4(p,\ba) \right\} g_\mr{x}(p) \label{eq:full_delfx} \\
    \Delta F_4(p,\ba) &= \left\{C_{2\mr{x}}\left[(1-\ba) - C_\eta p\right] + C_{\ba\ba}(1-\ba)^2 \right. \nonumber \\
    & \left. +C_{p\ba}p(1-\ba) + C_{pp}p^2 \right\}\Delta F_4^{\mr{damp}}(p,\ba) \\
    \Delta F_4^{\mr{damp}}(p,\ba) &= \frac{2\ba^2}{1+\ba^4} \exp\left[-\frac{(1-\ba)^2 }{d_{\ba 4}^2} - \frac{p^2}{d_{p4}^4}\right] \label{eq:delf4}\\
  	C_{\ba\ba} &= \frac{73}{5000} - \frac{\Delta f_4}{2}[h_\mr{0x}-1] \approx -0.0593531 \label{eq:c_ba_ba}\\
  	C_{p\ba} &= \frac{511}{13500} - \frac{73}{1500}\eta - \Delta f_2[C_\eta C_{2\mr{x}} + \mu] \nonumber \\
    & \approx 0.0402684 \label{eq:c_p_ba}\\
  	C_{pp} &= \frac{146}{2025}\left\{\eta\frac{3}{4} + \frac{2}{3}\right\}^2 - \frac{73}{405}\left\{\eta\frac{3}{4} + \frac{2}{3}\right\} \nonumber \\
    & + \frac{\left(C_\eta C_{2\mr{x}} + \mu\right)^2}{k_1} \approx -0.0880769. \label{eq:c_p_p}
  \end{align}
Further damping functions, $\Delta F_4^{\mr{damp}}$, are included to prevent the correction terms dominating as $(1 - \bar\alpha)$ and $p$ become large, introducing $d_{\bar\alpha 4}$ and $d_{p4}$ as additional parameters. These are again set to recover the appropriate norms in Section \ref{sec:settingParams}. For the fourth order expansion, the integration by parts substitution of Eq. \ref{eq:oma_to_p} cannot be applied, and hence $(1 - \bar\alpha)$ cannot be removed.

\subsubsection{Correlation}

%% broke this long stretch of equations up because of the excessive whitespace using a single align command
The SCAN model of the correlation energy per electron $\varepsilon\smc = \epsilon\smc/n$ is
\begin{equation}
    \varepsilon_\mr{c}^{\mr{SCAN}} = \varepsilon_\mr{c}^1 + f_\mr{c}^{\mr{SCAN}}(\alpha)\left[\varepsilon_\mr{c}^0 - \varepsilon_\mr{c}^1\right],
\end{equation}
with the $\alpha=0$ correlation energy per electron given by
\begin{align}
    \varepsilon_\mr{c}^0 &= \left(\varepsilon_\mr{c}^\mr{LDA0} + H_\mr{c}^0\right)g_\mr{c}(\zeta), \\
    \varepsilon_\mr{c}^\mr{LDA0} &= -\frac{b_{1\mr{c}}}{1 + b_{2\mr{c}}\sqrt{\rs} + b_{3\mr{c}}\rs}, \\
    H_\mr{c}^0 &= b_{1\mr{c}}\ln\{1 + w_0[ 1 - g_\infty(p)]\}, \\
    w_0 &= \exp[-\varepsilon_\mr{c}^{\mr{LDA0}}/b_{1\mr{c}}] - 1, \\
    g_\infty(p) &= (1 + 4\chi_\infty p)^{-1/4},\\
    g_\mr{c}(\zeta) &= \{ 1 - 2.3631[d_\mr{x}(\zeta) - 1]\}(1 - \zeta^{12}),
\end{align}
with $d_\mr{x}(\zeta) = [(1 + \zeta)^{4/3} + (1 - \zeta)^{4/3}]/2$.

Similarly, the $\alpha=1$ limit is given by
\begin{align}
    \varepsilon_\mr{c}^{1} &=  \varepsilon_\mr{c}^{\mr{LSDA}} + H_1, \\
    H_\mr{c}^1 &= \gamma\phi^3\ln\{ 1 + w_1[1 - g(y)]\}, \\
    w_1 &= \exp\left[-\frac{\varepsilon_\mr{c}^{\mr{LSDA}}}{\gamma\phi^3}\right] - 1, \\
    g(y) &= (1 + 4y)^{-1/4}, \\
    y &= \frac{\beta(\rs)}{\gamma w_1}t^2, \\
    \beta(\rs) &= \beta_\mr{MB}\frac{1 + A \rs}{1 + B \rs},
\end{align}
where $b_\mr{1c} = 0.0285764$, $b_\mr{2c} = 0.0889$, $b_\mr{3c} = 0.125541$, $\chi_\infty = 0.128026$ $\gamma = 0.0310907$, $\beta_{\mr{MB}} = 0.066725$, $A = 0.1$, and $B = 0.1778$. $\varepsilon_\mr{c}^\mr{LSDA}$ is the local spin-density approximation for correlation from Ref. \cite{Perdew1992a}.

As r++SCAN takes the same correlation model, the violation of Eq. \ref{eq:interp_const} by the rSCAN interpolation function breaks GE2C. The GE2C correction terms are restored to r++SCAN by replacing $g(y)$ in $\epsilon_{\mr{c}}^1$ with,
\begin{align}
    g(y, \Delta y) =& \left [1 + 4(y - \Delta y)\right ]^{-1/4}, \label{eq:g(y_dy)}\\
    \Delta y =& \frac{C_{2\mr{c}}}{27 \gamma d_s(\zeta) \phi^3 w_1} \left\{20\rs\left[g_\mr{c}(\zeta)\frac{\partial \varepsilon_c^{\text{LDA0}}}{\partial \rs} - \frac{\partial \varepsilon_c^{\text{LSDA}}}{\partial \rs} \right] \right. \nonumber \\
    & \left. - 45\eta\left[\varepsilon_c^{\text{LDA0}}g_\mr{c}(\zeta) - \varepsilon_c^{\text{LSDA}}\right] \vphantom{\frac{\partial \varepsilon_c^{\text{LSDA}}}{\partial \rs}} \right\} p \exp[-p^2/d_{p2}^4] \label{eq:del_y}
\end{align}
where the damping function $\exp[-p^2/d_{p2}^4]$ is the same as in Eq. \ref{eq:x_change}. Similarly to exchange, we restore the second order slowly-varying density gradient expansion when,
\begin{equation}
    C_{2\mr{c}} = \Delta f_2 \approx -0.711402.
\end{equation}
The derivation of this expression is shown in supplemental material A 4.%\ref{SM-AP:r2_c_deriv}.

Making these replacements to r++SCAN gives the ``r$^{2}$SCAN'' functional which only breaks GE4X. Including the full correction of Eq. \ref{eq:full_delfx} gives the ``r$^4$SCAN'' functional, which obeys all the exact constraints SCAN does. For convenience, a collected definition of the working equations for all new functionals is given in supplemental material C.%\ref{SM-AP:eqns}.

\subsubsection{Summary of Changes}

Here, we summarize the changes made from SCAN for each of the functionals.

\begin{enumerate}
    \item{rSCAN replaces $\alpha$ with $\alpha^\prime$, which contains two regularization parameters, $\tau_r=10^{-4}$ and $\alpha_r = 10^{-3}$. It also replaces the SCAN interpolation function with a polynomial between $0 \leq \alpha^\prime \leq 2.5$.}

    \item{r++SCAN evolves from rSCAN, by replacing $\alpha^\prime$ with $\bar\alpha$ that uses only a single regularization parameter, $\eta = 0.001$.}

    \item{r$^2$SCAN inherits all the changes of r++SCAN. Additionally, for exchange, it replaces $x(p, \alpha)$ in $h_{1\mr{x}}$ with Eq. \ref{eq:x_change}. For correlation, it replaces $g(y)$ with Eqs. \ref{eq:g(y_dy)} and \ref{eq:del_y} in $\epsilon_\mr{c}^1$.}

    \item{r$^4$SCAN inherits all the changes from r$^2$SCAN. Additionally, for exchange, it replaces $F_\mr{x}$ with Eq. \ref{eq:full_delfx}, which introduces $\Delta F_4$ of Eq. \ref{eq:delf4}.}
\end{enumerate}

\section{\label{sec:numerics}Numerical Challenges}

\begin{figure}
    \centering
    \includegraphics[width=0.45\textwidth]{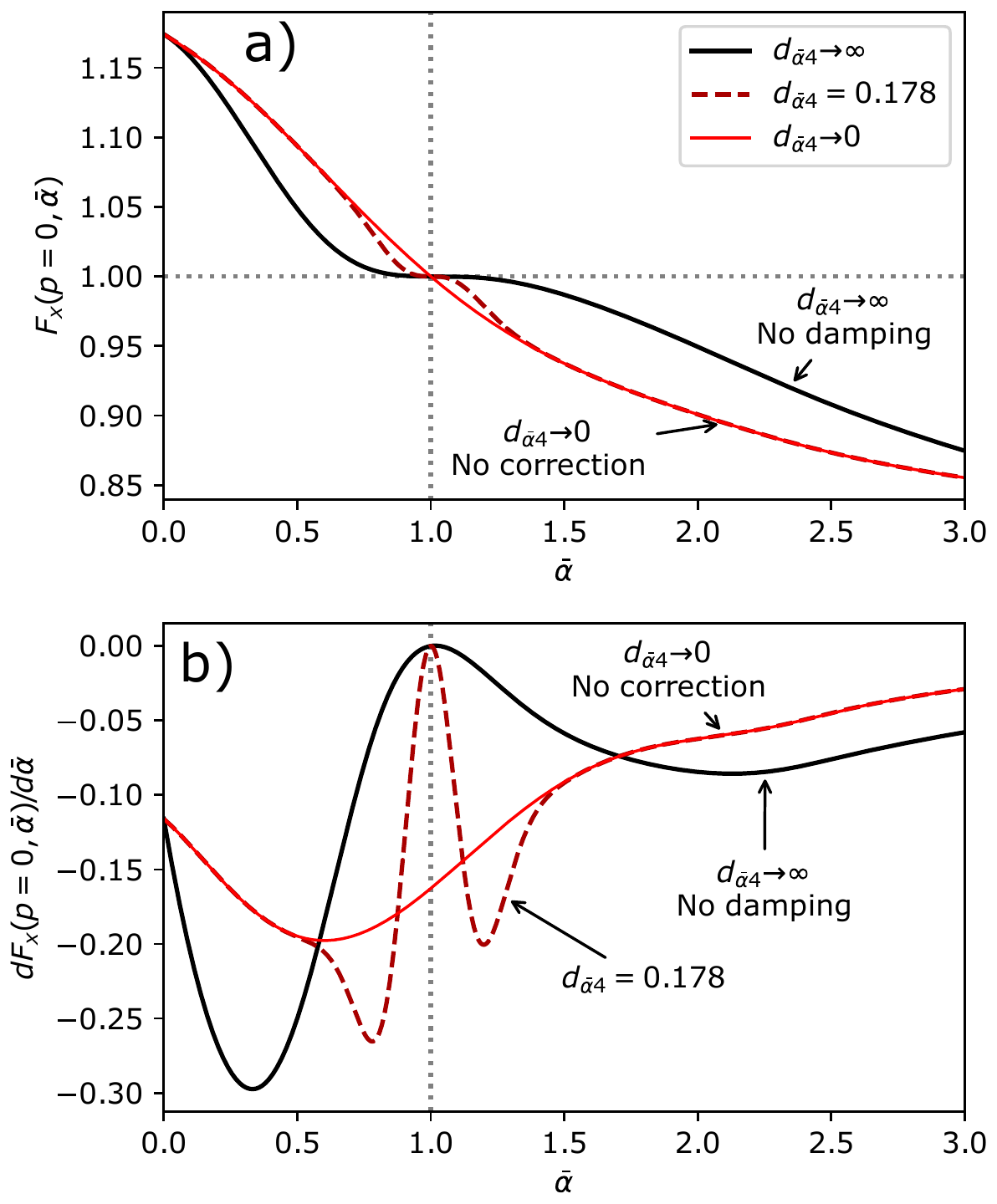}
    \caption{a) Exchange enhancement factor for r$^4$SCAN as a function of $\bar\alpha$ at $p = 0$. The uncorrected enhancement with $d_{\bar\alpha 4} \to 0$ (red) is contrasted against the un-damped corrections with $d_{\bar\alpha 4} \to \infty$ (black), the proposed r$^4$SCAN damping of $d_{\bar\alpha 4} = 0.178$ (dashed, dark red). b) Derivative of the exchange enhancement with respect to $\bar\alpha$ at $p = 0$ for the same conditions.}
    \label{fig:fx_osc}
\end{figure}

The corrections to restore the slowly-varying density gradient expansions for exchange and correlation contain terms linear in $(1 - \bar\alpha)$. These terms are necessary to restore the 2nd or 4th order gradient expansion, for example, of Eq. \ref{eq:GE_x} for exchange, if the integration in parts is not used as we did for \rrscan in Eq. \ref{eq:oma_to_p}. These corrections inevitably twist the slope of the interpolation function $f(\bar\alpha)$ to that of $F_{\mr{x}}^{GE}$ with respect to $\bar\alpha$ around $\bar\alpha \to 1$ as $|\nabla n|\to 0$, illustrated in Fig. \ref{fig:fx_osc}. This introduces oscillations into the derivatives of the enhancement factor with respect to $\bar\alpha$, and hence into the overall exchange-correlation potential. Such oscillations are undesirable and reintroduce the numerical problems rSCAN regularizes away. As the gradient expansion constraint requires \begin{equation}
    \left.\frac{\partial F_{\mr{x}}}{\partial \bar\alpha}\right|_{\bar\alpha = 1, p = 0} \propto \left.\frac{df(\bar\alpha)}{d\bar\alpha}\right|_{\bar\alpha = 1},
\end{equation}
this oscillation in derivatives must be present, at least in the interpolation scheme discussed here, and cannot be removed by damping. Figure \ref{fig:fx_osc}, compares uncorrected exchange enhancement, $d_{\bar\alpha 4} \to 0$, the exchange enhancement with no damping on the correction terms, $d_{\bar\alpha 4} \to \infty$, and the $d_{\bar\alpha 4} = 0.178$ determined in Section \ref{sec:settingParams}, showing this effect.

These numerical problems are not present in r$^2$SCAN, as the corrections do not depend upon $(1-\ba)$. Thus, the corresponding oscillation in derivative is avoided, allowing r$^2$SCAN to recover GE2X and GE2C whilst maintaining a smooth potential. As the integration by parts is not possible to fourth order, r$^4$SCAN necessarily suffers an oscillatory XC potential in order to recover GE4X.

\section{\label{sec:settingParams} Determining parameters}

The regularization of the $\bar\alpha$ indicator in r++SCAN, r$^2$SCAN, and r$^4$SCAN is controlled by the parameter $\eta$, with larger values increasing regularization strength. We find performance is largely insensitive to $\eta$ within the range of $0 \le \eta \le 0.001$ and take the upper value of $\eta = 0.001$.

We introduce a single damping parameter, $d_{p2}$, in r$^2$SCAN through Eqs. \ref{eq:x_change} and \ref{eq:del_y}, and set it using the appropriate norms philosophy of the SCAN functional. The parameter was chosen to minimize the sum of the mean absolute percentage errors in XC energy for four rare gas atoms: Ne, Ar, Kr, and Xe (evaluated for spherical Hartree--Fock orbitals \cite{Clementi1974} relative to \cite{Becke1988, Chakravorty1993, McCarthy2011} reference energies), and four jellium surface formation energies with $\rs =$ 2, 3, 4, and 6 bohr (relative to reference energies from Ref. \cite{Wood2007}). As the parameters are not fit to any bound systems, we regard the resulting functionals as non-empirical.

\begin{figure}
    \centering
    \includegraphics[width=0.45\textwidth]{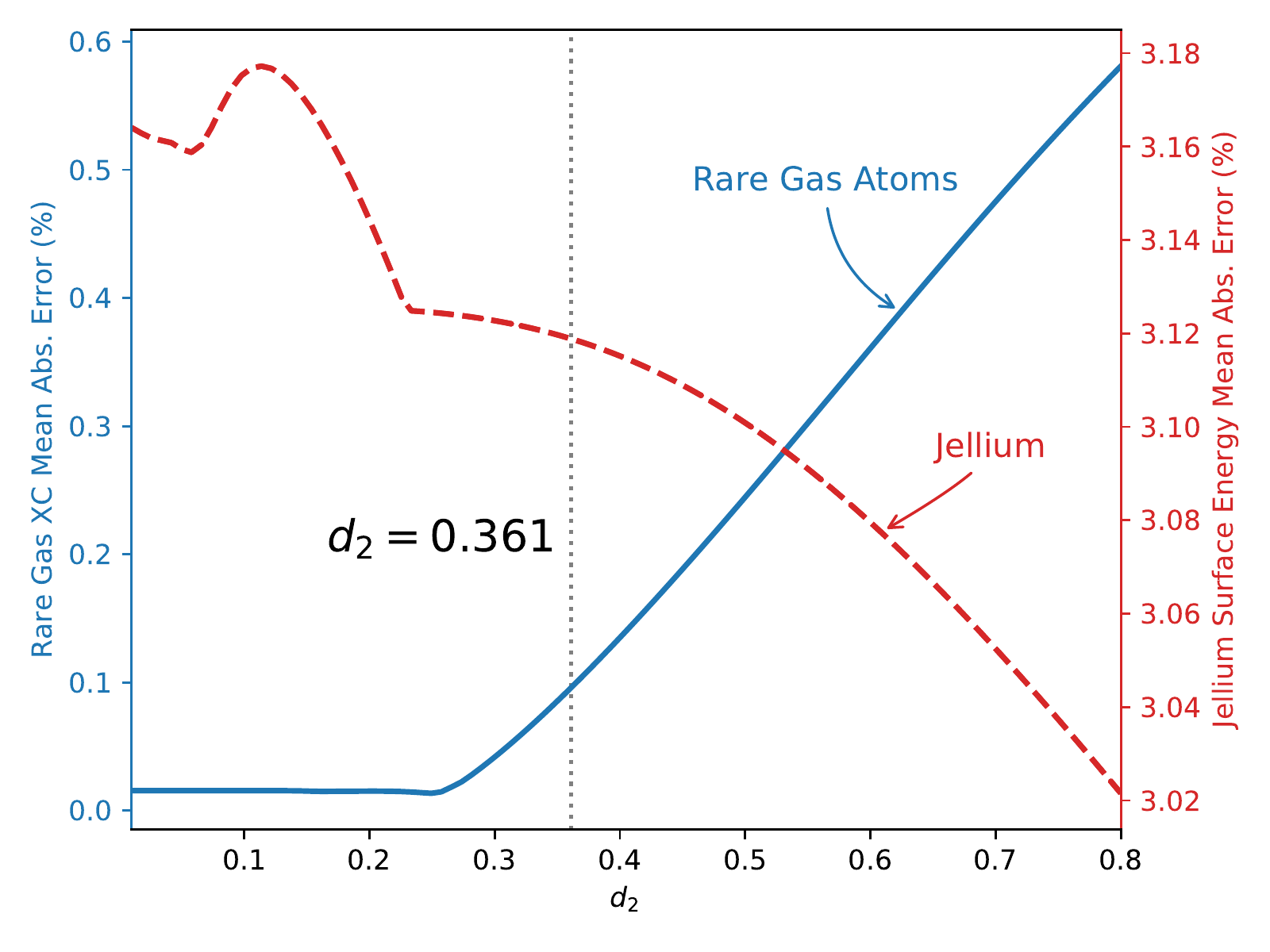}
    \caption{The mean absolute percentage error for (left axis, blue) the exchange-correlation energies of Ne, Ar, Kr, Xe \cite{Becke1988, Chakravorty1993, McCarthy2011}, and (right axis, red) the exchange-correlation jellium surface energy for $\rs = \{2,3,4,6\}$ \cite{Wood2007} as a function of second order gradient expansion damping parameter $d_{p2}$ for \rrscan. The optimal value is chosen as the largest for which the rare gas error is $< 0.1\%$ and the jellium error is $< 5\%$.}
    \label{fig:d2_Set}
\end{figure}

Objective error as a function of damping parameter, $d_{p2}$, is shown in Figure \ref{fig:d2_Set}. Setting the damping parameter too high degrades accuracy for the rare gas atoms (while mildly improving the jellium surface formation energies) as the gradient expansion terms dominate too far from $|\nabla n| \to 0$. Conversely, setting $d_{p2}$ too small degrades accuracy for the jellium surfaces, as the second order gradient expansion is not sufficiently corrected. As a sharper damping function causes sharper features in XC potential, we take the largest value for $d_{p2}$ which meets the accuracy threshold defined by SCAN: a mean absolute percentage error (MAPE) of $0.1\%$ for rare gas XC energies and $5\%$ for jellium surfaces. The optimizing value is found as $d_{p2} = 0.361$.

\begin{figure}[h!]
    \centering
    \includegraphics[width=0.36\textwidth]{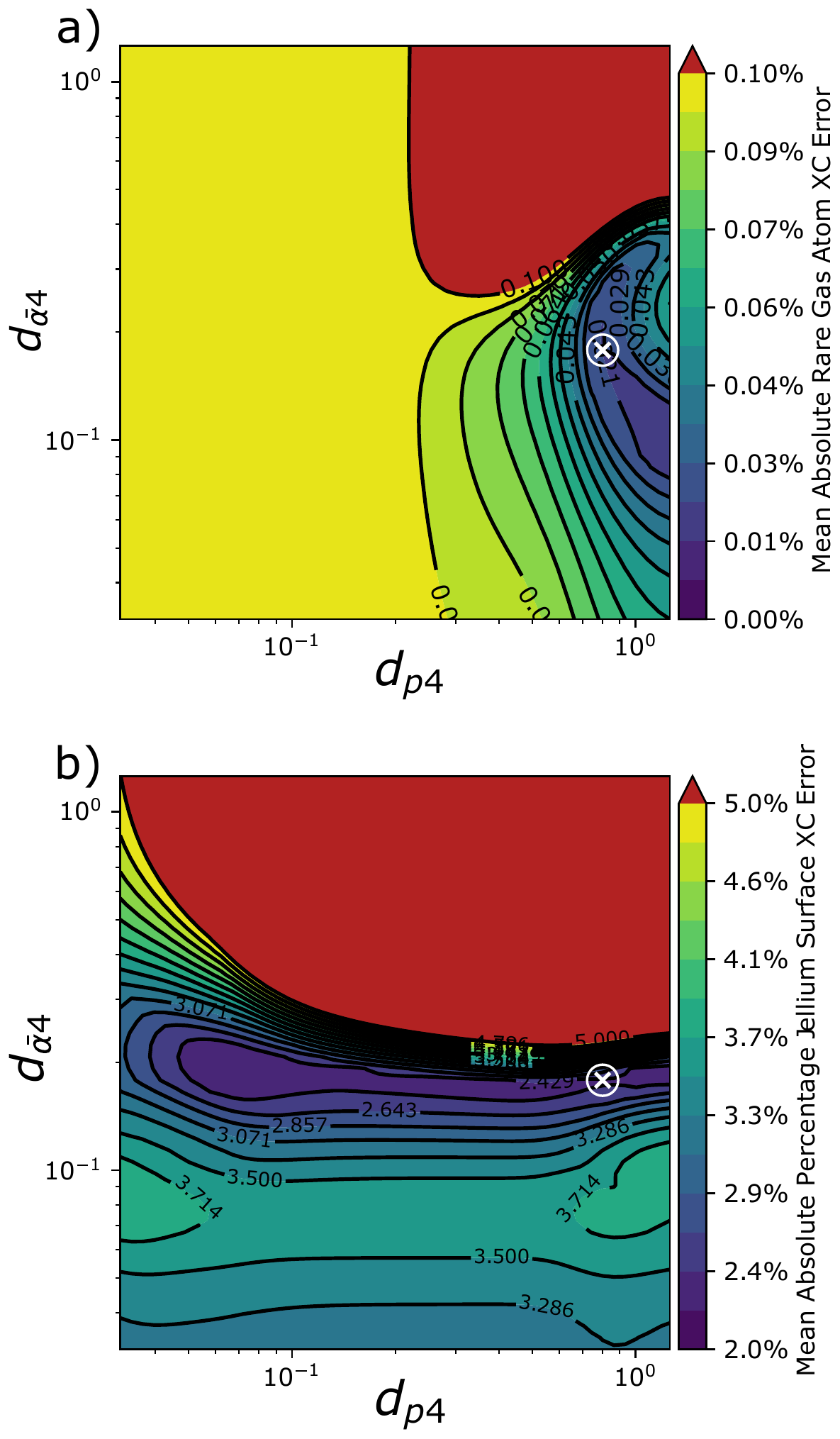}
    \caption{Mean absolute percentage error in rare gas XC energy (upper) and jellium surface exchange-correlation energy (lower) as a function of damping parameters $d_{\bar\alpha 4}$ and $d_{p4}$. Optimal parameters are identified by the white circled cross at $d_{p4} = 0.802, d_{\bar\alpha 4} = 0.178$.}
    \label{fig:setting_r4_params}
\end{figure}

Two additional parameters are introduced in r$^4$SCAN that control damping of the fourth-order gradient expansion terms. These were determined similarly as those which minimize a normalized sum of the rare gas and jellium surface mean absolute percentage errors. The minimizing parameters were found as $d_{\bar\alpha 4} = 0.178$ and $d_{p4} = 0.802$, as shown in Figure \ref{fig:setting_r4_params}.

\section{\label{sec:results} Results}

\subsection{Enhancement Factors and Derivatives\label{sec:Xe_osc}}

An important principle in functional design is to take an ``Occam's razor'' approach and determine a functional that is free of twists and kinks. In this way the functional avoids over-fitting to data and ensures smooth functional derivatives that are easy to render on numerical grids.

\begin{figure}[h]
    \centering
    \includegraphics[width=0.45\textwidth]{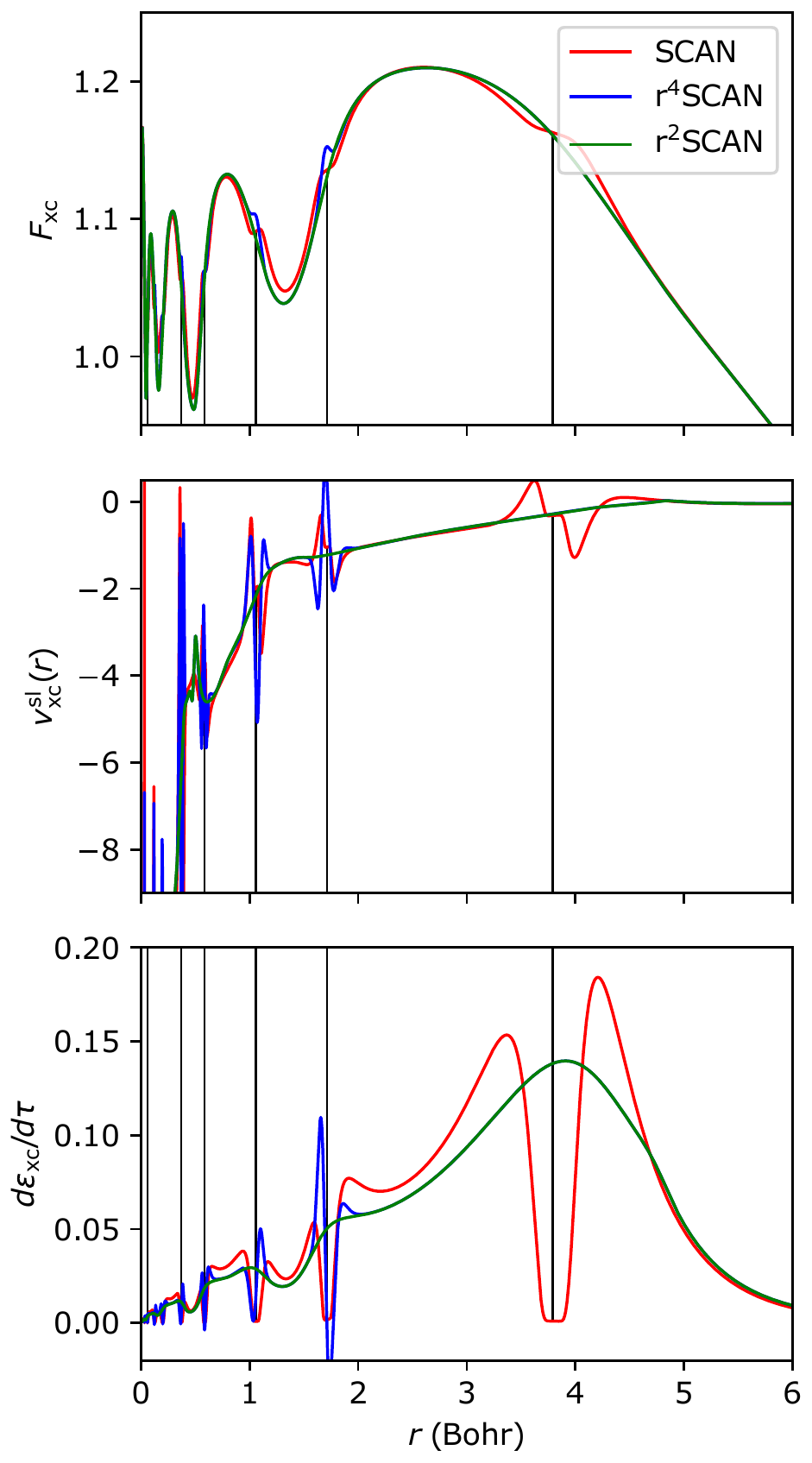}
    \caption{The XC enhancement factor (top), the multiplicative component of the XC potential (middle), and the non-local component, i.e., the derivative of the XC energy density with respect to the orbital dependent kinetic energy density, $\tau$ (bottom) in the Xe atom. Shown for the SCAN, \rrscan, and r$^4$SCAN functionals, calculated from reference Hartree--Fock Slater orbitals \cite{Clementi1974, Furness2021a}. Points where $\alpha = 1$ are shown by black vertical lines.}
    \label{fig:Xe_derivatives}
\end{figure}

Figure \ref{fig:Xe_derivatives} compares the XC enhancement factors of SCAN, r$^2$SCAN, and r$^4$SCAN for the Xenon atom. The SCAN enhancement factor shows sharp plateau like regions from the twists in its interpolation function around $\alpha = 1$. The smooth polynomial interpolation function removes these plateaus from r$^2$SCAN, though some twists are re-introduced in r$^4$SCAN by the $(1 - \bar\alpha)$ terms in the GE4X restoration.

The effect of twists in $F_\mr{xc}$ can be seen in the semi-local and non-local XC potential components of the XC potential, shown in Figure \ref{fig:Xe_derivatives}. The SCAN functional shows sharp oscillations around $\alpha = 1$ points and sharp drops in its non-local component. In contrast, the \rrscan functional is a smooth function of its ingredients, and hence has smooth semi-local and non-local components to its XC potential. While the potential components of \rrscan and r$^4$SCAN coincide for much of space they differ significantly around $\bar{\alpha} = 1$ points. Here the $(1-\bar\alpha)$ correction terms in r$^4$SCAN required for GE4X cause sharp oscillations that return the oscillatory behavior we aim to remove. As these terms cannot be removed by partial integration to the 4th order in $\nabla n$, we therefore conclude that GE4X is incompatible with functional smoothness under the present SCAN interpolation-based model: one must either twist the interpolation function to enforce Eq. \ref{eq:interp_const}, or include correcting terms that re-introduces oscillatory factors.

\subsection{Appropriate Norms}

An appropriate norm is defined in Ref. \cite{Sun2015} as ``systems for which semilocal functionals can be exact or extremely accurate''. Here, like SCAN, we take these as the exchange and correlation energies of four rare gas atoms (Ne, Ar, Kr, and Xe), the exchange and correlation surface energies of four jellium slabs ($\overline{\rs}=$ 2, 3, 4, and 6), and the interaction energy of Ar\mol{2} at repulsive inter-atomic distances ($R_{\mr{Ar-Ar}} =$ 1.6, 1.8, and 2.0 \AA). Table \ref{tab:norms} compares the accuracy of SCAN, rSCAN, r++SCAN, r$^2$SCAN, and r$^4$SCAN for the appropriate norms.

Care must be taken when computing the jellium surface energy for rSCAN as the uniform bulk density energy is changed by the $\alpha^\prime$ regularization. To evaluate the exchange-correlation contribution to the jellium surface formation energy, we compute \cite{Lang1970}
\begin{equation}
    \sigma_\mr{xc}(\overline{\rs}) = \int_{-\infty}^{\infty} [\varepsilon_\mr{xc}(n,\nabla n, \tau) - \varepsilon_\mr{xc}(\overline{n},0,\overline{\tau_U}) ]n(x) dx,
\end{equation}
where $\overline{\rs} = (4\pi \overline{n}/3)^{-1/3}$ is the density
parameter of the corresponding bulk jellium, and $\overline{\tau_U} =
3(3\pi^2)^{2/3}\overline{n}^{5/3}/10$ is its kinetic energy density. Here, we have assumed that the surface lies along the $x$ direction.
This ensures that the uniform density limit of a given functional is used, regardless of whether that uniform limit is the LSDA.
Building upon our previous example of the valence density in solid sodium: when $\rs=4$, the rSCAN exchange uniform density limit is
\[
    \epsilon_\mr{x}^\mr{rSCAN}(\overline{n},0,\overline{\tau_U})\bigg|_{\overline{\rs}=4} \approx 1.051 \epsilon_\mr{x}^\mr{LDA}(\overline{\rs}),
\]
making a substantial error over LDA exchange. For $\overline{\rs}=6$, this error is increased to roughly 14\%.

The importance of recovering the second order gradient expansion is seen in the relative accuracy of the functionals for the rare gas atoms. The two functionals which do not recover the gradient expansions (rSCAN and r++SCAN) have MAPEs of $\approx 0.25\%$ whereas the functionals that do have MAPEs of $\approx 0.1\%$. Restoring the fourth order gradient expansion for exchange improves accuracy further, though r$^2$SCAN already has similar accuracy to SCAN for all the appropriate norms suggesting GEX4 is less important for these systems. Outside of these differences, all functionals performed similarly for the appropriate norm systems.

% Appropriate norms table
\begin{table*}
    \centering
    \caption{Accuracy for appropriate norms. Rare gas and jellium surface exchange-correlation energies given in Hartrees ($E_h$), Ar\mol{2} interaction energies in kcal/mol. Benchmark data for rare gas atom exchange-correlation, jellium surface exchange-correlation, and  Ar\mol{2} interaction energy are from Refs. \cite{Becke1988, Chakravorty1993, McCarthy2011}, \cite{Wood2007}, and \cite{Patkowski2005} respectively. Error summaries are given as mean absolute percentage errors (MAPE). Full calculation details in main text.\label{tab:norms}}
    \begin{tabular}{ll|rrrrrr}
                           &                & \multicolumn{1}{c}{SCAN} & \multicolumn{1}{c}{rSCAN} & \multicolumn{1}{c}{r++SCAN} & \multicolumn{1}{c}{r$^2$SCAN} & \multicolumn{1}{c}{r$^4$SCAN} & \multicolumn{1}{c}{Benchmarks} \\
\hline
\hline
\multirow{3}{*}{Ne}        & $E_{\mr{x}}$   &      -12.164 (    0.48\%) &      -12.183 (    0.64\%) &      -12.176 (    0.58\%) &      -12.144 (    0.32\%) &      -12.146 (    0.34\%) &      -12.105 $E_h$ \\
                           & $E_{\mr{c}}$   &       -0.345 (  -11.81\%) &       -0.346 (  -11.53\%) &       -0.347 (  -11.36\%) &       -0.347 (  -11.24\%) &       -0.347 (  -11.24\%) &       -0.391 $E_h$ \\
                           & $E_{\mr{xc}}$  &      -12.508 (    0.10\%) &      -12.529 (    0.26\%) &      -12.522 (    0.21\%) &      -12.491 (   -0.04\%) &      -12.493 (   -0.03\%) &      -12.496 $E_h$ \\
\hline
\multirow{3}{*}{Ar}        & $E_{\mr{x}}$   &      -30.264 (    0.29\%) &      -30.295 (    0.40\%) &      -30.281 (    0.35\%) &      -30.182 (    0.02\%) &      -30.196 (    0.07\%) &      -30.175 $E_h$ \\
                           & $E_{\mr{c}}$   &       -0.690 (   -4.81\%) &       -0.695 (   -4.24\%) &       -0.696 (   -4.06\%) &       -0.697 (   -3.90\%) &       -0.697 (   -3.90\%) &       -0.725 $E_h$ \\
                           & $E_{\mr{xc}}$  &      -30.955 (    0.17\%) &      -30.990 (    0.29\%) &      -30.977 (    0.25\%) &      -30.879 (   -0.07\%) &      -30.893 (   -0.02\%) &      -30.901 $E_h$ \\
\hline
\multirow{3}{*}{Kr}        & $E_{\mr{x}}$   &      -94.071 (    0.25\%) &      -94.215 (    0.41\%) &      -94.186 (    0.37\%) &      -93.820 (   -0.02\%) &      -93.940 (    0.11\%) &      -93.834 $E_h$ \\
                           & $E_{\mr{c}}$   &       -1.756 (   -5.10\%) &       -1.765 (   -4.59\%) &       -1.768 (   -4.47\%) &       -1.770 (   -4.34\%) &       -1.770 (   -4.34\%) &       -1.850 $E_h$ \\
                           & $E_{\mr{xc}}$  &      -95.827 (    0.15\%) &      -95.980 (    0.31\%) &      -95.953 (    0.28\%) &      -95.590 (   -0.10\%) &      -95.710 (    0.03\%) &      -95.684 $E_h$ \\
\hline
\multirow{3}{*}{Xe}        & $E_{\mr{x}}$   &     -179.315 (    0.14\%) &     -179.614 (    0.31\%) &     -179.567 (    0.28\%) &     -178.827 (   -0.13\%) &     -179.136 (    0.04\%) &     -179.064 $E_h$ \\
                           & $E_{\mr{c}}$   &       -2.899 (   -3.43\%) &       -2.910 (   -3.07\%) &       -2.914 (   -2.96\%) &       -2.918 (   -2.82\%) &       -2.918 (   -2.82\%) &       -3.002 $E_h$ \\
                           & $E_{\mr{xc}}$  &     -182.214 (    0.08\%) &     -182.524 (    0.25\%) &     -182.480 (    0.23\%) &     -181.745 (   -0.18\%) &     -182.053 (   -0.01\%) &     -182.066 $E_h$ \\
\hline
\multicolumn{2}{c|}{MAPE[$E_{\mr{x}}$]} &     0.29\% &     0.44\% &     0.40\% &     0.12\% &     0.14\% & \\
\multicolumn{2}{c|}{MAPE[$E_{\mr{c}}$]} &     6.29\% &     5.86\% &     5.71\% &     5.58\% &     5.58\% & \\
\multicolumn{2}{c|}{MAPE[$E_{\mr{xc}}$]} &     0.13\% &     0.28\% &     0.24\% &     0.10\% &     0.02\% & \\
\hline
\hline
\multirow{3}{*}{$r_s$ 2}   & $E_{\mr{x}}$   &     2631.022 (   -0.27\%) &     2198.716 (   16.21\%) &     2259.204 (   13.90\%) &     2318.763 (   11.63\%) &     2419.739 (    7.78\%) &     2624.000 $E_h$ \\
                           & $E_{\mr{c}}$   &      811.437 (   -5.66\%) &      989.604 (  -28.85\%) &      971.939 (  -26.55\%) &      963.796 (  -25.49\%) &      963.796 (  -25.49\%) &      768.000 $E_h$ \\
                           & $E_{\mr{xc}}$  &     3442.459 (   -1.49\%) &     3188.320 (    6.00\%) &     3231.143 (    4.74\%) &     3282.559 (    3.23\%) &     3383.535 (    0.25\%) &     3392.000 $E_h$ \\
\hline
\multirow{3}{*}{$r_s$ 3}   & $E_{\mr{x}}$   &      489.080 (    7.02\%) &      352.217 (   33.04\%) &      394.134 (   25.07\%) &      412.474 (   21.58\%) &      424.755 (   19.25\%) &      526.000 $E_h$ \\
                           & $E_{\mr{c}}$   &      299.340 (  -23.69\%) &      361.776 (  -49.49\%) &      342.788 (  -41.65\%) &      339.987 (  -40.49\%) &      339.987 (  -40.49\%) &      242.000 $E_h$ \\
                           & $E_{\mr{xc}}$  &      788.419 (   -2.66\%) &      713.992 (    7.03\%) &      736.922 (    4.05\%) &      752.461 (    2.02\%) &      764.741 (    0.42\%) &      768.000 $E_h$ \\
\hline
\multirow{3}{*}{$r_s$ 4}   & $E_{\mr{x}}$   &      126.894 (   19.18\%) &       82.262 (   47.60\%) &       92.568 (   41.04\%) &      100.800 (   35.80\%) &      102.789 (   34.53\%) &      157.000 $E_h$ \\
                           & $E_{\mr{c}}$   &      146.586 (  -40.95\%) &      152.932 (  -47.05\%) &      162.499 (  -56.25\%) &      161.117 (  -54.92\%) &      161.117 (  -54.92\%) &      104.000 $E_h$ \\
                           & $E_{\mr{xc}}$  &      273.480 (   -4.78\%) &      235.194 (    9.89\%) &      255.068 (    2.27\%) &      261.917 (   -0.35\%) &      263.906 (   -1.11\%) &      261.000 $E_h$ \\
\hline
\multirow{3}{*}{$r_s$ 6}   & $E_{\mr{x}}$   &        6.264 (   71.53\%) &       30.717 (  -39.62\%) &       -1.577 (  107.17\%) &        1.210 (   94.50\%) &        1.604 (   92.71\%) &       22.000 $E_h$ \\
                           & $E_{\mr{c}}$   &       52.651 (  -69.84\%) &       20.481 (   33.93\%) &       56.057 (  -80.83\%) &       55.508 (  -79.06\%) &       55.508 (  -79.06\%) &       31.000 $E_h$ \\
                           & $E_{\mr{xc}}$  &       58.915 (  -11.16\%) &       51.199 (    3.40\%) &       54.480 (   -2.79\%) &       56.719 (   -7.02\%) &       57.112 (   -7.76\%) &       53.000 $E_h$ \\
\hline
\multicolumn{2}{c|}{MAPE[$E_{\mr{x}}$]} &    24.50\% &    34.12\% &    46.79\% &    40.88\% &    38.57\% & \\
\multicolumn{2}{c|}{MAPE[$E_{\mr{c}}$]} &    35.03\% &    39.83\% &    51.32\% &    49.99\% &    49.99\% & \\
\multicolumn{2}{c|}{MAPE[$E_{\mr{xc}}$]} &     5.02\% &     6.58\% &     3.46\% &     3.15\% &     2.39\% & \\
\hline
\hline
\multicolumn{2}{c|}{Ar$_2$ 1.6\AA} &      360.936 (   -1.19\%) &      361.031 (   -1.17\%) &      360.200 (   -1.39\%) &      362.516 (   -0.76\%) &      360.616 (   -1.28\%) &      365.292 kcal/mol\\
\multicolumn{2}{c|}{Ar$_2$ 1.8\AA} &      195.723 (   -1.28\%) &      197.023 (   -0.62\%) &      196.277 (   -1.00\%) &      198.109 (   -0.07\%) &      196.431 (   -0.92\%) &      198.255 kcal/mol\\
\multicolumn{2}{c|}{Ar$_2$ 2.0\AA} &      102.465 (   -0.73\%) &      103.577 (    0.35\%) &      102.889 (   -0.32\%) &      104.242 (    1.00\%) &      103.151 (   -0.06\%) &      103.215 kcal/mol\\
\multicolumn{2}{c|}{MAPE} &     1.07\% &     0.71\% &     0.90\% &     0.61\% &     0.75\% & \\
\end{tabular}

\end{table*}

\subsection{Atomization Energies}

The work of Ref. \cite{Mejia-Rodriguez2019} shows the performance of rSCAN is relatively poor for atomization energy prediction, as measured by its increased error for the G3 test set of molecules \cite{Curtiss2000}. Table \ref{tab:G3_stats} compares the errors for this test set for all the functionals derived above. Consistent with Ref. \cite{Mejia-Rodriguez2019} we find a large error from rSCAN, with this error only slightly corrected in r++SCAN. As in Ref. \citenum{Furness2020c}, restoration of GE2X and GE2C in r$^2$SCAN restores the good accuracy of SCAN, showing the importance of these constraints for atomization energies. The good accuracy of r$^2$SCAN suggests that recovering GE4X is not essential for accurately predicting atomization energies.

\begin{table}
    \centering
    \caption{Summary of atomization energy errors (in kcal/mol) for the G3 test set \cite{Curtiss2000} using the most dense numerical integration grid (\textsc{Turbomole} level 7).}
    \begin{tabular}{l|rrrrr}
        & SCAN      & rSCAN     & r++SCAN   & r$^2$SCAN & r$^4$SCAN \\
        \hline
        ME  & -5.036    & -14.010   & -12.912   & -5.042    & -6.939 \\
        MAE &  6.121    &  14.258   &  13.239   &  5.866    &  7.716
    \end{tabular}
    \label{tab:G3_stats}
\end{table}

\begin{figure}
    \centering
    \includegraphics[width=0.45\textwidth]{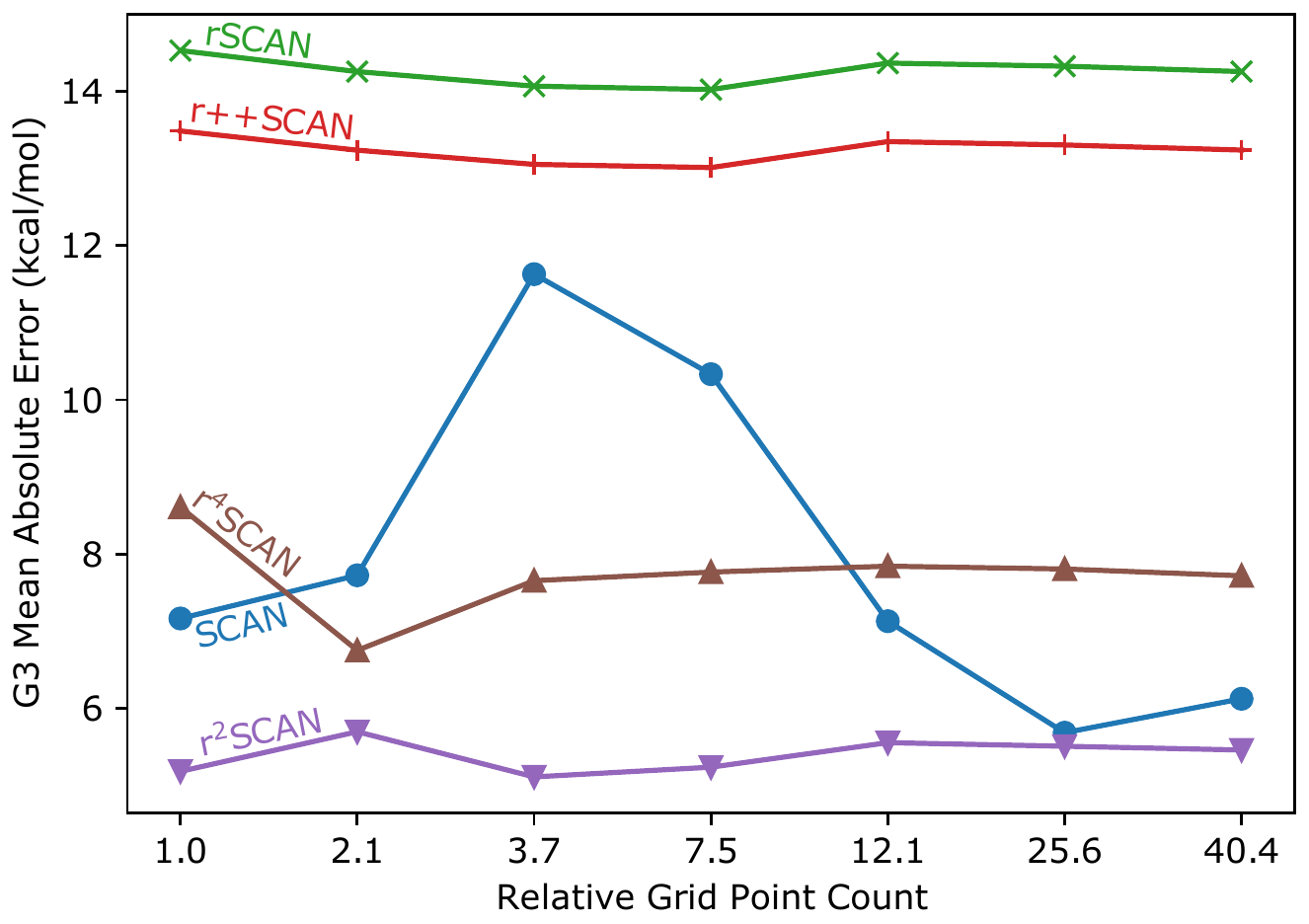}
    \caption{Mean absolute percentage error of atomization energies (kcal/mol) for the G3 set of 226 molecules \cite{Curtiss2000} as a function of increasing numerical integration grid density expressed relative to the smallest grid. The grids chosen were defined by the default \textsc{Turbomole} grid levels. The mean number of grid points per atom over the G3 set is about 1,326 for the smallest (gridsize=1) and 53,590 for the largest (gridsize=7) grid shown here.
}
    \label{fig:G3_progression}
\end{figure}

The improved numerical performance of the functionals is illustrated by examining the convergence of atomization energy predictions as a function of numerical grid density, as shown in Figure \ref{fig:G3_progression}. The original SCAN functional shows wild variation with changing grid density, with a range of over 6 kcal/mol in mean absolute error! While there is some indication that a convergence is approached for the most dense grids, the results from more computationally efficient grids are problematic and clearly untrustworthy.

All four regularized functionals show very fast grid convergence, with all sparse grids giving close agreement to the dense grids. Given the sharp oscillations in Figure \ref{fig:Xe_derivatives} and analysis of Section \ref{sec:numerics}, it is somewhat unexpected that r$^4$SCAN shows good convergence with grid density. We attribute the improved performance over SCAN to the reduction in plateau like behavior of $F_\mr{xc}$ but caution that grid convergence behavior will likely be more system dependent for r$^4$SCAN than r++SCAN and r$^2$SCAN as a result of the oscillations.

It is similarly unexpected to find that the accuracy of \rrscan is degraded by the inclusion of the GE4X in r$^4$SCAN. This supports our previous conclusion that GE4X is not important for the properties tested here and elsewhere\cite{Ehlert2021b, Grimme2021}. We take the mild degradation of G3 accuracy from extending to r$^4$SCAN as further evidence that this method of including GE4X into interpolation based meta-GGA functionals is problematic. The inclusion of two additional fitting parameters beyond those in \rrscan may also contribute.

\subsection{Further Testing}

Beyond atomization energies, we have tested the accuracy of the progression of SCAN-like functionals for the interaction energies of closed shell complexes (S22)\cite{Jurecka2006} and reaction barrier heights (BH76)\cite{Zhao2005}, summarized in Table \ref{tab:S22_BH76}. Additionally, Table \ref{tab:LC20_table} summarizes accuracy for the LC20 set of lattice constants for solids \cite{Sun2011}, obtained by fitting the stabilized jellium equation of state \cite{Staroverov2004} to single point energy calculations at a range of lattice volumes. All five functionals gave comparable good accuracy across all three test sets, showing that SCAN's good performance is not significantly changed by the regularizations or exact constraint restoration for these properties.

\begin{table}
    \caption{Mean error (ME) and Mean absolute error (MAE) of SCAN\cite{Sun2015}, rSCAN\cite{Bartok2019}, r++SCAN, \rrscan\cite{Furness2020c}, and r$^4$SCAN for the S22 set of 22 interaction energies between closed shell complexes\cite{Jurecka2006}, and the BH76 set of 76 chemical barrier heights\cite{Zhao2005}. Full data is presented in supplemental material Tables
    VI and VII. %\ref{SM-tab:full_S22} and \ref{SM-tab:BH76_full}.
    All calculations use the most dense integration grid (\textsc{Turbomole} level 7).}
    \label{tab:S22_BH76}
    \centering
    \begin{tabular}{cc|rrrrr}
                            &       & SCAN   & rSCAN     & r++SCAN   & r$^2$SCAN    & r$^4$SCAN \\
    \hline
    \multirow{2}{*}{S22}    & ME    & -0.524 & -1.153    & -0.554    & -0.937       & -0.874    \\
                            & MAE   & 0.786  & 1.273     & 0.846     & 1.057        & 1.015     \\
    \hline
    \multirow{2}{*}{BH76}   & ME    & -7.653 & -7.365    & -7.488    & -7.125       & -7.463    \\
                            & MAE   & 7.724  & 7.434     & 7.556     & 7.182        & 7.527     \\
    \end{tabular}
\end{table}

\begin{table}
    \caption{Error in lattice constant prediction for the LC20 lattice constant test set \cite{Sun2011}. Errors are in $\text{\AA}$ relative to zero-point anharmonic expansion corrected experimental data from Ref. \cite{Hao2012}. Lattice constants were obtained by fitting the stabilized jellium equation of state \cite{Staroverov2004} to single point energy calculations at a range of lattice volumes.}
    \label{tab:LC20_table}
    \centering
    \begin{tabular}{c|rrrrr}
    \hline
    \hline
        & SCAN      & rSCAN     & r++SCAN   & r$^2$SCAN & r$^4$SCAN \\ \hline
        Ag & 0.012 & 0.028 & 0.026 & 0.034 & 0.017 \\
        Al & -0.012 & -0.027 & -0.031 & -0.032 & -0.014 \\
        Ba & 0.049 & 0.100 & 0.046 & 0.076 & 0.062 \\
        C & -0.004 & -0.001 & -0.001 & 0.005 & 0.002 \\
        Ca & -0.009 & 0.017 & -0.011 & 0.018 & 0.016 \\
        Cu & -0.030 & -0.023 & -0.025 & -0.020 & -0.027 \\
        GaAs & 0.020 & 0.031 & 0.026 & 0.029 & 0.017 \\
        Ge & 0.029 & 0.043 & 0.038 & 0.039 & 0.028 \\
        Li & 0.011 & 0.021 & 0.006 & 0.016 & 0.010 \\
        LiCl & 0.016 & 0.021 & 0.017 & 0.034 & 0.032 \\
        LiF & 0.004 & 0.008 & 0.008 & 0.021 & 0.020 \\
        MgO & 0.018 & 0.019 & 0.018 & 0.027 & 0.024 \\
        NaCl & 0.010 & 0.025 & 0.015 & 0.036 & 0.036 \\
        NaF & 0.006 & 0.015 & 0.012 & 0.028 & 0.028 \\
        Na & -0.007 & 0.024 & -0.012 & 0.004 & 0.031 \\
        Pd & 0.016 & 0.026 & 0.026 & 0.032 & 0.019 \\
        Rh & -0.005 & 0.006 & 0.005 & 0.008 & -0.005 \\
        Si & 0.005 & 0.012 & 0.009 & 0.018 & 0.016 \\
        SiC & 0.002 & -0.001 & -0.002 & 0.006 & 0.007 \\
        Sr & 0.039 & 0.064 & 0.019 & 0.056 & 0.095 \\
    \hline
      ME & 0.009 & 0.020 & 0.009 & 0.022 & 0.021 \\
      MAE & 0.015 & 0.025 & 0.017 & 0.027 & 0.025 \\
    \hline
    \hline
    \end{tabular}
\end{table}

\section{\label{sec:conclusions} Conclusions}

To summarize, we have shown how exact constraints obeyed by SCAN are broken by the regularizations in rSCAN. Through this analysis we have shown how the exact constraint adherence can be restored and how this can be achieved without sacrificing the good numerical performance of rSCAN. This results in three new functionals with increasing exact constraint adherence: r++SCAN, r$^2$SCAN, and r$^4$SCAN. Additional parameters introduced to the new functionals are set without reference to any real bonded systems, thus we can still regard the resulting functionals as non-empirical and expect them to be applicable to a wide range of systems. Figure \ref{fig:G3_progression} suggests that restoring GEX4 in r$^4$SCAN gives similar accuracy to SCAN with some improvement in grid efficiency. We therefore expect the new r$^2$SCAN functional to remain the preferred choice for situations where the accuracy of SCAN is desired but its use is prohibited by poor numerical performance \cite{Kingsbury2021,Ning2021}.

Further improvement over \rrscan might be achieved by a smoother and fuller incorporation of the fourth-order density-gradient terms for the exchange energy in a SCAN-like functional. Work on this is underway.

Figure \ref{fig:G3_progression} shows that rSCAN and r++SCAN, which lose the correct second-order gradient expansions for densities that vary slowly over space, also lose accuracy for atomization energies of molecules, and that the restoration of this limit in SCAN, \rrscan, and \rfscan also restores accurate atomization energies. This result is in line with arguments made in Ref. \cite{Kaplan2020}.

Experience with SCAN and \rrscan (and with atomic densities \cite{Medvedev2017}) suggests that smoothness at fixed electron number could be elevated to the status of an 18$^{\text{th}}$ exact constraint that a meta-GGA can satisfy, or at least to the status of a construction principle: By Occam's Razor, the simplest assumption, consistent with what is known, should be preferred.

\begin{acknowledgments}

J.F., J.N., and J.S. acknowledge the support of the U.S. DOE, Office of Science, Basic Energy Sciences Grant No. DE-SC0019350. J.S. also acknowledges the support  of the US National Science Foundation under Grant No. DMR-2042618.
A.D.K. acknowledges the support of the U.S. Department of Energy, Office of Science, Basic Energy Sciences, through Grant No. DE-SC0012575 to the Energy Frontier Research Center: Center for Complex Materials from First Principles, and also support from Temple University.
JPP  acknowledges the support  of the US National  Science Foundation under Grant No. DMR-1939528.
We thank Albert Bart\'ok-Partay and Daniel Mej\'ia-Rodr\'iguez for their invaluable discussions around the ideas presented here. J.P.P. and J.S. thank Natalie Holzwarth for pointing out that the SCAN exchange-correlation potential for an atom diverges in the tail of the density, making pseudo-potential construction difficult.

\end{acknowledgments}

\section*{Materials availability}

\rrscan and \rfscan subroutines are made freely available at \url{https://gitlab.com/dhamil/r2scan-subroutines}. The data that support the findings of this study are available from the corresponding author upon reasonable request.

%\clearpage
%\bibliographystyle{apsrev4-2}
%\bibliography{r2SCAN_Paper.bib}% Produces the bibliography via BibTeX.

\begin{thebibliography}{78}%
\makeatletter
\providecommand \@ifxundefined [1]{%
 \@ifx{#1\undefined}
}%
\providecommand \@ifnum [1]{%
 \ifnum #1\expandafter \@firstoftwo
 \else \expandafter \@secondoftwo
 \fi
}%
\providecommand \@ifx [1]{%
 \ifx #1\expandafter \@firstoftwo
 \else \expandafter \@secondoftwo
 \fi
}%
\providecommand \natexlab [1]{#1}%
\providecommand \enquote  [1]{``#1''}%
\providecommand \bibnamefont  [1]{#1}%
\providecommand \bibfnamefont [1]{#1}%
\providecommand \citenamefont [1]{#1}%
\providecommand \href@noop [0]{\@secondoftwo}%
\providecommand \href [0]{\begingroup \@sanitize@url \@href}%
\providecommand \@href[1]{\@@startlink{#1}\@@href}%
\providecommand \@@href[1]{\endgroup#1\@@endlink}%
\providecommand \@sanitize@url [0]{\catcode `\\12\catcode `\$12\catcode
  `\&12\catcode `\#12\catcode `\^12\catcode `\_12\catcode `\%12\relax}%
\providecommand \@@startlink[1]{}%
\providecommand \@@endlink[0]{}%
\providecommand \url  [0]{\begingroup\@sanitize@url \@url }%
\providecommand \@url [1]{\endgroup\@href {#1}{\urlprefix }}%
\providecommand \urlprefix  [0]{URL }%
\providecommand \Eprint [0]{\href }%
\providecommand \doibase [0]{https://doi.org/}%
\providecommand \selectlanguage [0]{\@gobble}%
\providecommand \bibinfo  [0]{\@secondoftwo}%
\providecommand \bibfield  [0]{\@secondoftwo}%
\providecommand \translation [1]{[#1]}%
\providecommand \BibitemOpen [0]{}%
\providecommand \bibitemStop [0]{}%
\providecommand \bibitemNoStop [0]{.\EOS\space}%
\providecommand \EOS [0]{\spacefactor3000\relax}%
\providecommand \BibitemShut  [1]{\csname bibitem#1\endcsname}%
\let\auto@bib@innerbib\@empty
%</preamble>
\bibitem [{\citenamefont {Perdew}\ and\ \citenamefont
  {Schmidt}(2001)}]{Perdew2001}%
  \BibitemOpen
  \bibfield  {author} {\bibinfo {author} {\bibfnamefont {J.~P.}\ \bibnamefont
  {Perdew}}\ and\ \bibinfo {author} {\bibfnamefont {K.}~\bibnamefont
  {Schmidt}},\ }\href {https://doi.org/10.1063/1.1390175} {\bibfield  {journal}
  {\bibinfo  {journal} {AIP Conf. Proc.}\ }\textbf {\bibinfo {volume} {577}},\
  \bibinfo {pages} {1} (\bibinfo {year} {2001})}\BibitemShut {NoStop}%
\bibitem [{\citenamefont {Becke}(1983)}]{Becke1983}%
  \BibitemOpen
  \bibfield  {author} {\bibinfo {author} {\bibfnamefont {A.~D.}\ \bibnamefont
  {Becke}},\ }\href {https://doi.org/https://doi.org/10.1002/qua.560230605}
  {\bibfield  {journal} {\bibinfo  {journal} {Int. J. Quantum Chem.}\ }\textbf
  {\bibinfo {volume} {23}},\ \bibinfo {pages} {1915} (\bibinfo {year}
  {1983})},\ \Eprint
  {https://arxiv.org/abs/https://onlinelibrary.wiley.com/doi/pdf/10.1002/qua.560230605}
  {https://onlinelibrary.wiley.com/doi/pdf/10.1002/qua.560230605} \BibitemShut
  {NoStop}%
\bibitem [{\citenamefont {Becke}\ and\ \citenamefont
  {Roussel}(1989)}]{Becke1989}%
  \BibitemOpen
  \bibfield  {author} {\bibinfo {author} {\bibfnamefont {A.~D.}\ \bibnamefont
  {Becke}}\ and\ \bibinfo {author} {\bibfnamefont {M.~R.}\ \bibnamefont
  {Roussel}},\ }\href {https://doi.org/10.1103/PhysRevA.39.3761} {\bibfield
  {journal} {\bibinfo  {journal} {Phys. Rev. A}\ }\textbf {\bibinfo {volume}
  {39}},\ \bibinfo {pages} {3761} (\bibinfo {year} {1989})}\BibitemShut
  {NoStop}%
\bibitem [{\citenamefont {Perdew}(1985)}]{Perdew1985}%
  \BibitemOpen
  \bibfield  {author} {\bibinfo {author} {\bibfnamefont {J.~P.}\ \bibnamefont
  {Perdew}},\ }\href {https://doi.org/10.1103/PhysRevLett.55.1665} {\bibfield
  {journal} {\bibinfo  {journal} {Phys. Rev. Lett.}\ }\textbf {\bibinfo
  {volume} {55}},\ \bibinfo {pages} {1665} (\bibinfo {year} {1985})},\ \bibinfo
  {note} {\textit{ibid.}, \textbf{55}, 2370 (1985)}\BibitemShut {NoStop}%
\bibitem [{\citenamefont {Sun}\ \emph {et~al.}(2015{\natexlab{a}})\citenamefont
  {Sun}, \citenamefont {Ruzsinszky},\ and\ \citenamefont {Perdew}}]{Sun2015}%
  \BibitemOpen
  \bibfield  {author} {\bibinfo {author} {\bibfnamefont {J.}~\bibnamefont
  {Sun}}, \bibinfo {author} {\bibfnamefont {A.}~\bibnamefont {Ruzsinszky}},\
  and\ \bibinfo {author} {\bibfnamefont {J.~P.}\ \bibnamefont {Perdew}},\
  }\href {https://doi.org/10.1103/PhysRevLett.115.036402} {\bibfield  {journal}
  {\bibinfo  {journal} {Phys. Rev. Lett.}\ }\textbf {\bibinfo {volume} {115}},\
  \bibinfo {pages} {036402} (\bibinfo {year} {2015}{\natexlab{a}})}\BibitemShut
  {NoStop}%
\bibitem [{\citenamefont {Mezei}\ \emph {et~al.}(2018)\citenamefont {Mezei},
  \citenamefont {Csonka},\ and\ \citenamefont {Kallay}}]{Mezei2018}%
  \BibitemOpen
  \bibfield  {author} {\bibinfo {author} {\bibfnamefont {P.~D.}\ \bibnamefont
  {Mezei}}, \bibinfo {author} {\bibfnamefont {G.~I.}\ \bibnamefont {Csonka}},\
  and\ \bibinfo {author} {\bibfnamefont {M.}~\bibnamefont {Kallay}},\
  }\bibfield  {journal} {\bibinfo  {journal} {J. Chem. Theory Comput.}\ }\href
  {https://doi.org/10.1021/acs.jctc.8b00072} {10.1021/acs.jctc.8b00072}
  (\bibinfo {year} {2018})\BibitemShut {NoStop}%
\bibitem [{\citenamefont {Aschebrock}\ and\ \citenamefont
  {K{\"{u}}mmel}(2019)}]{Aschebrock2019}%
  \BibitemOpen
  \bibfield  {author} {\bibinfo {author} {\bibfnamefont {T.}~\bibnamefont
  {Aschebrock}}\ and\ \bibinfo {author} {\bibfnamefont {S.}~\bibnamefont
  {K{\"{u}}mmel}},\ }\href {https://doi.org/10.1103/PhysRevResearch.1.033082}
  {\bibfield  {journal} {\bibinfo  {journal} {Phys. Rev. Res.}\ }\textbf
  {\bibinfo {volume} {1}},\ \bibinfo {pages} {033082} (\bibinfo {year}
  {2019})}\BibitemShut {NoStop}%
\bibitem [{\citenamefont {Neupane}\ \emph {et~al.}(2021)\citenamefont
  {Neupane}, \citenamefont {Tang}, \citenamefont {Nepal}, \citenamefont
  {Adhikari},\ and\ \citenamefont {Ruzsinszky}}]{Neupane2021}%
  \BibitemOpen
  \bibfield  {author} {\bibinfo {author} {\bibfnamefont {B.}~\bibnamefont
  {Neupane}}, \bibinfo {author} {\bibfnamefont {H.}~\bibnamefont {Tang}},
  \bibinfo {author} {\bibfnamefont {N.~K.}\ \bibnamefont {Nepal}}, \bibinfo
  {author} {\bibfnamefont {S.}~\bibnamefont {Adhikari}},\ and\ \bibinfo
  {author} {\bibfnamefont {A.}~\bibnamefont {Ruzsinszky}},\ }\href
  {http://arxiv.org/abs/2104.04053} {\bibfield  {journal} {\bibinfo  {journal}
  {arXiv}\ ,\ \bibinfo {pages} {1}} (\bibinfo {year} {2021})},\ \Eprint
  {https://arxiv.org/abs/2104.04053} {arXiv:2104.04053} \BibitemShut {NoStop}%
\bibitem [{\citenamefont {Kitchaev}\ \emph {et~al.}(2016)\citenamefont
  {Kitchaev}, \citenamefont {Peng}, \citenamefont {Liu}, \citenamefont {Sun},
  \citenamefont {Perdew},\ and\ \citenamefont {Ceder}}]{Kitchaev2016}%
  \BibitemOpen
  \bibfield  {author} {\bibinfo {author} {\bibfnamefont {D.~A.}\ \bibnamefont
  {Kitchaev}}, \bibinfo {author} {\bibfnamefont {H.}~\bibnamefont {Peng}},
  \bibinfo {author} {\bibfnamefont {Y.}~\bibnamefont {Liu}}, \bibinfo {author}
  {\bibfnamefont {J.}~\bibnamefont {Sun}}, \bibinfo {author} {\bibfnamefont
  {J.~P.}\ \bibnamefont {Perdew}},\ and\ \bibinfo {author} {\bibfnamefont
  {G.}~\bibnamefont {Ceder}},\ }\href
  {https://doi.org/10.1103/PhysRevB.93.045132} {\bibfield  {journal} {\bibinfo
  {journal} {Phys. Rev. B}\ }\textbf {\bibinfo {volume} {93}},\ \bibinfo
  {pages} {045132} (\bibinfo {year} {2016})}\BibitemShut {NoStop}%
\bibitem [{\citenamefont {Sun}\ \emph {et~al.}(2016)\citenamefont {Sun},
  \citenamefont {Remsing}, \citenamefont {Zhang}, \citenamefont {Sun},
  \citenamefont {Ruzsinszky}, \citenamefont {Peng}, \citenamefont {Yang},
  \citenamefont {Paul}, \citenamefont {Waghmare}, \citenamefont {Wu},
  \citenamefont {Klein},\ and\ \citenamefont {Perdew}}]{Sun2016}%
  \BibitemOpen
  \bibfield  {author} {\bibinfo {author} {\bibfnamefont {J.}~\bibnamefont
  {Sun}}, \bibinfo {author} {\bibfnamefont {R.~C.}\ \bibnamefont {Remsing}},
  \bibinfo {author} {\bibfnamefont {Y.}~\bibnamefont {Zhang}}, \bibinfo
  {author} {\bibfnamefont {Z.}~\bibnamefont {Sun}}, \bibinfo {author}
  {\bibfnamefont {A.}~\bibnamefont {Ruzsinszky}}, \bibinfo {author}
  {\bibfnamefont {H.}~\bibnamefont {Peng}}, \bibinfo {author} {\bibfnamefont
  {Z.}~\bibnamefont {Yang}}, \bibinfo {author} {\bibfnamefont {A.}~\bibnamefont
  {Paul}}, \bibinfo {author} {\bibfnamefont {U.}~\bibnamefont {Waghmare}},
  \bibinfo {author} {\bibfnamefont {X.}~\bibnamefont {Wu}}, \bibinfo {author}
  {\bibfnamefont {M.~L.}\ \bibnamefont {Klein}},\ and\ \bibinfo {author}
  {\bibfnamefont {J.~P.}\ \bibnamefont {Perdew}},\ }\href
  {https://doi.org/10.1038/nchem.2535} {\bibfield  {journal} {\bibinfo
  {journal} {Nat. Chem.}\ }\textbf {\bibinfo {volume} {8}},\ \bibinfo {pages}
  {831} (\bibinfo {year} {2016})}\BibitemShut {NoStop}%
\bibitem [{\citenamefont {Peng}\ and\ \citenamefont {Perdew}(2017)}]{Peng2017}%
  \BibitemOpen
  \bibfield  {author} {\bibinfo {author} {\bibfnamefont {H.}~\bibnamefont
  {Peng}}\ and\ \bibinfo {author} {\bibfnamefont {J.~P.}\ \bibnamefont
  {Perdew}},\ }\href {https://doi.org/10.1103/PhysRevB.96.100101} {\bibfield
  {journal} {\bibinfo  {journal} {Phys. Rev. B}\ }\textbf {\bibinfo {volume}
  {96}},\ \bibinfo {pages} {100101(R)} (\bibinfo {year} {2017})}\BibitemShut
  {NoStop}%
\bibitem [{\citenamefont {Zhang}\ \emph {et~al.}(2017)\citenamefont {Zhang},
  \citenamefont {Sun}, \citenamefont {Perdew},\ and\ \citenamefont
  {Wu}}]{Zhang2017}%
  \BibitemOpen
  \bibfield  {author} {\bibinfo {author} {\bibfnamefont {Y.}~\bibnamefont
  {Zhang}}, \bibinfo {author} {\bibfnamefont {J.}~\bibnamefont {Sun}}, \bibinfo
  {author} {\bibfnamefont {J.~P.}\ \bibnamefont {Perdew}},\ and\ \bibinfo
  {author} {\bibfnamefont {X.}~\bibnamefont {Wu}},\ }\href
  {https://doi.org/10.1103/PhysRevB.96.035143} {\bibfield  {journal} {\bibinfo
  {journal} {Phys. Rev. B}\ }\textbf {\bibinfo {volume} {96}},\ \bibinfo
  {pages} {035143} (\bibinfo {year} {2017})}\BibitemShut {NoStop}%
\bibitem [{\citenamefont {Chen}\ \emph {et~al.}(2017)\citenamefont {Chen},
  \citenamefont {Ko}, \citenamefont {Remsing}, \citenamefont {{Calegari
  Andrade}}, \citenamefont {Santra}, \citenamefont {Sun}, \citenamefont
  {Selloni}, \citenamefont {Car}, \citenamefont {Klein}, \citenamefont
  {Perdew},\ and\ \citenamefont {Wu}}]{Chen2017}%
  \BibitemOpen
  \bibfield  {author} {\bibinfo {author} {\bibfnamefont {M.}~\bibnamefont
  {Chen}}, \bibinfo {author} {\bibfnamefont {H.-Y.}\ \bibnamefont {Ko}},
  \bibinfo {author} {\bibfnamefont {R.~C.}\ \bibnamefont {Remsing}}, \bibinfo
  {author} {\bibfnamefont {M.~F.}\ \bibnamefont {{Calegari Andrade}}}, \bibinfo
  {author} {\bibfnamefont {B.}~\bibnamefont {Santra}}, \bibinfo {author}
  {\bibfnamefont {Z.}~\bibnamefont {Sun}}, \bibinfo {author} {\bibfnamefont
  {A.}~\bibnamefont {Selloni}}, \bibinfo {author} {\bibfnamefont
  {R.}~\bibnamefont {Car}}, \bibinfo {author} {\bibfnamefont {M.~L.}\
  \bibnamefont {Klein}}, \bibinfo {author} {\bibfnamefont {J.~P.}\ \bibnamefont
  {Perdew}},\ and\ \bibinfo {author} {\bibfnamefont {X.}~\bibnamefont {Wu}},\
  }\href {https://doi.org/10.1073/pnas.1712499114} {\bibfield  {journal}
  {\bibinfo  {journal} {Proc. Natl. Acad. Sci. U. S. A.}\ }\textbf {\bibinfo
  {volume} {114}},\ \bibinfo {pages} {10846} (\bibinfo {year} {2017})},\
  \Eprint {https://arxiv.org/abs/1709.10493} {arXiv:1709.10493} \BibitemShut
  {NoStop}%
\bibitem [{\citenamefont {Furness}\ \emph {et~al.}(2018)\citenamefont
  {Furness}, \citenamefont {Zhang}, \citenamefont {Lane}, \citenamefont {Buda},
  \citenamefont {Barbiellini}, \citenamefont {Markiewicz}, \citenamefont
  {Bansil},\ and\ \citenamefont {Sun}}]{Furness2018}%
  \BibitemOpen
  \bibfield  {author} {\bibinfo {author} {\bibfnamefont {J.~W.}\ \bibnamefont
  {Furness}}, \bibinfo {author} {\bibfnamefont {Y.}~\bibnamefont {Zhang}},
  \bibinfo {author} {\bibfnamefont {C.}~\bibnamefont {Lane}}, \bibinfo {author}
  {\bibfnamefont {I.~G.}\ \bibnamefont {Buda}}, \bibinfo {author}
  {\bibfnamefont {B.}~\bibnamefont {Barbiellini}}, \bibinfo {author}
  {\bibfnamefont {R.~S.}\ \bibnamefont {Markiewicz}}, \bibinfo {author}
  {\bibfnamefont {A.}~\bibnamefont {Bansil}},\ and\ \bibinfo {author}
  {\bibfnamefont {J.}~\bibnamefont {Sun}},\ }\href
  {https://doi.org/10.1038/s42005-018-0009-4} {\bibfield  {journal} {\bibinfo
  {journal} {Commun. Phys.}\ }\textbf {\bibinfo {volume} {1}},\ \bibinfo
  {pages} {1} (\bibinfo {year} {2018})}\BibitemShut {NoStop}%
\bibitem [{\citenamefont {Lane}\ \emph {et~al.}(2018)\citenamefont {Lane},
  \citenamefont {Furness}, \citenamefont {Buda}, \citenamefont {Zhang},
  \citenamefont {Markiewicz}, \citenamefont {Barbiellini}, \citenamefont
  {Sun},\ and\ \citenamefont {Bansil}}]{Lane2018}%
  \BibitemOpen
  \bibfield  {author} {\bibinfo {author} {\bibfnamefont {C.}~\bibnamefont
  {Lane}}, \bibinfo {author} {\bibfnamefont {J.~W.}\ \bibnamefont {Furness}},
  \bibinfo {author} {\bibfnamefont {I.~G.}\ \bibnamefont {Buda}}, \bibinfo
  {author} {\bibfnamefont {Y.}~\bibnamefont {Zhang}}, \bibinfo {author}
  {\bibfnamefont {R.~S.}\ \bibnamefont {Markiewicz}}, \bibinfo {author}
  {\bibfnamefont {B.}~\bibnamefont {Barbiellini}}, \bibinfo {author}
  {\bibfnamefont {J.}~\bibnamefont {Sun}},\ and\ \bibinfo {author}
  {\bibfnamefont {A.}~\bibnamefont {Bansil}},\ }\href
  {https://doi.org/10.1103/PhysRevB.98.125140} {\bibfield  {journal} {\bibinfo
  {journal} {Phys. Rev. B}\ }\textbf {\bibinfo {volume} {98}},\ \bibinfo
  {pages} {125140} (\bibinfo {year} {2018})}\BibitemShut {NoStop}%
\bibitem [{\citenamefont {{Sai Gautam}}\ and\ \citenamefont
  {Carter}(2018)}]{SaiGautam2018}%
  \BibitemOpen
  \bibfield  {author} {\bibinfo {author} {\bibfnamefont {G.}~\bibnamefont {{Sai
  Gautam}}}\ and\ \bibinfo {author} {\bibfnamefont {E.~A.}\ \bibnamefont
  {Carter}},\ }\href {https://doi.org/10.1103/PhysRevMaterials.2.095401}
  {\bibfield  {journal} {\bibinfo  {journal} {Phys. Rev. Mater.}\ }\textbf
  {\bibinfo {volume} {2}},\ \bibinfo {pages} {095401} (\bibinfo {year}
  {2018})}\BibitemShut {NoStop}%
\bibitem [{\citenamefont {Zhang}\ \emph {et~al.}(2019)\citenamefont {Zhang},
  \citenamefont {Furness}, \citenamefont {Xiao},\ and\ \citenamefont
  {Sun}}]{Zhang2019}%
  \BibitemOpen
  \bibfield  {author} {\bibinfo {author} {\bibfnamefont {Y.}~\bibnamefont
  {Zhang}}, \bibinfo {author} {\bibfnamefont {J.~W.}\ \bibnamefont {Furness}},
  \bibinfo {author} {\bibfnamefont {B.}~\bibnamefont {Xiao}},\ and\ \bibinfo
  {author} {\bibfnamefont {J.}~\bibnamefont {Sun}},\ }\href
  {https://doi.org/10.1063/1.5055623} {\bibfield  {journal} {\bibinfo
  {journal} {J. Chem. Phys.}\ }\textbf {\bibinfo {volume} {150}},\ \bibinfo
  {pages} {014105} (\bibinfo {year} {2019})}\BibitemShut {NoStop}%
\bibitem [{\citenamefont {Zhang}\ \emph
  {et~al.}(2020{\natexlab{a}})\citenamefont {Zhang}, \citenamefont {Lane},
  \citenamefont {Furness}, \citenamefont {Barbiellini}, \citenamefont {Perdew},
  \citenamefont {Markiewicz}, \citenamefont {Bansil},\ and\ \citenamefont
  {Sun}}]{Zhang2020b}%
  \BibitemOpen
  \bibfield  {author} {\bibinfo {author} {\bibfnamefont {Y.}~\bibnamefont
  {Zhang}}, \bibinfo {author} {\bibfnamefont {C.}~\bibnamefont {Lane}},
  \bibinfo {author} {\bibfnamefont {J.~W.}\ \bibnamefont {Furness}}, \bibinfo
  {author} {\bibfnamefont {B.}~\bibnamefont {Barbiellini}}, \bibinfo {author}
  {\bibfnamefont {J.~P.}\ \bibnamefont {Perdew}}, \bibinfo {author}
  {\bibfnamefont {R.~S.}\ \bibnamefont {Markiewicz}}, \bibinfo {author}
  {\bibfnamefont {A.}~\bibnamefont {Bansil}},\ and\ \bibinfo {author}
  {\bibfnamefont {J.}~\bibnamefont {Sun}},\ }\href
  {https://doi.org/10.1073/pnas.1910411116} {\bibfield  {journal} {\bibinfo
  {journal} {Proc. Natl. Acad. Sci. U.S.A.}\ }\textbf {\bibinfo {volume}
  {117}},\ \bibinfo {pages} {68} (\bibinfo {year} {2020}{\natexlab{a}})},\
  \Eprint
  {https://arxiv.org/abs/https://www.pnas.org/content/117/1/68.full.pdf}
  {https://www.pnas.org/content/117/1/68.full.pdf} \BibitemShut {NoStop}%
\bibitem [{\citenamefont {Pulkkinen}\ \emph {et~al.}(2020)\citenamefont
  {Pulkkinen}, \citenamefont {Barbiellini}, \citenamefont {Nokelainen},
  \citenamefont {Sokolovskiy}, \citenamefont {Baigutlin}, \citenamefont
  {Miroshkina}, \citenamefont {Zagrebin}, \citenamefont {Buchelnikov},
  \citenamefont {Lane}, \citenamefont {Markiewicz}, \citenamefont {Bansil},
  \citenamefont {Sun}, \citenamefont {Pussi},\ and\ \citenamefont
  {L\"ahderanta}}]{Pulkkinen2020}%
  \BibitemOpen
  \bibfield  {author} {\bibinfo {author} {\bibfnamefont {A.}~\bibnamefont
  {Pulkkinen}}, \bibinfo {author} {\bibfnamefont {B.}~\bibnamefont
  {Barbiellini}}, \bibinfo {author} {\bibfnamefont {J.}~\bibnamefont
  {Nokelainen}}, \bibinfo {author} {\bibfnamefont {V.}~\bibnamefont
  {Sokolovskiy}}, \bibinfo {author} {\bibfnamefont {D.}~\bibnamefont
  {Baigutlin}}, \bibinfo {author} {\bibfnamefont {O.}~\bibnamefont
  {Miroshkina}}, \bibinfo {author} {\bibfnamefont {M.}~\bibnamefont
  {Zagrebin}}, \bibinfo {author} {\bibfnamefont {V.}~\bibnamefont
  {Buchelnikov}}, \bibinfo {author} {\bibfnamefont {C.}~\bibnamefont {Lane}},
  \bibinfo {author} {\bibfnamefont {R.~S.}\ \bibnamefont {Markiewicz}},
  \bibinfo {author} {\bibfnamefont {A.}~\bibnamefont {Bansil}}, \bibinfo
  {author} {\bibfnamefont {J.}~\bibnamefont {Sun}}, \bibinfo {author}
  {\bibfnamefont {K.}~\bibnamefont {Pussi}},\ and\ \bibinfo {author}
  {\bibfnamefont {E.}~\bibnamefont {L\"ahderanta}},\ }\href
  {https://doi.org/10.1103/PhysRevB.101.075115} {\bibfield  {journal} {\bibinfo
   {journal} {Phys. Rev. B}\ }\textbf {\bibinfo {volume} {101}},\ \bibinfo
  {pages} {075115} (\bibinfo {year} {2020})}\BibitemShut {NoStop}%
\bibitem [{\citenamefont {Zhang}\ \emph
  {et~al.}(2020{\natexlab{b}})\citenamefont {Zhang}, \citenamefont {Singh},
  \citenamefont {Lane}, \citenamefont {Kidd}, \citenamefont {Zhang},
  \citenamefont {Barbiellini}, \citenamefont {Markiewicz}, \citenamefont
  {Bansil},\ and\ \citenamefont {Sun}}]{Zhang2020a}%
  \BibitemOpen
  \bibfield  {author} {\bibinfo {author} {\bibfnamefont {R.}~\bibnamefont
  {Zhang}}, \bibinfo {author} {\bibfnamefont {B.}~\bibnamefont {Singh}},
  \bibinfo {author} {\bibfnamefont {C.}~\bibnamefont {Lane}}, \bibinfo {author}
  {\bibfnamefont {J.}~\bibnamefont {Kidd}}, \bibinfo {author} {\bibfnamefont
  {Y.}~\bibnamefont {Zhang}}, \bibinfo {author} {\bibfnamefont
  {B.}~\bibnamefont {Barbiellini}}, \bibinfo {author} {\bibfnamefont {R.~S.}\
  \bibnamefont {Markiewicz}}, \bibinfo {author} {\bibfnamefont
  {A.}~\bibnamefont {Bansil}},\ and\ \bibinfo {author} {\bibfnamefont
  {J.}~\bibnamefont {Sun}},\ }\Eprint {https://arxiv.org/abs/2003.11052}
  {arXiv:2003.11052 [cond-mat.str-el]}  (\bibinfo {year}
  {2020}{\natexlab{b}})\BibitemShut {NoStop}%
\bibitem [{\citenamefont {Ning}\ \emph {et~al.}(2021)\citenamefont {Ning},
  \citenamefont {Furness},\ and\ \citenamefont {Sun}}]{Ning2021}%
  \BibitemOpen
  \bibfield  {author} {\bibinfo {author} {\bibfnamefont {J.}~\bibnamefont
  {Ning}}, \bibinfo {author} {\bibfnamefont {J.~W.}\ \bibnamefont {Furness}},\
  and\ \bibinfo {author} {\bibfnamefont {J.}~\bibnamefont {Sun}},\ }\Eprint
  {https://arxiv.org/abs/2107.11850} {arXiv:2107.11850 [cond-mat.mtrl-sci]}
  (\bibinfo {year} {2021})\BibitemShut {NoStop}%
\bibitem [{\citenamefont {Bart{\'{o}}k}\ and\ \citenamefont
  {Yates}(2019{\natexlab{a}})}]{Bartok2019}%
  \BibitemOpen
  \bibfield  {author} {\bibinfo {author} {\bibfnamefont {A.~P.}\ \bibnamefont
  {Bart{\'{o}}k}}\ and\ \bibinfo {author} {\bibfnamefont {J.~R.}\ \bibnamefont
  {Yates}},\ }\href {https://doi.org/10.1063/1.5094646} {\bibfield  {journal}
  {\bibinfo  {journal} {J. Chem. Phys.}\ }\textbf {\bibinfo {volume} {150}},\
  \bibinfo {pages} {161101} (\bibinfo {year} {2019}{\natexlab{a}})}\BibitemShut
  {NoStop}%
\bibitem [{\citenamefont {Furness}\ and\ \citenamefont
  {Sun}(2019)}]{Furness2019}%
  \BibitemOpen
  \bibfield  {author} {\bibinfo {author} {\bibfnamefont {J.~W.}\ \bibnamefont
  {Furness}}\ and\ \bibinfo {author} {\bibfnamefont {J.}~\bibnamefont {Sun}},\
  }\href {https://doi.org/10.1103/PhysRevB.99.041119} {\bibfield  {journal}
  {\bibinfo  {journal} {Phys. Rev. B}\ }\textbf {\bibinfo {volume} {99}},\
  \bibinfo {pages} {041119} (\bibinfo {year} {2019})}\BibitemShut {NoStop}%
\bibitem [{\citenamefont {Mej{\'{i}}a-Rodr{\'{i}}guez}\ and\ \citenamefont
  {Trickey}(2019)}]{Mejia-Rodriguez2019}%
  \BibitemOpen
  \bibfield  {author} {\bibinfo {author} {\bibfnamefont {D.}~\bibnamefont
  {Mej{\'{i}}a-Rodr{\'{i}}guez}}\ and\ \bibinfo {author} {\bibfnamefont
  {S.~B.}\ \bibnamefont {Trickey}},\ }\href {https://doi.org/10.1063/1.5120408}
  {\bibfield  {journal} {\bibinfo  {journal} {J. Chem. Phys.}\ }\textbf
  {\bibinfo {volume} {151}},\ \bibinfo {pages} {207101} (\bibinfo {year}
  {2019})}\BibitemShut {NoStop}%
\bibitem [{\citenamefont {Bart{\'{o}}k}\ and\ \citenamefont
  {Yates}(2019{\natexlab{b}})}]{Bartok2019a}%
  \BibitemOpen
  \bibfield  {author} {\bibinfo {author} {\bibfnamefont {A.~P.}\ \bibnamefont
  {Bart{\'{o}}k}}\ and\ \bibinfo {author} {\bibfnamefont {J.~R.}\ \bibnamefont
  {Yates}},\ }\href {https://doi.org/10.1063/1.5128484} {\bibfield  {journal}
  {\bibinfo  {journal} {J. Chem. Phys.}\ }\textbf {\bibinfo {volume} {151}},\
  \bibinfo {pages} {207102} (\bibinfo {year} {2019}{\natexlab{b}})}\BibitemShut
  {NoStop}%
\bibitem [{\citenamefont {Furness}\ \emph
  {et~al.}(2020{\natexlab{a}})\citenamefont {Furness}, \citenamefont {Kaplan},
  \citenamefont {Ning}, \citenamefont {Perdew},\ and\ \citenamefont
  {Sun}}]{Furness2020c}%
  \BibitemOpen
  \bibfield  {author} {\bibinfo {author} {\bibfnamefont {J.~W.}\ \bibnamefont
  {Furness}}, \bibinfo {author} {\bibfnamefont {A.~D.}\ \bibnamefont {Kaplan}},
  \bibinfo {author} {\bibfnamefont {J.}~\bibnamefont {Ning}}, \bibinfo {author}
  {\bibfnamefont {J.~P.}\ \bibnamefont {Perdew}},\ and\ \bibinfo {author}
  {\bibfnamefont {J.}~\bibnamefont {Sun}},\ }\href
  {https://doi.org/10.1021/acs.jpclett.0c02405} {\bibfield  {journal} {\bibinfo
   {journal} {J. Phys. Chem. Lett.}\ }\textbf {\bibinfo {volume} {11}},\
  \bibinfo {pages} {8208} (\bibinfo {year} {2020}{\natexlab{a}})}\BibitemShut
  {NoStop}%
\bibitem [{\citenamefont {Mejia-Rodriguez}\ and\ \citenamefont
  {Trickey}(2020)}]{Mejia-Rodriguez2020f}%
  \BibitemOpen
  \bibfield  {author} {\bibinfo {author} {\bibfnamefont {D.}~\bibnamefont
  {Mejia-Rodriguez}}\ and\ \bibinfo {author} {\bibfnamefont {S.~B.}\
  \bibnamefont {Trickey}},\ }\href
  {https://doi.org/10.1103/PhysRevB.102.121109} {\bibfield  {journal} {\bibinfo
   {journal} {Phys. Rev. B}\ }\textbf {\bibinfo {volume} {102}},\ \bibinfo
  {pages} {121109(R)} (\bibinfo {year} {2020})}\BibitemShut {NoStop}%
\bibitem [{\citenamefont {Meji{\'{a}}-Rodr{\'{i}}guez}\ and\ \citenamefont
  {Trickey}(2020)}]{Mejia-Rodriguez2020g}%
  \BibitemOpen
  \bibfield  {author} {\bibinfo {author} {\bibfnamefont {D.}~\bibnamefont
  {Meji{\'{a}}-Rodr{\'{i}}guez}}\ and\ \bibinfo {author} {\bibfnamefont
  {S.~B.}\ \bibnamefont {Trickey}},\ }\href
  {https://doi.org/10.1021/acs.jpca.0c08883} {\bibfield  {journal} {\bibinfo
  {journal} {J. Phys. Chem. A}\ }\textbf {\bibinfo {volume} {124}},\ \bibinfo
  {pages} {9889} (\bibinfo {year} {2020})}\BibitemShut {NoStop}%
\bibitem [{\citenamefont {Ehlert}\ \emph {et~al.}(2021)\citenamefont {Ehlert},
  \citenamefont {Huniar}, \citenamefont {Ning}, \citenamefont {Furness},
  \citenamefont {Sun}, \citenamefont {Kaplan}, \citenamefont {Perdew},\ and\
  \citenamefont {Brandenburg}}]{Ehlert2021b}%
  \BibitemOpen
  \bibfield  {author} {\bibinfo {author} {\bibfnamefont {S.}~\bibnamefont
  {Ehlert}}, \bibinfo {author} {\bibfnamefont {U.}~\bibnamefont {Huniar}},
  \bibinfo {author} {\bibfnamefont {J.}~\bibnamefont {Ning}}, \bibinfo {author}
  {\bibfnamefont {J.~W.}\ \bibnamefont {Furness}}, \bibinfo {author}
  {\bibfnamefont {J.}~\bibnamefont {Sun}}, \bibinfo {author} {\bibfnamefont
  {A.~D.}\ \bibnamefont {Kaplan}}, \bibinfo {author} {\bibfnamefont {J.~P.}\
  \bibnamefont {Perdew}},\ and\ \bibinfo {author} {\bibfnamefont {J.~G.}\
  \bibnamefont {Brandenburg}},\ }\href {https://doi.org/10.1063/5.0041008}
  {\bibfield  {journal} {\bibinfo  {journal} {J. Chem. Phys.}\ }\textbf
  {\bibinfo {volume} {154}},\ \bibinfo {pages} {061101} (\bibinfo {year}
  {2021})}\BibitemShut {NoStop}%
\bibitem [{\citenamefont {Grimme}\ \emph {et~al.}(2021)\citenamefont {Grimme},
  \citenamefont {Hansen}, \citenamefont {Ehlert},\ and\ \citenamefont
  {Mewes}}]{Grimme2021}%
  \BibitemOpen
  \bibfield  {author} {\bibinfo {author} {\bibfnamefont {S.}~\bibnamefont
  {Grimme}}, \bibinfo {author} {\bibfnamefont {A.}~\bibnamefont {Hansen}},
  \bibinfo {author} {\bibfnamefont {S.}~\bibnamefont {Ehlert}},\ and\ \bibinfo
  {author} {\bibfnamefont {J.-M.}\ \bibnamefont {Mewes}},\ }\href
  {https://doi.org/10.1063/5.0040021} {\bibfield  {journal} {\bibinfo
  {journal} {J. Chem. Phys.}\ }\textbf {\bibinfo {volume} {154}},\ \bibinfo
  {pages} {064103} (\bibinfo {year} {2021})}\BibitemShut {NoStop}%
\bibitem [{\citenamefont {Goerigk}\ \emph {et~al.}(2017)\citenamefont
  {Goerigk}, \citenamefont {Hansen}, \citenamefont {Bauer}, \citenamefont
  {Ehrlich}, \citenamefont {Najibi},\ and\ \citenamefont
  {Grimme}}]{Goerigk2017}%
  \BibitemOpen
  \bibfield  {author} {\bibinfo {author} {\bibfnamefont {L.}~\bibnamefont
  {Goerigk}}, \bibinfo {author} {\bibfnamefont {A.}~\bibnamefont {Hansen}},
  \bibinfo {author} {\bibfnamefont {C.}~\bibnamefont {Bauer}}, \bibinfo
  {author} {\bibfnamefont {S.}~\bibnamefont {Ehrlich}}, \bibinfo {author}
  {\bibfnamefont {A.}~\bibnamefont {Najibi}},\ and\ \bibinfo {author}
  {\bibfnamefont {S.}~\bibnamefont {Grimme}},\ }\href
  {https://doi.org/10.1039/C7CP04913G} {\bibfield  {journal} {\bibinfo
  {journal} {Phys. Chem. Chem. Phys.}\ }\textbf {\bibinfo {volume} {19}},\
  \bibinfo {pages} {32184} (\bibinfo {year} {2017})}\BibitemShut {NoStop}%
\bibitem [{\citenamefont {Santra}\ and\ \citenamefont
  {Martin}(2021)}]{Santra2021}%
  \BibitemOpen
  \bibfield  {author} {\bibinfo {author} {\bibfnamefont {G.}~\bibnamefont
  {Santra}}\ and\ \bibinfo {author} {\bibfnamefont {J.~M.~L.}\ \bibnamefont
  {Martin}},\ }\href {https://doi.org/10.1021/acs.jctc.0c01055} {\bibfield
  {journal} {\bibinfo  {journal} {J. Chem. Theory Comput.}\ }\textbf {\bibinfo
  {volume} {17}},\ \bibinfo {pages} {1368–1379} (\bibinfo {year}
  {2021})}\BibitemShut {NoStop}%
\bibitem [{\citenamefont {Song}\ \emph {et~al.}(2021)\citenamefont {Song},
  \citenamefont {Vuckovic}, \citenamefont {Sim},\ and\ \citenamefont
  {Burke}}]{Song2021}%
  \BibitemOpen
  \bibfield  {author} {\bibinfo {author} {\bibfnamefont {S.}~\bibnamefont
  {Song}}, \bibinfo {author} {\bibfnamefont {S.}~\bibnamefont {Vuckovic}},
  \bibinfo {author} {\bibfnamefont {E.}~\bibnamefont {Sim}},\ and\ \bibinfo
  {author} {\bibfnamefont {K.}~\bibnamefont {Burke}},\ }\href
  {https://doi.org/10.1021/acs.jpclett.0c03545} {\bibfield  {journal} {\bibinfo
   {journal} {The Journal of Physical Chemistry Letters}\ }\textbf {\bibinfo
  {volume} {12}},\ \bibinfo {pages} {800} (\bibinfo {year} {2021})},\ \bibinfo
  {note} {pMID: 33411542},\ \Eprint
  {https://arxiv.org/abs/https://doi.org/10.1021/acs.jpclett.0c03545}
  {https://doi.org/10.1021/acs.jpclett.0c03545} \BibitemShut {NoStop}%
\bibitem [{\citenamefont {Dasgupta}\ \emph {et~al.}(2021)\citenamefont
  {Dasgupta}, \citenamefont {Lambros}, \citenamefont {Perdew},\ and\
  \citenamefont {Paesani}}]{Dasgupta2021}%
  \BibitemOpen
  \bibfield  {author} {\bibinfo {author} {\bibfnamefont {S.}~\bibnamefont
  {Dasgupta}}, \bibinfo {author} {\bibfnamefont {E.}~\bibnamefont {Lambros}},
  \bibinfo {author} {\bibfnamefont {J.}~\bibnamefont {Perdew}},\ and\ \bibinfo
  {author} {\bibfnamefont {F.}~\bibnamefont {Paesani}},\ }\bibfield  {journal}
  {\bibinfo  {journal} {chemrXiv}\ }\href
  {https://doi.org/10.33774/chemrxiv-2021-hstgf-v2}
  {10.33774/chemrxiv-2021-hstgf-v2} (\bibinfo {year} {2021})\BibitemShut
  {NoStop}%
\bibitem [{\citenamefont {Becke}\ and\ \citenamefont
  {Edgecombe}(1990)}]{Becke1990}%
  \BibitemOpen
  \bibfield  {author} {\bibinfo {author} {\bibfnamefont {A.~D.}\ \bibnamefont
  {Becke}}\ and\ \bibinfo {author} {\bibfnamefont {K.~E.}\ \bibnamefont
  {Edgecombe}},\ }\href {https://doi.org/10.1063/1.458517} {\bibfield
  {journal} {\bibinfo  {journal} {J. Chem. Phys.}\ }\textbf {\bibinfo {volume}
  {92}},\ \bibinfo {pages} {5397} (\bibinfo {year} {1990})}\BibitemShut
  {NoStop}%
\bibitem [{\citenamefont {Ruzsinszky}\ \emph {et~al.}(2012)\citenamefont
  {Ruzsinszky}, \citenamefont {Sun}, \citenamefont {Xiao},\ and\ \citenamefont
  {Csonka}}]{Ruzsinszky2012}%
  \BibitemOpen
  \bibfield  {author} {\bibinfo {author} {\bibfnamefont {A.}~\bibnamefont
  {Ruzsinszky}}, \bibinfo {author} {\bibfnamefont {J.}~\bibnamefont {Sun}},
  \bibinfo {author} {\bibfnamefont {B.}~\bibnamefont {Xiao}},\ and\ \bibinfo
  {author} {\bibfnamefont {G.~I.}\ \bibnamefont {Csonka}},\ }\href
  {https://doi.org/10.1021/ct300269u} {\bibfield  {journal} {\bibinfo
  {journal} {J. Chem. Theory Comput.}\ }\textbf {\bibinfo {volume} {8}},\
  \bibinfo {pages} {2078} (\bibinfo {year} {2012})}\BibitemShut {NoStop}%
\bibitem [{\citenamefont {Sun}\ \emph {et~al.}(2012)\citenamefont {Sun},
  \citenamefont {Xiao},\ and\ \citenamefont {Ruzsinszky}}]{Sun2012}%
  \BibitemOpen
  \bibfield  {author} {\bibinfo {author} {\bibfnamefont {J.}~\bibnamefont
  {Sun}}, \bibinfo {author} {\bibfnamefont {B.}~\bibnamefont {Xiao}},\ and\
  \bibinfo {author} {\bibfnamefont {A.}~\bibnamefont {Ruzsinszky}},\ }\href
  {https://doi.org/10.1063/1.4742312} {\bibfield  {journal} {\bibinfo
  {journal} {J. Chem. Phys.}\ }\textbf {\bibinfo {volume} {137}},\ \bibinfo
  {pages} {051101} (\bibinfo {year} {2012})},\ \Eprint
  {https://arxiv.org/abs/https://doi.org/10.1063/1.4742312}
  {https://doi.org/10.1063/1.4742312} \BibitemShut {NoStop}%
\bibitem [{\citenamefont {Sun}\ \emph {et~al.}(2013)\citenamefont {Sun},
  \citenamefont {Xiao}, \citenamefont {Fang}, \citenamefont {Haunschild},
  \citenamefont {Hao}, \citenamefont {Ruzsinszky}, \citenamefont {Csonka},
  \citenamefont {Scuseria},\ and\ \citenamefont {Perdew}}]{Sun2013a}%
  \BibitemOpen
  \bibfield  {author} {\bibinfo {author} {\bibfnamefont {J.}~\bibnamefont
  {Sun}}, \bibinfo {author} {\bibfnamefont {B.}~\bibnamefont {Xiao}}, \bibinfo
  {author} {\bibfnamefont {Y.}~\bibnamefont {Fang}}, \bibinfo {author}
  {\bibfnamefont {R.}~\bibnamefont {Haunschild}}, \bibinfo {author}
  {\bibfnamefont {P.}~\bibnamefont {Hao}}, \bibinfo {author} {\bibfnamefont
  {A.}~\bibnamefont {Ruzsinszky}}, \bibinfo {author} {\bibfnamefont {G.~I.}\
  \bibnamefont {Csonka}}, \bibinfo {author} {\bibfnamefont {G.~E.}\
  \bibnamefont {Scuseria}},\ and\ \bibinfo {author} {\bibfnamefont {J.~P.}\
  \bibnamefont {Perdew}},\ }\href
  {https://doi.org/10.1103/PhysRevLett.111.106401} {\bibfield  {journal}
  {\bibinfo  {journal} {Phys. Rev. Lett.}\ }\textbf {\bibinfo {volume} {111}},\
  \bibinfo {pages} {106401} (\bibinfo {year} {2013})}\BibitemShut {NoStop}%
\bibitem [{\citenamefont {Furness}\ \emph
  {et~al.}(2020{\natexlab{b}})\citenamefont {Furness}, \citenamefont
  {Sengupta}, \citenamefont {Ning}, \citenamefont {Ruzsinszky},\ and\
  \citenamefont {Sun}}]{Furness2020a}%
  \BibitemOpen
  \bibfield  {author} {\bibinfo {author} {\bibfnamefont {J.~W.}\ \bibnamefont
  {Furness}}, \bibinfo {author} {\bibfnamefont {N.}~\bibnamefont {Sengupta}},
  \bibinfo {author} {\bibfnamefont {J.}~\bibnamefont {Ning}}, \bibinfo {author}
  {\bibfnamefont {A.}~\bibnamefont {Ruzsinszky}},\ and\ \bibinfo {author}
  {\bibfnamefont {J.}~\bibnamefont {Sun}},\ }\href
  {https://doi.org/10.1063/5.0008014} {\bibfield  {journal} {\bibinfo
  {journal} {J. Chem. Phys.}\ }\textbf {\bibinfo {volume} {152}},\ \bibinfo
  {pages} {244112} (\bibinfo {year} {2020}{\natexlab{b}})},\ \Eprint
  {https://arxiv.org/abs/https://doi.org/10.1063/5.0008014}
  {https://doi.org/10.1063/5.0008014} \BibitemShut {NoStop}%
\bibitem [{\citenamefont {Levy}\ and\ \citenamefont {Perdew}(1985)}]{Levy1985}%
  \BibitemOpen
  \bibfield  {author} {\bibinfo {author} {\bibfnamefont {M.}~\bibnamefont
  {Levy}}\ and\ \bibinfo {author} {\bibfnamefont {J.~P.}\ \bibnamefont
  {Perdew}},\ }\href {https://doi.org/10.1103/PhysRevA.32.2010} {\bibfield
  {journal} {\bibinfo  {journal} {Phys. Rev. A}\ }\textbf {\bibinfo {volume}
  {32}},\ \bibinfo {pages} {2010} (\bibinfo {year} {1985})}\BibitemShut
  {NoStop}%
\bibitem [{\citenamefont {Levy}(1991)}]{Levy1991}%
  \BibitemOpen
  \bibfield  {author} {\bibinfo {author} {\bibfnamefont {M.}~\bibnamefont
  {Levy}},\ }\href {https://doi.org/10.1103/PhysRevA.43.4637} {\bibfield
  {journal} {\bibinfo  {journal} {Phys. Rev. A}\ }\textbf {\bibinfo {volume}
  {43}},\ \bibinfo {pages} {4637} (\bibinfo {year} {1991})}\BibitemShut
  {NoStop}%
\bibitem [{\citenamefont {Levy}\ and\ \citenamefont {Perdew}(1993)}]{Levy1993}%
  \BibitemOpen
  \bibfield  {author} {\bibinfo {author} {\bibfnamefont {M.}~\bibnamefont
  {Levy}}\ and\ \bibinfo {author} {\bibfnamefont {J.~P.}\ \bibnamefont
  {Perdew}},\ }\href {https://doi.org/10.1103/PhysRevB.48.11638} {\bibfield
  {journal} {\bibinfo  {journal} {Phys. Rev. B}\ }\textbf {\bibinfo {volume}
  {48}},\ \bibinfo {pages} {11638} (\bibinfo {year} {1993})}\BibitemShut
  {NoStop}%
\bibitem [{\citenamefont {G\"orling}\ and\ \citenamefont
  {Levy}(1992)}]{Gorling1992}%
  \BibitemOpen
  \bibfield  {author} {\bibinfo {author} {\bibfnamefont {A.}~\bibnamefont
  {G\"orling}}\ and\ \bibinfo {author} {\bibfnamefont {M.}~\bibnamefont
  {Levy}},\ }\href {https://doi.org/10.1103/PhysRevA.45.1509} {\bibfield
  {journal} {\bibinfo  {journal} {Phys. Rev. A}\ }\textbf {\bibinfo {volume}
  {45}},\ \bibinfo {pages} {1509} (\bibinfo {year} {1992})}\BibitemShut
  {NoStop}%
\bibitem [{\citenamefont {Pollack}\ and\ \citenamefont
  {Perdew}(2000)}]{Pollack2000}%
  \BibitemOpen
  \bibfield  {author} {\bibinfo {author} {\bibfnamefont {L.}~\bibnamefont
  {Pollack}}\ and\ \bibinfo {author} {\bibfnamefont {J.~P.}\ \bibnamefont
  {Perdew}},\ }\href {https://doi.org/10.1088/0953-8984/12/7/308} {\bibfield
  {journal} {\bibinfo  {journal} {J. Phys. Condens. Matter}\ }\textbf {\bibinfo
  {volume} {12}},\ \bibinfo {pages} {1239} (\bibinfo {year}
  {2000})}\BibitemShut {NoStop}%
\bibitem [{\citenamefont {Chiodo}\ \emph {et~al.}(2012)\citenamefont {Chiodo},
  \citenamefont {Constantin}, \citenamefont {Fabiano},\ and\ \citenamefont
  {Della~Sala}}]{Chiodo2012}%
  \BibitemOpen
  \bibfield  {author} {\bibinfo {author} {\bibfnamefont {L.}~\bibnamefont
  {Chiodo}}, \bibinfo {author} {\bibfnamefont {L.~A.}\ \bibnamefont
  {Constantin}}, \bibinfo {author} {\bibfnamefont {E.}~\bibnamefont
  {Fabiano}},\ and\ \bibinfo {author} {\bibfnamefont {F.}~\bibnamefont
  {Della~Sala}},\ }\href {https://doi.org/10.1103/PhysRevLett.108.126402}
  {\bibfield  {journal} {\bibinfo  {journal} {Phys. Rev. Lett.}\ }\textbf
  {\bibinfo {volume} {108}},\ \bibinfo {pages} {126402} (\bibinfo {year}
  {2012})}\BibitemShut {NoStop}%
\bibitem [{\citenamefont {Perdew}\ \emph {et~al.}(2014)\citenamefont {Perdew},
  \citenamefont {Ruzsinszky}, \citenamefont {Sun},\ and\ \citenamefont
  {Burke}}]{Perdew2014}%
  \BibitemOpen
  \bibfield  {author} {\bibinfo {author} {\bibfnamefont {J.~P.}\ \bibnamefont
  {Perdew}}, \bibinfo {author} {\bibfnamefont {A.}~\bibnamefont {Ruzsinszky}},
  \bibinfo {author} {\bibfnamefont {J.}~\bibnamefont {Sun}},\ and\ \bibinfo
  {author} {\bibfnamefont {K.}~\bibnamefont {Burke}},\ }\href
  {https://doi.org/10.1063/1.4870763} {\bibfield  {journal} {\bibinfo
  {journal} {J. Chem. Phys.}\ }\textbf {\bibinfo {volume} {140}},\ \bibinfo
  {pages} {18A533} (\bibinfo {year} {2014})},\ \Eprint
  {https://arxiv.org/abs/https://doi.org/10.1063/1.4870763}
  {https://doi.org/10.1063/1.4870763} \BibitemShut {NoStop}%
\bibitem [{\citenamefont {Furness}\ and\ \citenamefont
  {Lehtola}(2021)}]{Furness2021a}%
  \BibitemOpen
  \bibfield  {author} {\bibinfo {author} {\bibfnamefont {J.~W.}\ \bibnamefont
  {Furness}}\ and\ \bibinfo {author} {\bibfnamefont {S.}~\bibnamefont
  {Lehtola}},\ }\href {https://github.com/JFurness1/AtomicOrbitals} {\bibinfo
  {title} {{Hartree-Fock Orbitals for Spherical Atoms - a python toolbox}}}
  (\bibinfo {year} {2021})\BibitemShut {NoStop}%
\bibitem [{\citenamefont {Clementi}\ and\ \citenamefont
  {Roetti}(1974)}]{Clementi1974}%
  \BibitemOpen
  \bibfield  {author} {\bibinfo {author} {\bibfnamefont {E.}~\bibnamefont
  {Clementi}}\ and\ \bibinfo {author} {\bibfnamefont {C.}~\bibnamefont
  {Roetti}},\ }\href {https://doi.org/10.1016/S0092-640X(74)80016-1} {\bibfield
   {journal} {\bibinfo  {journal} {At. Data Nucl. Data Tables}\ }\textbf
  {\bibinfo {volume} {14}},\ \bibinfo {pages} {177} (\bibinfo {year}
  {1974})}\BibitemShut {NoStop}%
\bibitem [{\citenamefont {Yang}\ \emph {et~al.}(2016)\citenamefont {Yang},
  \citenamefont {Peng}, \citenamefont {Sun},\ and\ \citenamefont
  {Perdew}}]{Yang2016}%
  \BibitemOpen
  \bibfield  {author} {\bibinfo {author} {\bibfnamefont {Z.-h.}\ \bibnamefont
  {Yang}}, \bibinfo {author} {\bibfnamefont {H.}~\bibnamefont {Peng}}, \bibinfo
  {author} {\bibfnamefont {J.}~\bibnamefont {Sun}},\ and\ \bibinfo {author}
  {\bibfnamefont {J.~P.}\ \bibnamefont {Perdew}},\ }\href
  {https://doi.org/10.1103/PhysRevB.93.205205} {\bibfield  {journal} {\bibinfo
  {journal} {Phys. Rev. B}\ }\textbf {\bibinfo {volume} {93}},\ \bibinfo
  {pages} {205205} (\bibinfo {year} {2016})}\BibitemShut {NoStop}%
\bibitem [{\citenamefont {Kirzhnits}(1957)}]{Kirzhnits1957}%
  \BibitemOpen
  \bibfield  {author} {\bibinfo {author} {\bibfnamefont {D.~A.}\ \bibnamefont
  {Kirzhnits}},\ }\href@noop {} {\bibfield  {journal} {\bibinfo  {journal} {J.
  Exp. Theor. Phys.}\ }\textbf {\bibinfo {volume} {5}},\ \bibinfo {pages} {64}
  (\bibinfo {year} {1957})}\BibitemShut {NoStop}%
\bibitem [{\citenamefont {Svendsen}\ and\ \citenamefont {von
  Barth}(1996)}]{Svendsen1996}%
  \BibitemOpen
  \bibfield  {author} {\bibinfo {author} {\bibfnamefont {P.}~\bibnamefont
  {Svendsen}}\ and\ \bibinfo {author} {\bibfnamefont {U.}~\bibnamefont {von
  Barth}},\ }\href {https://doi.org/10.1103/PhysRevB.54.17402} {\bibfield
  {journal} {\bibinfo  {journal} {Phys. Rev. B}\ }\textbf {\bibinfo {volume}
  {54}},\ \bibinfo {pages} {17402} (\bibinfo {year} {1996})}\BibitemShut
  {NoStop}%
\bibitem [{\citenamefont {Ma}\ and\ \citenamefont {Brueckner}(1968)}]{Ma1968}%
  \BibitemOpen
  \bibfield  {author} {\bibinfo {author} {\bibfnamefont {S.-K.}\ \bibnamefont
  {Ma}}\ and\ \bibinfo {author} {\bibfnamefont {K.~A.}\ \bibnamefont
  {Brueckner}},\ }\href {https://doi.org/10.1103/PhysRev.165.18} {\bibfield
  {journal} {\bibinfo  {journal} {Phys. Rev.}\ }\textbf {\bibinfo {volume}
  {165}},\ \bibinfo {pages} {18} (\bibinfo {year} {1968})}\BibitemShut
  {NoStop}%
\bibitem [{\citenamefont {Wang}\ and\ \citenamefont {Perdew}(1991)}]{Wang1991}%
  \BibitemOpen
  \bibfield  {author} {\bibinfo {author} {\bibfnamefont {Y.}~\bibnamefont
  {Wang}}\ and\ \bibinfo {author} {\bibfnamefont {J.~P.}\ \bibnamefont
  {Perdew}},\ }\href {https://doi.org/10.1103/PhysRevB.43.8911} {\bibfield
  {journal} {\bibinfo  {journal} {Phys. Rev. B}\ }\textbf {\bibinfo {volume}
  {43}},\ \bibinfo {pages} {8911} (\bibinfo {year} {1991})}\BibitemShut
  {NoStop}%
\bibitem [{\citenamefont {Perdew}\ \emph {et~al.}(1996)\citenamefont {Perdew},
  \citenamefont {Burke},\ and\ \citenamefont {Ernzerhof}}]{Perdew1996}%
  \BibitemOpen
  \bibfield  {author} {\bibinfo {author} {\bibfnamefont {J.~P.}\ \bibnamefont
  {Perdew}}, \bibinfo {author} {\bibfnamefont {K.}~\bibnamefont {Burke}},\ and\
  \bibinfo {author} {\bibfnamefont {M.}~\bibnamefont {Ernzerhof}},\ }\href@noop
  {} {\bibfield  {journal} {\bibinfo  {journal} {Phys. Rev. Lett.}\ }\textbf
  {\bibinfo {volume} {77}},\ \bibinfo {pages} {3865} (\bibinfo {year}
  {1996})}\BibitemShut {NoStop}%
\bibitem [{\citenamefont {Sun}\ \emph {et~al.}(2015{\natexlab{b}})\citenamefont
  {Sun}, \citenamefont {Perdew},\ and\ \citenamefont {Ruzsinszky}}]{Sun2015a}%
  \BibitemOpen
  \bibfield  {author} {\bibinfo {author} {\bibfnamefont {J.}~\bibnamefont
  {Sun}}, \bibinfo {author} {\bibfnamefont {J.~P.}\ \bibnamefont {Perdew}},\
  and\ \bibinfo {author} {\bibfnamefont {A.}~\bibnamefont {Ruzsinszky}},\
  }\href {https://doi.org/10.1073/pnas.1423145112} {\bibfield  {journal}
  {\bibinfo  {journal} {Proc. Natl. Acad. Sci.}\ }\textbf {\bibinfo {volume}
  {112}},\ \bibinfo {pages} {685} (\bibinfo {year}
  {2015}{\natexlab{b}})}\BibitemShut {NoStop}%
\bibitem [{\citenamefont {Perdew}\ and\ \citenamefont
  {Wang}(1992)}]{Perdew1992a}%
  \BibitemOpen
  \bibfield  {author} {\bibinfo {author} {\bibfnamefont {J.~P.}\ \bibnamefont
  {Perdew}}\ and\ \bibinfo {author} {\bibfnamefont {Y.}~\bibnamefont {Wang}},\
  }\href {https://doi.org/10.1103/PhysRevB.45.13244} {\bibfield  {journal}
  {\bibinfo  {journal} {Phys. Rev. B}\ }\textbf {\bibinfo {volume} {45}},\
  \bibinfo {pages} {13244} (\bibinfo {year} {1992})}\BibitemShut {NoStop}%
\bibitem [{\citenamefont {Becke}(1988)}]{Becke1988}%
  \BibitemOpen
  \bibfield  {author} {\bibinfo {author} {\bibfnamefont {A.~D.}\ \bibnamefont
  {Becke}},\ }\href {https://doi.org/10.1103/PhysRevA.38.3098} {\bibfield
  {journal} {\bibinfo  {journal} {Phys. Rev. A}\ }\textbf {\bibinfo {volume}
  {38}},\ \bibinfo {pages} {3098} (\bibinfo {year} {1988})}\BibitemShut
  {NoStop}%
\bibitem [{\citenamefont {Chakravorty}\ \emph {et~al.}(1993)\citenamefont
  {Chakravorty}, \citenamefont {Gwaltney},\ and\ \citenamefont
  {Davidson}}]{Chakravorty1993}%
  \BibitemOpen
  \bibfield  {author} {\bibinfo {author} {\bibfnamefont {S.~J.}\ \bibnamefont
  {Chakravorty}}, \bibinfo {author} {\bibfnamefont {S.~R.}\ \bibnamefont
  {Gwaltney}},\ and\ \bibinfo {author} {\bibfnamefont {E.~R.}\ \bibnamefont
  {Davidson}},\ }\href {https://doi.org/10.1103/PhysRevA.47.3649} {\bibfield
  {journal} {\bibinfo  {journal} {Phys. Rev. A}\ }\textbf {\bibinfo {volume}
  {47}},\ \bibinfo {pages} {3649} (\bibinfo {year} {1993})}\BibitemShut
  {NoStop}%
\bibitem [{\citenamefont {McCarthy}\ and\ \citenamefont
  {Thakkar}(2011)}]{McCarthy2011}%
  \BibitemOpen
  \bibfield  {author} {\bibinfo {author} {\bibfnamefont {S.~P.}\ \bibnamefont
  {McCarthy}}\ and\ \bibinfo {author} {\bibfnamefont {A.~J.}\ \bibnamefont
  {Thakkar}},\ }\href {https://doi.org/10.1063/1.3547262} {\bibfield  {journal}
  {\bibinfo  {journal} {J. Chem. Phys.}\ }\textbf {\bibinfo {volume} {134}},\
  \bibinfo {pages} {044102} (\bibinfo {year} {2011})}\BibitemShut {NoStop}%
\bibitem [{\citenamefont {Wood}\ \emph {et~al.}(2007)\citenamefont {Wood},
  \citenamefont {Hine}, \citenamefont {Foulkes},\ and\ \citenamefont
  {Garc{\'{i}}a-Gonz{\'{a}}lez}}]{Wood2007}%
  \BibitemOpen
  \bibfield  {author} {\bibinfo {author} {\bibfnamefont {B.}~\bibnamefont
  {Wood}}, \bibinfo {author} {\bibfnamefont {N.~D.~M.}\ \bibnamefont {Hine}},
  \bibinfo {author} {\bibfnamefont {W.~M.~C.}\ \bibnamefont {Foulkes}},\ and\
  \bibinfo {author} {\bibfnamefont {P.}~\bibnamefont
  {Garc{\'{i}}a-Gonz{\'{a}}lez}},\ }\href
  {https://doi.org/10.1103/PhysRevB.76.035403} {\bibfield  {journal} {\bibinfo
  {journal} {Phys. Rev. B}\ }\textbf {\bibinfo {volume} {76}},\ \bibinfo
  {pages} {035403} (\bibinfo {year} {2007})}\BibitemShut {NoStop}%
\bibitem [{\citenamefont {Lang}\ and\ \citenamefont {Kohn}(1970)}]{Lang1970}%
  \BibitemOpen
  \bibfield  {author} {\bibinfo {author} {\bibfnamefont {N.~D.}\ \bibnamefont
  {Lang}}\ and\ \bibinfo {author} {\bibfnamefont {W.}~\bibnamefont {Kohn}},\
  }\href {https://doi.org/10.1103/PhysRevB.1.4555} {\bibfield  {journal}
  {\bibinfo  {journal} {Phys. Rev. B}\ }\textbf {\bibinfo {volume} {1}},\
  \bibinfo {pages} {4555} (\bibinfo {year} {1970})}\BibitemShut {NoStop}%
\bibitem [{\citenamefont {Patkowski}\ \emph {et~al.}(2005)\citenamefont
  {Patkowski}, \citenamefont {Murdachaew}, \citenamefont {Fou},\ and\
  \citenamefont {Szalewicz}}]{Patkowski2005}%
  \BibitemOpen
  \bibfield  {author} {\bibinfo {author} {\bibfnamefont {K.}~\bibnamefont
  {Patkowski}}, \bibinfo {author} {\bibfnamefont {G.}~\bibnamefont
  {Murdachaew}}, \bibinfo {author} {\bibfnamefont {C.~M.}\ \bibnamefont
  {Fou}},\ and\ \bibinfo {author} {\bibfnamefont {K.}~\bibnamefont
  {Szalewicz}},\ }\href {https://doi.org/10.1080/00268970500130241} {\bibfield
  {journal} {\bibinfo  {journal} {Mol. Phys.}\ }\textbf {\bibinfo {volume}
  {103}},\ \bibinfo {pages} {2031} (\bibinfo {year} {2005})}\BibitemShut
  {NoStop}%
\bibitem [{\citenamefont {Curtiss}\ \emph {et~al.}(2000)\citenamefont
  {Curtiss}, \citenamefont {Raghavachari}, \citenamefont {Redfern},\ and\
  \citenamefont {Pople}}]{Curtiss2000}%
  \BibitemOpen
  \bibfield  {author} {\bibinfo {author} {\bibfnamefont {L.~A.}\ \bibnamefont
  {Curtiss}}, \bibinfo {author} {\bibfnamefont {K.}~\bibnamefont
  {Raghavachari}}, \bibinfo {author} {\bibfnamefont {P.~C.}\ \bibnamefont
  {Redfern}},\ and\ \bibinfo {author} {\bibfnamefont {J.~A.}\ \bibnamefont
  {Pople}},\ }\href {https://doi.org/10.1063/1.481336} {\bibfield  {journal}
  {\bibinfo  {journal} {J. Chem. Phys.}\ }\textbf {\bibinfo {volume} {112}},\
  \bibinfo {pages} {7374} (\bibinfo {year} {2000})}\BibitemShut {NoStop}%
\bibitem [{\citenamefont {Jure{\v{c}}ka}\ \emph {et~al.}(2006)\citenamefont
  {Jure{\v{c}}ka}, \citenamefont {{\v{S}}poner}, \citenamefont
  {{\v{C}}ern{\'{y}}},\ and\ \citenamefont {Hobza}}]{Jurecka2006}%
  \BibitemOpen
  \bibfield  {author} {\bibinfo {author} {\bibfnamefont {P.}~\bibnamefont
  {Jure{\v{c}}ka}}, \bibinfo {author} {\bibfnamefont {J.}~\bibnamefont
  {{\v{S}}poner}}, \bibinfo {author} {\bibfnamefont {J.}~\bibnamefont
  {{\v{C}}ern{\'{y}}}},\ and\ \bibinfo {author} {\bibfnamefont
  {P.}~\bibnamefont {Hobza}},\ }\href {https://doi.org/10.1039/b600027d}
  {\bibfield  {journal} {\bibinfo  {journal} {Phys. Chem. Chem. Phys.}\
  }\textbf {\bibinfo {volume} {8}},\ \bibinfo {pages} {1985} (\bibinfo {year}
  {2006})}\BibitemShut {NoStop}%
\bibitem [{\citenamefont {Zhao}\ \emph {et~al.}(2005)\citenamefont {Zhao},
  \citenamefont {Gonz{\'{a}}lez-Garda},\ and\ \citenamefont
  {Truhlar}}]{Zhao2005}%
  \BibitemOpen
  \bibfield  {author} {\bibinfo {author} {\bibfnamefont {Y.}~\bibnamefont
  {Zhao}}, \bibinfo {author} {\bibfnamefont {N.}~\bibnamefont
  {Gonz{\'{a}}lez-Garda}},\ and\ \bibinfo {author} {\bibfnamefont {D.~G.}\
  \bibnamefont {Truhlar}},\ }\href {https://doi.org/10.1021/jp045141s}
  {\bibfield  {journal} {\bibinfo  {journal} {J. Phys. Chem. A}\ }\textbf
  {\bibinfo {volume} {109}},\ \bibinfo {pages} {2012} (\bibinfo {year}
  {2005})}\BibitemShut {NoStop}%
\bibitem [{\citenamefont {Sun}\ \emph {et~al.}(2011)\citenamefont {Sun},
  \citenamefont {Marsman}, \citenamefont {Csonka}, \citenamefont {Ruzsinszky},
  \citenamefont {Hao}, \citenamefont {Kim}, \citenamefont {Kresse},\ and\
  \citenamefont {Perdew}}]{Sun2011}%
  \BibitemOpen
  \bibfield  {author} {\bibinfo {author} {\bibfnamefont {J.}~\bibnamefont
  {Sun}}, \bibinfo {author} {\bibfnamefont {M.}~\bibnamefont {Marsman}},
  \bibinfo {author} {\bibfnamefont {G.~I.}\ \bibnamefont {Csonka}}, \bibinfo
  {author} {\bibfnamefont {A.}~\bibnamefont {Ruzsinszky}}, \bibinfo {author}
  {\bibfnamefont {P.}~\bibnamefont {Hao}}, \bibinfo {author} {\bibfnamefont
  {Y.~S.}\ \bibnamefont {Kim}}, \bibinfo {author} {\bibfnamefont
  {G.}~\bibnamefont {Kresse}},\ and\ \bibinfo {author} {\bibfnamefont {J.~P.}\
  \bibnamefont {Perdew}},\ }\href {https://doi.org/10.1103/PhysRevB.84.035117}
  {\bibfield  {journal} {\bibinfo  {journal} {Phys. Rev. B}\ }\textbf {\bibinfo
  {volume} {84}},\ \bibinfo {pages} {035117} (\bibinfo {year}
  {2011})}\BibitemShut {NoStop}%
\bibitem [{\citenamefont {Staroverov}\ \emph {et~al.}(2004)\citenamefont
  {Staroverov}, \citenamefont {Scuseria}, \citenamefont {Tao},\ and\
  \citenamefont {Perdew}}]{Staroverov2004}%
  \BibitemOpen
  \bibfield  {author} {\bibinfo {author} {\bibfnamefont {V.~N.}\ \bibnamefont
  {Staroverov}}, \bibinfo {author} {\bibfnamefont {G.~E.}\ \bibnamefont
  {Scuseria}}, \bibinfo {author} {\bibfnamefont {J.}~\bibnamefont {Tao}},\ and\
  \bibinfo {author} {\bibfnamefont {J.~P.}\ \bibnamefont {Perdew}},\ }\href
  {https://doi.org/10.1103/PhysRevB.69.075102} {\bibfield  {journal} {\bibinfo
  {journal} {Phys. Rev. B}\ }\textbf {\bibinfo {volume} {69}},\ \bibinfo
  {pages} {075102} (\bibinfo {year} {2004})}\BibitemShut {NoStop}%
\bibitem [{\citenamefont {Hao}\ \emph {et~al.}(2012)\citenamefont {Hao},
  \citenamefont {Fang}, \citenamefont {Sun}, \citenamefont {Csonka},
  \citenamefont {Philipsen},\ and\ \citenamefont {Perdew}}]{Hao2012}%
  \BibitemOpen
  \bibfield  {author} {\bibinfo {author} {\bibfnamefont {P.}~\bibnamefont
  {Hao}}, \bibinfo {author} {\bibfnamefont {Y.}~\bibnamefont {Fang}}, \bibinfo
  {author} {\bibfnamefont {J.}~\bibnamefont {Sun}}, \bibinfo {author}
  {\bibfnamefont {G.~I.}\ \bibnamefont {Csonka}}, \bibinfo {author}
  {\bibfnamefont {P.~H.~T.}\ \bibnamefont {Philipsen}},\ and\ \bibinfo {author}
  {\bibfnamefont {J.~P.}\ \bibnamefont {Perdew}},\ }\href
  {https://doi.org/10.1103/PhysRevB.85.014111} {\bibfield  {journal} {\bibinfo
  {journal} {Phys. Rev. B}\ }\textbf {\bibinfo {volume} {85}},\ \bibinfo
  {pages} {014111} (\bibinfo {year} {2012})}\BibitemShut {NoStop}%
\bibitem [{\citenamefont {Kingsbury}\ \emph {et~al.}(2021)\citenamefont
  {Kingsbury}, \citenamefont {Gupta}, \citenamefont {Bartel}, \citenamefont
  {Munro}, \citenamefont {Dwaraknath}, \citenamefont {Horton},\ and\
  \citenamefont {Persson}}]{Kingsbury2021}%
  \BibitemOpen
  \bibfield  {author} {\bibinfo {author} {\bibfnamefont {R.}~\bibnamefont
  {Kingsbury}}, \bibinfo {author} {\bibfnamefont {A.}~\bibnamefont {Gupta}},
  \bibinfo {author} {\bibfnamefont {C.~J.}\ \bibnamefont {Bartel}}, \bibinfo
  {author} {\bibfnamefont {J.~M.}\ \bibnamefont {Munro}}, \bibinfo {author}
  {\bibfnamefont {S.}~\bibnamefont {Dwaraknath}}, \bibinfo {author}
  {\bibfnamefont {M.}~\bibnamefont {Horton}},\ and\ \bibinfo {author}
  {\bibfnamefont {K.~A.}\ \bibnamefont {Persson}},\ }\Eprint
  {https://arxiv.org/abs/10.33774/chemrxiv-2021-gwm9m-v2}
  {chemrXiv:10.33774/chemrxiv-2021-gwm9m-v2}  (\bibinfo {year}
  {2021})\BibitemShut {NoStop}%
\bibitem [{\citenamefont {Kaplan}\ \emph {et~al.}(2020)\citenamefont {Kaplan},
  \citenamefont {Santra}, \citenamefont {Bhattarai}, \citenamefont {Wagle},
  \citenamefont {Chowdhury}, \citenamefont {Bhetwal}, \citenamefont {Yu},
  \citenamefont {Tang}, \citenamefont {Burke}, \citenamefont {Levy},\ and\
  \citenamefont {Perdew}}]{Kaplan2020}%
  \BibitemOpen
  \bibfield  {author} {\bibinfo {author} {\bibfnamefont {A.~D.}\ \bibnamefont
  {Kaplan}}, \bibinfo {author} {\bibfnamefont {B.}~\bibnamefont {Santra}},
  \bibinfo {author} {\bibfnamefont {P.}~\bibnamefont {Bhattarai}}, \bibinfo
  {author} {\bibfnamefont {K.}~\bibnamefont {Wagle}}, \bibinfo {author}
  {\bibfnamefont {S.~T. u.~R.}\ \bibnamefont {Chowdhury}}, \bibinfo {author}
  {\bibfnamefont {P.}~\bibnamefont {Bhetwal}}, \bibinfo {author} {\bibfnamefont
  {J.}~\bibnamefont {Yu}}, \bibinfo {author} {\bibfnamefont {H.}~\bibnamefont
  {Tang}}, \bibinfo {author} {\bibfnamefont {K.}~\bibnamefont {Burke}},
  \bibinfo {author} {\bibfnamefont {M.}~\bibnamefont {Levy}},\ and\ \bibinfo
  {author} {\bibfnamefont {J.~P.}\ \bibnamefont {Perdew}},\ }\href
  {https://doi.org/10.1063/5.0017805} {\bibfield  {journal} {\bibinfo
  {journal} {J. Chem. Phys.}\ }\textbf {\bibinfo {volume} {153}},\ \bibinfo
  {pages} {074114} (\bibinfo {year} {2020})},\ \Eprint
  {https://arxiv.org/abs/https://doi.org/10.1063/5.0017805}
  {https://doi.org/10.1063/5.0017805} \BibitemShut {NoStop}%
\bibitem [{\citenamefont {Medvedev}\ \emph {et~al.}(2017)\citenamefont
  {Medvedev}, \citenamefont {Bushmarinov}, \citenamefont {Sun}, \citenamefont
  {Perdew},\ and\ \citenamefont {Lyssenko}}]{Medvedev2017}%
  \BibitemOpen
  \bibfield  {author} {\bibinfo {author} {\bibfnamefont {M.~G.}\ \bibnamefont
  {Medvedev}}, \bibinfo {author} {\bibfnamefont {I.~S.}\ \bibnamefont
  {Bushmarinov}}, \bibinfo {author} {\bibfnamefont {J.}~\bibnamefont {Sun}},
  \bibinfo {author} {\bibfnamefont {J.~P.}\ \bibnamefont {Perdew}},\ and\
  \bibinfo {author} {\bibfnamefont {K.~A.}\ \bibnamefont {Lyssenko}},\ }\href
  {https://doi.org/10.1126/science.aah5975} {\bibfield  {journal} {\bibinfo
  {journal} {Science}\ }\textbf {\bibinfo {volume} {355}},\ \bibinfo {pages}
  {49} (\bibinfo {year} {2017})},\ \Eprint
  {https://arxiv.org/abs/https://science.sciencemag.org/content/355/6320/49.full.pdf}
  {https://science.sciencemag.org/content/355/6320/49.full.pdf} \BibitemShut
  {NoStop}%
\bibitem [{\citenamefont {Brack}\ \emph {et~al.}(1976)\citenamefont {Brack},
  \citenamefont {Jennings},\ and\ \citenamefont {Chu}}]{Brack1976}%
  \BibitemOpen
  \bibfield  {author} {\bibinfo {author} {\bibfnamefont {M.}~\bibnamefont
  {Brack}}, \bibinfo {author} {\bibfnamefont {B.~K.}\ \bibnamefont
  {Jennings}},\ and\ \bibinfo {author} {\bibfnamefont {Y.~H.}\ \bibnamefont
  {Chu}},\ }\href {https://doi.org/10.1016/0370-2693(76)90519-0} {\bibfield
  {journal} {\bibinfo  {journal} {Phys. Lett. B}\ }\textbf {\bibinfo {volume}
  {65}},\ \bibinfo {pages} {1} (\bibinfo {year} {1976})}\BibitemShut {NoStop}%
\bibitem [{\citenamefont {Oliver}\ and\ \citenamefont
  {Perdew}(1979)}]{Oliver1979}%
  \BibitemOpen
  \bibfield  {author} {\bibinfo {author} {\bibfnamefont {G.~L.}\ \bibnamefont
  {Oliver}}\ and\ \bibinfo {author} {\bibfnamefont {J.~P.}\ \bibnamefont
  {Perdew}},\ }\href {https://doi.org/10.1103/PhysRevA.20.397} {\bibfield
  {journal} {\bibinfo  {journal} {Phys. Rev. A}\ }\textbf {\bibinfo {volume}
  {20}},\ \bibinfo {pages} {397} (\bibinfo {year} {1979})}\BibitemShut
  {NoStop}%
\bibitem [{\citenamefont {Tao}\ \emph {et~al.}(2003)\citenamefont {Tao},
  \citenamefont {Perdew}, \citenamefont {Staroverov},\ and\ \citenamefont
  {Scuseria}}]{Tao2003}%
  \BibitemOpen
  \bibfield  {author} {\bibinfo {author} {\bibfnamefont {J.}~\bibnamefont
  {Tao}}, \bibinfo {author} {\bibfnamefont {J.~P.}\ \bibnamefont {Perdew}},
  \bibinfo {author} {\bibfnamefont {V.~N.}\ \bibnamefont {Staroverov}},\ and\
  \bibinfo {author} {\bibfnamefont {G.~E.}\ \bibnamefont {Scuseria}},\ }\href
  {https://doi.org/10.1103/PhysRevLett.91.146401} {\bibfield  {journal}
  {\bibinfo  {journal} {Phys. Rev. Lett.}\ }\textbf {\bibinfo {volume} {91}},\
  \bibinfo {pages} {146401} (\bibinfo {year} {2003})}\BibitemShut {NoStop}%
\bibitem [{\citenamefont {Almbladh}\ and\ \citenamefont {von
  Barth}(1985)}]{Almbladh1985}%
  \BibitemOpen
  \bibfield  {author} {\bibinfo {author} {\bibfnamefont {C.-O.}\ \bibnamefont
  {Almbladh}}\ and\ \bibinfo {author} {\bibfnamefont {U.}~\bibnamefont {von
  Barth}},\ }\href {https://doi.org/10.1103/PhysRevB.31.3231} {\bibfield
  {journal} {\bibinfo  {journal} {Phys. Rev. B}\ }\textbf {\bibinfo {volume}
  {31}},\ \bibinfo {pages} {3231} (\bibinfo {year} {1985})}\BibitemShut
  {NoStop}%
\bibitem [{\citenamefont {Kaplan}\ \emph {et~al.}(2018)\citenamefont {Kaplan},
  \citenamefont {Wagle},\ and\ \citenamefont {Perdew}}]{Kaplan2018}%
  \BibitemOpen
  \bibfield  {author} {\bibinfo {author} {\bibfnamefont {A.~D.}\ \bibnamefont
  {Kaplan}}, \bibinfo {author} {\bibfnamefont {K.}~\bibnamefont {Wagle}},\ and\
  \bibinfo {author} {\bibfnamefont {J.~P.}\ \bibnamefont {Perdew}},\ }\href
  {https://doi.org/10.1103/PhysRevB.98.085147} {\bibfield  {journal} {\bibinfo
  {journal} {Phys. Rev. B}\ }\textbf {\bibinfo {volume} {98}},\ \bibinfo
  {pages} {085147} (\bibinfo {year} {2018})}\BibitemShut {NoStop}%
\bibitem [{\citenamefont {Dunning}(1989)}]{Dunning1989}%
  \BibitemOpen
  \bibfield  {author} {\bibinfo {author} {\bibfnamefont {T.~H.}\ \bibnamefont
  {Dunning}},\ }\href {https://doi.org/10.1063/1.456153} {\bibfield  {journal}
  {\bibinfo  {journal} {J. Chem. Phys.}\ }\textbf {\bibinfo {volume} {90}},\
  \bibinfo {pages} {1007} (\bibinfo {year} {1989})}\BibitemShut {NoStop}%
\bibitem [{\citenamefont {Woon}\ and\ \citenamefont
  {Dunning}(1995)}]{Woon1995}%
  \BibitemOpen
  \bibfield  {author} {\bibinfo {author} {\bibfnamefont {D.~E.}\ \bibnamefont
  {Woon}}\ and\ \bibinfo {author} {\bibfnamefont {T.~H.}\ \bibnamefont
  {Dunning}},\ }\href {https://doi.org/10.1063/1.470645} {\bibfield  {journal}
  {\bibinfo  {journal} {J. Chem. Phys.}\ }\textbf {\bibinfo {volume} {103}},\
  \bibinfo {pages} {4572} (\bibinfo {year} {1995})}\BibitemShut {NoStop}%
\end{thebibliography}

\clearpage
\begin{widetext}
  \section*{Supplemental Material: Construction of meta-GGA functionals through restoration of exact constraint adherence to regularized SCAN functionals.}
\end{widetext}

\appendix

\section{The slowly-varying limit of \rrscan and \rfscan \label{AP:exc_deriv}}

This Appendix sketches the derivation of \rrscan and \rfscan. We will presume that both functionals have the
structure of Eq. 1, %\ref{MT-eq:scan_framework},
where the interpolation functions $f_\mr{x/c}(\v{r})$, are taken from rSCAN. As discussed in the main text, this functional is termed \rppscan.

From the starting point of \rppscan, we will derive the corrections needed to restore the second order gradient expansions for exchange and correlation (\rrscan), and those for the fourth-order gradient expansion for exchange (\rfscan). Thus \rfscan can be viewed as a correction to \rrscan exchange, and we begin with \rrscan.

\subsection{The gradient expansion of $\ba$}

The gradient expansion of $\tau[n]$, the one-body, spin-unpolarized kinetic energy density, was derived in Ref. \cite{Brack1976}
\begin{equation}
  \tau[n] = \tauu(n) + \frac{1}{6} \lan + \frac{1}{72}\frac{|\gn|^2}{n} + \mo{4}
\end{equation}
(in Hartree atomic units). Here, $\mo{4}$ indicates that the next term in the series of higher order is of the form $|\gn|^4$, $\lan\, |\gn|^2$, $(\lan)^2$, etc. The gradient expansion is more useful in terms of dimensionless (length-scale invariant) variables
\begin{align}
  p(n) &= \left[ \frac{|\gn|}{2 k_{\mr{F}}n}\right]^2 = \frac{3}{40}\frac{|\gn|^2}{\tauu(n)n} \\
  q(n) &= \frac{\lan}{4 k^2_{\mr{F}}n} = \frac{3}{40}\frac{\lan}{\tauu(n)},
\end{align}
where we have used
\begin{align}
  \tauu(n) &= \frac{3}{10}k_{\mr{F}}^2 n \\
  k_{\mr{F}} &= (3 \pi^2 n)^{1/3}.
\end{align}
Then the gradient expansion of $\tau\sus$ can be cast as
\begin{equation}
  \tau[n] = \tauu(n)\left[1 + \frac{20}{9}q(n) + \frac{5}{27}p(n) \right] + \mo{4}.
\end{equation}

The integrated kinetic energy scales with the spin-densities in the same manner as the exchange energy \cite{Oliver1979}
\begin{equation}
  T[\nup,\ndn] = \frac{1}{2}\{ T[2\nup]+T[2\ndn]\}.
\end{equation}
This implies a local spin-scaling relation
\begin{equation}
  \tau[\nup,\ndn] = \frac{1}{2}\{ \tau[2\nup]+\tau[2\ndn] \}.
\end{equation}
We will seek a gradient expansion in terms of $n$ and $\zeta$, where
\begin{equation}
  \zeta = \frac{\nup - \ndn}{\nup + \ndn},
\end{equation}
rather than the individual spin-densities. After simplification, one finds that
\begin{align}
  \tau(\nup,\ndn) &= \tauu(n)d_s(\zeta) \left[  1 + \frac{20}{9d_s(\zeta)}q + \frac{5}{27d_s(\zeta)} p \right. \nonumber \\
  & \left. + \frac{5}{27} \frac{\xi^2}{d_s(\zeta)(1-\zeta^2)} \right] + \mo{4}, \label{eq:tau_spin_res}
\end{align}
where
\begin{equation}
  d_s(\zeta) = [(1 + \zeta)^{5/3} + (1 - \zeta)^{5/3} ]/2
\end{equation}
describes the spin-scaling of the uniform electron gas kinetic energy density, and
\begin{equation}
  \xi = \frac{|\nabla \zeta|}{2 k_{\mr{F}}},
\end{equation}
which also appeared in TPSS \cite{Tao2003}.

The spin resolved $\ba$ tends to
\begin{align}
  \ba(\nup,\ndn) &= \frac{\tau(\nup,\ndn) - \tau_W}{\tauu(n)d_s(\zeta) + \eta \tau_W} \\
  & = \left[\frac{\tau}{\tauu(n)d_s(\zeta)} - \frac{5}{3d_s(\zeta)}p \right]\left[1 + \frac{5\eta}{3d_s(\zeta)}p \right]^{-1}.
\end{align}
After performing a Taylor expansion in $p$, and inserting Eq. \ref{eq:tau_spin_res} for $\tau(\nup,\ndn)$, we find
\begin{align}
  \ba(\nup,\ndn) &= 1 + \frac{20}{9d_s(\zeta)}q - \frac{5(8 +9 \eta )}{27 d_s(\zeta)}p \nonumber \\
  & + \frac{5}{27d_s(\zeta)(1-\zeta^2)}\xi^2 + \mo{4}. \label{eq:ba_ge_pol}
\end{align}
In the special case where $\zeta=0$ (needed for the exchange energy),
\begin{equation}
  \ba(n) =  1 + \frac{20}{9}q - \frac{5(8 +9 \eta )}{27 }p + \mo{4}. \label{eq:ba_ge_unp}
\end{equation}

\subsection{Exchange, second order gradient expansion \label{AP:r2_x_deriv}}

For any spin-unpolarized exchange energy density $\epsilon_{\mr{x}}(n)$, the spin-scaled exchange energy density is \cite{Oliver1979}
\begin{equation}
  \epsilon_{\mr{x}}(\nup,\ndn) = \frac{1}{2}[\epsilon_{\mr{x}}(2\nup) + \epsilon_{\mr{x}}(2\ndn)].
\end{equation}
Therefore we need only consider the spin-unpolarized exchange energy density, and apply the spin-scaling relationship as needed.

We start with an explicit expression for the \rrscan exchange enhancement factor
\begin{equation}
  F^{\text{\rrscan}}\smx(p,\ba) = \{h\smx^1(p) + f\smx(\ba)[h\smx^0 - h\smx^1(p)] \}g\smx(p),
\end{equation}
with $h\smx^0=1.174$. The function
\begin{equation}
  g\smx(p) = 1-\exp(-a\smx/p^{1/4}) = 1 + \mo{\infty} \label{eq:gx_taylor}
\end{equation}
in the slowly-varying limit (its derivatives of all order vanish in the limit $p\to 0$). In \rrscan, we take
\begin{equation}
  h\smx^1(p) = 1 + k_1 - k_1\{1 + [D\smx \exp(-p^2/d_{p2}^4) + \muak \}]p/k_1 \}^{-1},
\end{equation}
which has the following Taylor series:
\begin{equation}
  h\smx^1(p) = 1 + (D\smx + \muak) p - \frac{(D\smx + \muak)^2}{k_1}p^2 + \mo{6}. \label{eq:hx_taylor}
\end{equation}
The \rrscan interpolation function is taken from rSCAN, and has the structure
\begin{equation}
  f\smx(\ba) = \left\{ \begin{array}{ll} \sum_{i=0}^7c_{\mr{x},i}\ba^i, & \ba <= 2.5 \\
  -c_{\mr{dx}}^{\mr{SCAN}}\exp\left[\frac{c_{\mr{2x}}^{\mr{SCAN}}}{1-\ba}\right], & \ba > 2.5
  \end{array} \right.,
\end{equation}
with Taylor series
\begin{align}
  f\smx(\ba) &= (\ba -1)\sum_{i=1}^7 i c_{\mr{x},i} + \frac{1}{2}(\ba - 1)^2\sum_{i=2}^7 i (i-1) c_{\mr{x},i} \nonumber \\
  & + \mathcal{O}[(\ba-1)^3] \label{eq:fx_taylor}
\end{align}
in the slowly-varying limit. It's important here to note that $\ba$ has a gradient expansion to much higher order than $\mo{2}$, as we derived in Eqs. \ref{eq:ba_ge_pol} and \ref{eq:ba_ge_unp}. Therefore, the term of lowest order in $(1-\ba)$ is $\mo{2}$, and the term of lowest order in $(1-\ba)^2$ is $\mo{4}$.

Inserting the Taylor series of Eqs. \ref{eq:gx_taylor}, \ref{eq:hx_taylor}, and \ref{eq:fx_taylor} into $F\smx^{\text{\rrscan}}$, we find, to $\mo{4}$,
\begin{align}
  & F\smx^{\text{\rrscan}}(p,\ba) = 1 + (D\smx + \muak) p \nonumber \\
  &+ \left[(h\smx^0-1)\sum_{i=1}^7 i c_{\mr{x},i}\right](\ba -1) - \frac{(D\smx + \muak)^2}{k_1}p^2 \nonumber \\
  & -\left[(D\smx + \muak) \sum_{i=1}^7 i c_{\mr{x},i}\right](\ba -1)p \nonumber \\
  & +\left[\frac{h\smx^0-1}{2}\sum_{i=2}^7 i(i-1) c_{\mr{x},i}\right](\ba -1)^2 + \mo{6}, \label{eq:fx_rr_taylor}
\end{align}
with the terms written in (generally) increasing order. For \rrscan, we demand that the exchange enhancement factor recover the exchange gradient expansion \cite{Kirzhnits1957, Svendsen1996}
\begin{equation}
  F^{\text{GE}}\smx(p,q) = 1 + \muak p + \frac{146}{2025}q^2 - \frac{73}{405}pq + \mo{6} \label{eq:ge4x_pq}
\end{equation}
only to second order, i.e. $ 1 + \muak p$. Therefore, we can ignore all terms of order higher than $\mo{2}$ that are written explicitly in Eq. \ref{eq:fx_rr_taylor}, but also those that are included implicitly in $(\ba-1)$,
\begin{widetext}
\begin{equation}
  F\smx^{\text{\rrscan}}(p,\ba) = 1 + \left[D\smx - \frac{5(8 +9 \eta )}{27 }(h\smx^0-1)\sum_{i=1}^7 i c_{\mr{x},i} +\muak\right] p + \frac{20(h\smx^0-1)}{9}\left(\sum_{i=1}^7 i c_{\mr{x},i}\right)q + \mo{4},
\end{equation}
where we have evaluated $(\ba-1)$ using Eq. \ref{eq:ba_ge_unp}.

To eliminate the term linear in $q$, we perform an integration by parts. The ``gauge variance'' of the exchange energy density implies that two exchange enhancement factors can yield the same integrated exchange energy, but different exchange energy densities,
\begin{equation}
  \int_{\Omega} F\smx \epsilon\smx^{\text{LSDA}}d^3 r = \int_{\Omega} [\widetilde{F}\smx \epsilon\smx^{\text{LSDA}} + n^{-4/3}\nabla \cdot \bm{G}\smx]d^3 r = \int_{\Omega}\widetilde{F}\smx d^3r -\frac{3}{4\pi}(3\pi^2)^{1/3} \int_{\text{bdy}\,\Omega} \bm{G}\cdot d\bm{S}.
\end{equation}
The second equality is a straightforward application of the divergence theorem. Provided that the gauge function $\bm{G}\smx$ vanishes sufficiently rapidly over the bounding surface $\text{bdy}\,\Omega$ of the integration volume $\Omega$, the surface integral vanishes. Note also that the LSDA exchange energy density is
\begin{equation}
  \epsilon\smx^{\text{LSDA}} = -\frac{3}{4\pi}(3\pi^2)^{1/3} n^{4/3}.
\end{equation}
Consider then
\begin{equation}
  q \,\epsilon\smx^{\text{LSDA}} = \left\{ \frac{p}{3} + n^{-4/3}\nabla \cdot \left[ \frac{\gn}{4(3\pi^2)^{2/3}n^{1/3} } \right] \right\}\epsilon\smx^{\text{LSDA}},
\end{equation}
thus
\begin{equation}
  %F\smx^{\text{\rrscan}}(p,\ba) = 1 + \left[D\smx - \frac{5(4 +9 \eta )}{27 }(h\smx^0-1)\sum_{i=1}^7 i c_{\mr{x},i} +\muak\right] p + n^{-4/3}\nabla \cdot \underbrace{\left\{ \left[ \frac{5(h\smx^0-1)}{9(3\pi^2)^{2/3}} \left(\sum_{i=1}^7 i c_{\mr{x},i}\right) \right]\frac{\gn}{n^{1/3}}\right\}}_{\bm{G}\smx} + \mo{4}.
  F\smx^{\text{\rrscan}}(p,\ba) = 1 + \left[D\smx - \frac{5(4 +9 \eta )}{27 }(h\smx^0-1)\sum_{i=1}^7 i c_{\mr{x},i} +\muak\right] p + n^{-4/3}\nabla \cdot \left\{ \left[ \frac{5(h\smx^0-1)}{9(3\pi^2)^{2/3}} \left(\sum_{i=1}^7 i c_{\mr{x},i}\right) \right]\frac{\gn}{n^{1/3}}\right\} + \mo{4}.
\end{equation}
\end{widetext}
The rightmost term in curly braces is the gauge function $\bm{G}\smx$. Except in certain situations, like the density tail of an atom (generally, outside the Kohn-Sham turning surface if one exists), where the gradient expansion does not apply, $n^{-1/3} \gn $ vanishes sufficiently rapidly at infinity. Moreover, we generalize $F\smx^{\text{\rrscan}}$ by cutting off the divergent gradient expansion terms and keeping only the integrated-by-parts expression
\begin{align}
  & \widetilde{F}\smx^{\text{\rrscan}}(p,\ba) = 1 + \left[D\smx - \frac{5(4 +9 \eta )}{27 }(h\smx^0-1)\sum_{i=1}^7 i c_{\mr{x},i} \right. \nonumber \\
  & +\muak \bigg] p + \mo{4},
\end{align}
allowing for validity even outside the Kohn-Sham turning surface.

Thus, we recover the correct second-order gradient expansion for exchange by demanding
\begin{equation}
  D\smx = \frac{5(4 +9 \eta )}{27 } (h\smx^0-1)\sum_{i=1}^7 i c_{\mr{x},i}
\end{equation}
which is the product of $C_{\eta}$ and $C_{2\mr{x}}$ defined
in Eqs. 53 and 57, %\ref{MT-eq:c_eta_expr} and \ref{MT-eq:c_2x_expr},
respectively,
\begin{align}
  C_{\eta} &= \frac{5(4 +9 \eta )}{27 }  \nonumber \\
  C_{2\mr{x}} &= (h\smx^0-1)\sum_{i=1}^7 i c_{\mr{x},i}. \nonumber
\end{align}

\subsection{Exchange, fourth-order gradient expansion \label{AP:r4_x_deriv}}

\rfscan prescribes an explicit correction to the \rrscan enhancement factor,
\begin{equation}
  F\smx^{\text{\rfscan}} = F\smx^{\text{\rrscan}} + \Delta F_4(p,\ba)g\smx(p)
\end{equation}
where $g\smx(p)$ is unchanged from \rrscan, and
\begin{align}
  \Delta F_4(p,\ba) &= \left\{ -C_{2\mr{x}}[ (\ba-1) + C_{\eta}p ] + C_{\ba \ba}(1-\ba)^2 \right. \nonumber \\
  & \left. + C_{p \ba}p(1-\ba) + C_{pp} p^2 \right\}\frac{2\ba^2}{1+\ba^4}\nonumber \\
  & \times \exp\left[-\frac{(1-\ba)^2}{d_{\ba4}^2} - \frac{p^2}{d_{p4}^4} \right].
\end{align}
Despite the complexity of $\Delta F_4$, the damping function used to modulate these corrections,  it has a simple Taylor series
\begin{align}
  \Delta F_4(p,\ba)  &= -C_{2\mr{x}}[ (\ba-1) + C_{\eta}p ] + C_{\ba \ba}(1-\ba)^2 \nonumber \\
  & + C_{p \ba}p(1-\ba) + C_{pp} p^2 + \mo{6}.
\end{align}
We will now take $D\smx = C_{\eta}C_{2\mr{x}}$ in the \rrscan exchange enhancement factor. Returning to Eq. \ref{eq:fx_rr_taylor} for the \rrscan exchange enhancement factor, and adding in the \rfscan corrections,
\begin{align}
  & F\smx^{\text{\rfscan}}(p,\ba) = 1 + \muak p +\left[C_{pp} - \frac{(C_{\eta}C_{2\mr{x}} + \muak)^2}{k_1} \right]p^2 \nonumber \\
  &  + \left[ C_{p \ba} + (C_{\eta}C_{2\mr{x}} + \muak) \sum_{i=1}^7 i c_{\mr{x},i}\right] (1 - \ba)p \nonumber \\
  &  +\left[C_{\ba \ba} + \frac{h\smx^0-1}{2}\sum_{i=2}^7 i(i-1) c_{\mr{x},i}\right](\ba -1)^2 + \mo{6}. \label{eq:fx_rf_taylor}
\end{align}
Now, again using Eq. \ref{eq:ba_ge_unp} for the gradient expansion of $\ba$ to second-order,
\begin{align}
  (1 - \ba)p &= -\frac{20}{9}p q + \frac{5(8 +9 \eta )}{27 }p^2 + \mo{6} \label{eq:p_ba_ge}\\
  (1 - \ba)^2 &= \frac{400}{81}q^2 + \left[\frac{5(8 +9 \eta )}{27 } \right]^2 p^2 \nonumber \\
  & - \frac{200(8 +9 \eta )}{243}p q + \mo{6}. \label{eq:ba_ba_ge}
\end{align}
The second-order gradient expansion of $\ba$ is valid here, because any higher order terms in $1-\ba$ will yield, to lowest order, sixth-order terms in these products. Using Eqs. \ref{eq:p_ba_ge} and \ref{eq:p_ba_ge}, one can show that
\begin{align}
  & \frac{73}{5000}(\ba-1)^2 + \left[\frac{511}{13500} -\frac{73}{1500}\eta \right] (1 - \ba )p \nonumber \\
  & + \left[\frac{146}{2025} \left(\frac{2}{3} + \frac{3\eta}{4} \right)^2 - \frac{73}{405}\left(\frac{2}{3} + \frac{3\eta}{4} \right) \right] p^2 \nonumber \\
  & = \frac{146}{2025} q^2 - \frac{73}{405} p q + \mo{6}, \label{eq:ge4_p_ba}
\end{align}
the same fourth-order terms as in Eq. \ref{eq:ge4x_pq}.

Then to recover the fourth-order gradient expansion for exchange in \rfscan, we equate \ref{eq:fx_rf_taylor} and \ref{eq:ge4_p_ba}, and find
\begin{align}
  C_{pp} &= \frac{(C_{\eta}C_{2\mr{x}} + \muak)^2}{k_1} + \frac{146}{2025} \left(\frac{2}{3} + \frac{3\eta}{4} \right)^2 \nonumber \\
  & - \frac{73}{405}\left(\frac{2}{3} + \frac{3\eta}{4} \right) \\
  C_{p \ba} &= \frac{511}{13500} -\frac{73}{1500}\eta - (C_{\eta}C_{2\mr{x}} + \muak) \sum_{i=1}^7 i c_{\mr{x},i} \\
  C_{\ba \ba} &= \frac{73}{5000} - \frac{h\smx^0-1}{2}\sum_{i=2}^7 i(i-1) c_{\mr{x},i},
\end{align}
as presented in Eqs. 61--63. %\ref{MT-eq:c_ba_ba}--\ref{MT-eq:c_p_p}.

\subsection{Correlation, second-order gradient expansion \label{AP:r2_c_deriv}}

The gradient expansion for the correlation energy per electron is known only to second order \cite{Ma1968,Wang1991,Perdew1996,Sun2015}
\begin{equation}
  \varepsilon\smc(r_{\mathrm{s}},\zeta,t) = \varepsilon\smc^{\text{LSDA}}(\rs,\zeta) + \beta(\rs)\phi^3(\zeta) t^2.
\end{equation}
The density-dependent function $\beta(\rs)$ is known only for small values of $\rs$, and we take the parameterization used in Ref. \cite{Sun2015},
\begin{equation}
  \beta(\rs) = \beta_{\mr{MB}}\frac{1 + 0.1 \rs}{1 + 0.1778 \rs},
\end{equation}
constructed to cancel with the second-order gradient expansion term for exchange in the limit $\rs \to \infty$.
Two other quantities enter this expansion: the spin-scaling function
\begin{equation}
  \phi(\zeta) = [(1 + \zeta)^{2/3} + (1 - \zeta)^{2/3}]/2,
\end{equation}
and dimensionless density gradient on the length scale of the Thomas-Fermi wavevector
\begin{equation}
  t^2 = \left( \frac{3\pi^2}{16} \right)^{2/3}\frac{p}{\phi^2(\zeta)\rs}.
\end{equation}

In both \rrscan and \rfscan, we propose that the correlation energy per electron is
\begin{align}
  \varepsilon^{\text{\rrscan}}\smc(\rs,\zeta,p,\ba) &= \varepsilon\smc^1(\rs,\zeta,p) + f\smc(\ba)[ \varepsilon\smc^0(\rs,\zeta,p) \nonumber \\
  & - \varepsilon\smc^1(\rs,\zeta,p)]
\end{align}
with $f\smc(\ba)$ taken from rSCAN. It has a Taylor series about $\ba = 1$ that is identical in structure (but not value) to the Taylor series for $f\smx(\ba)$. The individual energies per electron are
\begin{align}
  \varepsilon\smc^0(\rs,\zeta,p) &= [\varepsilon\smc^{\text{LDA0}}(\rs,\zeta) + H_0(\rs,\zeta,p)]g\smc(\zeta) \\
  \varepsilon\smc^1(\rs,\zeta,p) &= \varepsilon\smc^{\text{LSDA1}}(\rs,\zeta) + H_1(\rs,\zeta,p),
\end{align}
with $\varepsilon\smc^0$ unchanged from SCAN (see also Eqs. \ref{eq:ec0_scan}--\ref{eq:gc_zeta}). In \rrscan, we posit that
\begin{align}
  H_1(\rs,\zeta,p) &= \gamma \phi^3(\zeta)\ln \left\{ 1 + w_1\left[ 1 - g(y,\Delta y)\right]\right\} \\
  y &= \frac{\beta(\rs)}{\gamma w_1} t^2 \\
  g(y,\Delta y)&= [1 + 4(y - \Delta y)]^{-1/4} \\
  \Delta y &= D\smc p\exp[-p^2/d_{p2}^4],
\end{align}
with $d_{p2}$ unchanged from the exchange component of \rrscan. It can readily be seen that these have the following Taylor series:
\begin{align}
  \varepsilon\smc^0(\rs,\zeta,p) &= \varepsilon\smc^{\text{LDA0}}(\rs,\zeta)g\smc(\zeta) + \chi_{\infty}g\smc(\zeta) p + \mo{4} \\
  \varepsilon\smc^1(\rs,\zeta,p) &= \varepsilon\smc^{\text{LSDA1}}(\rs,\zeta) + \beta(\rs)\phi^3(\zeta)t^2 \nonumber \\
  & - \gamma \phi^3(\zeta)w_1 D\smc p + \mo{4}.
\end{align}
Then the full gradient expansion of the \rrscan correlation energy per electron is, after simplification,
\begin{align}
  & \varepsilon^{\text{\rrscan}}\smc(\rs,\zeta,p,\ba) = \varepsilon\smc^{\text{LSDA1}} + \beta(\rs)\phi^3(\zeta)t^2 \nonumber \\
  & - \gamma \phi^3(\zeta)w_1 D\smc p + \left(\sum_{i=1}^7 i c_{\mr{c},i}\right)(\ba - 1) + \mo{4},
\end{align}
where $\varepsilon\smc^{\text{LSDA0}} = \varepsilon\smc^{\text{LDA0}}g\smc(\zeta)$. Let
\begin{equation}
  \Delta f_{\mr{c}2} \equiv \sum_{i=1}^7 i c_{\mr{c},i}.
\end{equation}
We can now use Eq. \ref{eq:ba_ge_pol} for the gradient expansion of $\ba$ at arbitrary spin polarization to find that
\begin{widetext}
\begin{align}
  &\varepsilon^{\text{\rrscan}}\smc(\rs,\zeta,p,\ba) = \varepsilon\smc^{\text{LSDA1}}(\rs,\zeta) + \beta(\rs)\phi^3(\zeta)t^2 + \frac{20\Delta f_{\mr{c}2}}{9d_s(\zeta)}[\varepsilon\smc^{\text{LSDA0}}(\rs,\zeta)-\varepsilon\smc^{\text{LSDA1}}(\rs,\zeta)]q \nonumber \\
  & - \left\{\gamma \phi^3(\zeta)w_1 D\smc + \frac{5(8 +9 \eta )\Delta f_{\mr{c}2}}{27 d_s(\zeta)}[\varepsilon\smc^{\text{LSDA0}}(\rs,\zeta)-\varepsilon\smc^{\text{LSDA1}}(\rs,\zeta)] \right\} p \nonumber \\
  & + \frac{5\Delta f_{\mr{c}2}}{27d_s(\zeta)(1-\zeta^2)}[\varepsilon\smc^{\text{LSDA0}}(\rs,\zeta)-\varepsilon\smc^{\text{LSDA1}}(\rs,\zeta)]\xi^2 + \mo{4}.
\end{align}

We can eliminate the term linear in $q$ using a similar gauge variance principle for the correlation energy:
\begin{equation}
  \int_{\Omega} \varepsilon\smc n \, d^3 r = \int_{\Omega} [\widetilde{\varepsilon}\smc + n^{-1} \nabla \cdot \bm{G}\smc ]n \, d^3 r = \int_{\Omega} \widetilde{\varepsilon}\smc n \, d^3 r + \int_{\text{bdy}\,\Omega}\bm{G}\smc \cdot d\bm{S}.
\end{equation}
Again, provided that the gauge function $\bm{G}\smc$ vanishes sufficiently rapidly at the bounding surface $\text{bdy}\,\Omega$, we may replace the \rrscan correlation energy per electron with the integrated-by-parts expression $\widetilde{\varepsilon}\smc$. To do this, consider that for a general function $f(\rs,\zeta)$,
\begin{equation}
  \nabla f(\rs,\zeta) = -\frac{\rs}{3n}\frac{\partial f}{\partial \rs}\gn + \frac{\partial f}{\partial \zeta} \nabla \zeta,
\end{equation}
where we have used $\rs=[4\pi n/3]^{-3}$. Then
\begin{align}
  &\frac{\varepsilon\smc^{\text{LSDA0}}(\rs,\zeta)-\varepsilon\smc^{\text{LSDA1}}(\rs,\zeta)}{d_s(\zeta)}q \, n = \left\{ \frac{2}{3} \frac{\varepsilon\smc^{\text{LSDA0}}(\rs,\zeta)-\varepsilon\smc^{\text{LSDA1}}(\rs,\zeta)}{d_s(\zeta)}  + \frac{\rs}{3d_s(\zeta)} \left[ \frac{\partial\varepsilon\smc^{\text{LSDA0}}}{\partial \rs}- \frac{\partial\varepsilon\smc^{\text{LSDA1}}}{\partial \rs} \right] \right\} p \, n \nonumber \\
  & - \frac{\gn \cdot \nabla \zeta}{4(3\pi^2)n^{5/3}}\frac{\partial}{\partial \zeta}\left[\frac{\varepsilon\smc^{\text{LSDA0}}(\rs,\zeta)-\varepsilon\smc^{\text{LSDA1}}(\rs,\zeta)}{d_s(\zeta)} \right] n
  + n^{-1} \nabla \cdot \left[\frac{\varepsilon\smc^{\text{LSDA0}}(\rs,\zeta)-\varepsilon\smc^{\text{LSDA1}}(\rs,\zeta)}{d_s(\zeta)} \frac{\gn}{4(3\pi^2n)^{2/3}} \right].
  %& + n^{-1} \nabla \cdot \underbrace{\left[\frac{\varepsilon\smc^{\text{LSDA0}}(\rs,\zeta)-\varepsilon\smc^{\text{LSDA1}}(\rs,\zeta)}{d_s(\zeta)} \frac{\gn}{4(3\pi^2n)^{2/3}} \right]}_{\bm{G}\smc}.
\end{align}
The rightmost term in square brackets is the gauge function $\bm{G}\smc$. Then the integrated-by-parts \rrscan correlation energy per electron is
\begin{align}
  &\widetilde{\varepsilon}^{\text{\rrscan}}\smc(\rs,\zeta,p,\ba) = \varepsilon\smc^{\text{LSDA1}}(\rs,\zeta) + \beta(\rs)\phi^3(\zeta)t^2 \nonumber \\
  & - \left\{\gamma \phi^3(\zeta)w_1 D\smc + \frac{45\Delta f_{\mr{c}2} \eta}{27 d_s(\zeta)}[\varepsilon\smc^{\text{LSDA0}}(\rs,\zeta)-\varepsilon\smc^{\text{LSDA1}}(\rs,\zeta)] - \frac{20\Delta f_{\mr{c}2}\rs}{27d_s(\zeta)} \left[ \frac{\partial\varepsilon\smc^{\text{LSDA0}}}{\partial \rs} - \frac{\partial\varepsilon\smc^{\text{LSDA1}}}{\partial \rs} \right] \right\} p \nonumber \\
  & - \frac{5\gn \cdot \nabla \zeta}{9(3\pi^2)n^{5/3}}\frac{\partial}{\partial \zeta}\left[\frac{\varepsilon\smc^{\text{LSDA0}}(\rs,\zeta)-\varepsilon\smc^{\text{LSDA1}}(\rs,\zeta)}{d_s(\zeta)} \right] + \frac{5\Delta f_{\mr{c}2}}{27d_s(\zeta)(1-\zeta^2)}[\varepsilon\smc^{\text{LSDA0}}(\rs,\zeta)-\varepsilon\smc^{\text{LSDA1}}(\rs,\zeta)]\xi^2 + \mo{4}.
\end{align}

In \rrscan, we make the simplification that $\nabla \zeta\approx 0$. Thus $\xi\approx 0$, and
\begin{align}
  &\widetilde{\varepsilon}^{\text{\rrscan}}\smc(\rs,\zeta,p,\ba) = \varepsilon\smc^{\text{LSDA1}}(\rs,\zeta) + \beta(\rs)\phi^3(\zeta)t^2 \nonumber \\
  & - \left\{\gamma \phi^3(\zeta)w_1 D\smc + \frac{45\Delta f_{\mr{c}2} \eta}{27 d_s(\zeta)}[\varepsilon\smc^{\text{LSDA0}}(\rs,\zeta)-\varepsilon\smc^{\text{LSDA1}}(\rs,\zeta)] - \frac{20\Delta f_{\mr{c}2}\rs}{27d_s(\zeta)}\left[ \frac{\partial\varepsilon\smc^{\text{LSDA0}}}{\partial \rs} - \frac{\partial\varepsilon\smc^{\text{LSDA1}}}{\partial \rs} \right] \right\} p + \mo{4}.
\end{align}
To recover the second order gradient expansion for correlation, we take
\begin{equation}
  D\smc = \frac{\Delta f_{\mr{c}2}}{27 \gamma \phi^3(\zeta)d_s(\zeta) w_1(\rs,\zeta)}\left\{20\rs \left[ \frac{\partial\varepsilon\smc^{\text{LSDA0}}}{\partial \rs} - \frac{\partial\varepsilon\smc^{\text{LSDA1}}}{\partial \rs} \right] - 45\eta [\varepsilon\smc^{\text{LSDA0}}(\rs,\zeta)-\varepsilon\smc^{\text{LSDA1}}(\rs,\zeta)]\right\},
\end{equation}
which is the factor appearing Eq. 78.%\ref{MT-eq:del_y}.
\end{widetext}

\section{Further discussion of non-uniform coordinate scaling \label{AP:non_unif_scl}}

As described in Sec. II A, %\ref{MT-sec:theory_scaling}
a density $n$ and Kohn-Sham orbital $\phi_i$ that are non-uniformly scaled along one dimension, here the $x$ coordinate, have the form
\begin{align}
    n_\lambda^z(x,y,z) &= \lambda n(u,y,z) \\
    [\phi_i]_\lambda^z(x,y,z) &= \lambda^{1/2} \phi_i(u,y,z),
\end{align}
where $u = \lambda x$.
Using Eqs. 33--35, %\ref{MT-eq:nus_tau}--\ref{MT-eq:nus_tauu},
one can show that the iso-orbital indicator $\alpha$ scales as
\begin{align}
    \alpha_\lambda^x &= \frac{5}{12(3\pi^2)^{2/3}[n(u, y, z)]^{8/3}}\left\{ \lambda^{4/3}f_{\alpha 1}(u,y,z) \right. \\
    & \left. + \lambda^{-2/3} f_{\alpha 2}(u,y,z) \right\}, \nonumber
\end{align}
where
\begin{align}
    f_{\alpha1}(u,y,z) &= 4n(u, y, z)
    \left(\sum_i^{\text{occ.}} \left|\frac{\partial \phi_i(u, y, z)}{\partial u}\right|^2\right)
     \nonumber \\
    &  - \left(\frac{\partial n(u, y, z)}{\partial u}\right)^2  \\
  f_{\alpha2}(u,y,z) &=\left(\sum_i^{\text{occ.}} \left|\nabla_\perp \phi_i(u, y, z)\right|^2\right) \nonumber \\
   & - \left|\nabla_\perp n(u, y, z)\right|^2.
\end{align}
Similarly, the dimensionless gradient $p$ can be expressed as
\begin{align}
    \frac{5}{3} p_\lambda^x &= \frac{5}{12(3\pi^2)^{2/3}[n(u, y, z)]^{8/3}}
    \left\{ \lambda^{4/3}f_{p 1}(u,y,z) \right. \\
    & \left. + \lambda^{-2/3} f_{p 2}(u,y,z) \right\}, \nonumber
\end{align}
where
\begin{align}
    f_{p 1}(u,y,z)&= \left[\frac{\partial n(u, y, z)}{\partial u}\right]^2 \\
    f_{p 2}(u,y,z) &= \left|\nabla_\perp n(u, y, z)\right|^2.
\end{align}
Recall that the iso-orbital indicator used in r++SCAN, \rrscan, and \rfscan is
\begin{equation}
    \bar\alpha = \frac{\alpha}{1 + 5\eta p/3}.
\end{equation}

Now, in the limit $\lambda \to 0$, $\lambda^{-2/3} \gg 1 \gg \lambda^{4/3}$, and the leading order of the scaled $\alpha$ and scaled $p$ will be $\lambda^{-2/3}$.
Consequently,
\begin{equation}
    \lim_{\lambda \to 0} \bar\alpha = \frac{f_{\alpha 2}(u,y,z)}{\eta f_{ 2}(u,y,z)},
\end{equation}
which is independent of the scaling parameter (here we are assuming that a change of coordinates from $x \to u = \lambda x$ is used to evaluate all necessary integrals as well).

As we will show, the $\lambda \to \infty$ limit can result in two different scaling behaviors for $\alpha$.
It appears that systems that are finite in at least one dimension have $\alpha\sim\lambda^{-2/3}$ as $\lambda \to \infty$, which yields an iso-orbital ($\alpha=0$) character in the hard limit.
Completely extended systems (like the uniform electron gas) will have the other scaling, $\alpha \sim \lambda^{4/3}$ as $\lambda \to \infty$.

To understand why this might be, consider a generic isolated atom.
This is a prototype for finite systems, so we expect the ensuing analysis to hold qualitatively for related systems like molecules.
As is well-known, far from the nucleus, the density is dominated by the character of the highest-occupied (HO) Kohn-Sham orbital, and thus decays exponentially like \cite{Almbladh1985}
\begin{align}
    \phi_\mr{HO}(r) \sim e^{-\kappa r} \\
    n(r) \sim e^{-2\kappa r},
\end{align}
where $\kappa = \sqrt{-2I}$ and $I$ is the ionization potential.
Under the non-uniform coordinate scaling described here, $r \to r_u = \sqrt{u^2 + y^2 + z^2}$.
Thus as $\lambda \to \infty$, $r \to \infty$ as well and the density and HO Kohn-Sham orbital should tend to their respective asymptotic behaviors.
Then
\begin{align}
    \sum_i^{\text{occ.}} \left|\frac{\partial \phi_i(u, y, z)}{\partial u}\right|^2 &\sim \left|\frac{\partial \phi_\text{HO}(u, y, z)}{\partial u}\right|^2 \nonumber \\
    \sum_i^{\text{occ.}} \left|\frac{\partial \phi_i(u, y, z)}{\partial u}\right|^2 &\sim \kappa^2 e^{-2\kappa r_u} \frac{u^2}{r_u^2} \\
    \left(\frac{\partial n(u, y, z)}{\partial u}\right)^2 &\sim 4\kappa^2 e^{-4\kappa r_u} \frac{u^2}{r_u^2},
\end{align}
and the $f_{\alpha1}(u,y,z)$ and $f_{\alpha2}(u,y,z)$ functions vanish identically.
Thus we anticipate that finite systems have $\alpha \sim \lambda^{-2/3}$ for large $\lambda$, as the other scaling behavior cannot yield an iso-orbital asymptotic behavior.
Note that, in this case,
\begin{equation}
    \lim_{\lambda \to \infty} \bar\alpha \to \lambda^{-2} \frac{f_{\alpha2}(u,y,z)}{\eta f_{p1}(u,y,z)},
\end{equation}
which tends to zero like $\alpha$.

Additionally, systems that are finite only in one dimension can exhibit similar scaling behavior for $\alpha$.
Consider the quasi-two dimensional electron gas in the infinite barrier model \cite{Pollack2000}.
This system is a two-dimensional uniform electron gas in the $yz$ plane with a finite, small thickness along the $x$ axis.
Moreover, the kinetic energy density of this system can be shown to be \cite{Kaplan2018}
\begin{equation}
    \tau(x) = \tau_W(x) + \frac{1}{2 [\rs^\text{2D}]^2} n(x),
\end{equation}
where $\rs^\text{2D}$ is taken to be a constant Wigner-Seitz density parameter for the two-dimensional gas.
Under non-uniform coordinate scaling,
\begin{equation}
    \alpha_\lambda^x = \lambda^{-2/3} \frac{10}{6(3\pi^2)^{2/3} [\rs^\text{2D}]^2[n(u)]^{2/3}}.
\end{equation}

Finally, if the scaled $\alpha \sim \lambda^{4/3}$ as $\lambda \to \infty$,
\begin{equation}
    \lim_{\lambda \to \infty} \bar\alpha \to \frac{f_{\alpha1}(u,y,z)}{\eta f_{p1}(u,y,z)}.
\end{equation}
This is again independent of the scale parameter.
In a three-dimensional uniform electron gas, it is straightforward to show (using cylindrical coordinates) that
\begin{equation}
    \alpha_\lambda^x = \frac{1}{3}(\lambda^{4/3} + 2\lambda^{-2/3}).
\end{equation}
It should be emphasized that this example is only illustrative, and serves to demonstrate that such a $\lambda^{4/3}$ scaling is possible in the large $\lambda$ limit.
We cannot assert that all extended systems will exhibit similar scaling behavior for $\alpha$.

\onecolumngrid
\section{Working Equations\label{AP:eqns}}

The full equations required for implementing r$^4$SCAN and r${}^{4}$SCAN are given below. Note that by construction $\bar\alpha \ge 0$ (also $\alpha$ and $\alpha^\prime$). In pseudo-potential codes (e.g. VASP) or through rounding errors in very small density regions $\bar\alpha$ can become negative however, which can cause numerical problems for interpolation functions that do not consider this possibility. An additional condition was included to Eqs. \ref{eq:fx} and \ref{eq:fc} to consistently handle negative $\bar\alpha$ regions. These provisions were essential to reliably converge calculations in VASP, as neither extending the polynomial interpolation nor setting a constant $f(\bar\alpha < 0) = 1$ were sufficient.

A list of constants needed for both functionals follows the lists of equations.

\subsection{Exchange r${}^2$SCAN}
\begin{align}
    E\smx^\mathrm{r^2SCAN}[n_\uparrow,n_\downarrow] &= \frac{1}{2}\{ E\smx^\mathrm{r^2SCAN}[2n_\uparrow] + E\smx^\mathrm{r^2SCAN}[2n_\downarrow] \} \\
    E\smx^\mathrm{r^2SCAN}[n] &= \int \varepsilon_\mr{x}^\mathrm{r^2SCAN} n \, d^3 r \\
    \varepsilon_\mr{x}^\mathrm{r^2SCAN} &= \varepsilon_\mr{x}^{\mr{LDA}}(\rs)F_\mr{x}^\mathrm{r^2SCAN}(p, \bar\alpha)\\
    \varepsilon_\mr{x}^{\mr{LDA}}(\rs) &= -\frac{\frac{3}{4\pi}\left(\frac{9\pi}{4}\right)^{1/3}}{\rs}\\
	F_\mathrm{x}^\mathrm{r^2SCAN}(p, \bar\alpha) &= \left\{h_\mathrm{1x}(p) + f_\mr{x}(\bar{\alpha})\left [h_\mathrm{0x} - h_\mathrm{1x}(p)\right ]\right\} g_\mathrm{x}(p)\\
	\bar{\alpha}(p, \alpha) &= \frac{\alpha}{1 + \eta \frac{5}{3}p} = \frac{\tau - \tau_{\mathrm{W}}}{\tau_{\mathrm{U}} + \eta\tau_{\mathrm{W}}} \\
% 	f(\bar{\alpha}) &= 1 - c\frac{\bar\alpha^2}{\bar\alpha^2 + c - 1} \\
    f_\mr{x}(\bar\alpha) &= \begin{cases}
        \exp\left[-\frac{c_{1\mr{x}}^{\mr{SCAN}}\bar\alpha}{1 - \bar\alpha}\right] & \bar{\alpha} < 0 \\
        \sum_{i=0}^7c_{\mr{x}i}\bar\alpha^i & 0 \le \bar\alpha \leq 2.5 \\
        -c_{\mr{dx}}^{\mr{SCAN}}\exp\left[\frac{c_{\mr{2x}}^{\mr{SCAN}}}{1-\bar\alpha}\right] & \bar\alpha > 2.5
        \end{cases}\label{eq:fx}\\
	h_\mathrm{0x} &= 1 + k_0\\
	h_\mathrm{1x}(p) &= 1 + k_1 - \frac{k_1}{1 + \frac{x(p)}{k1}}\\
	x(p) &= \left(C_\eta C_2\exp[-p^2/d_{p2}^4] + \mu\right)p\\
	C_\eta &= \left[\frac{20}{27} + \eta\frac{5}{3}\right]\\
	C_2 &= -\sum_{i=1}^{7}ic_{\mr{x}i}[1 - h_{0\mr{x}}] \approx -0.162742 \\
	g_\mathrm{x}(p) &= 1 - \exp\left[\frac{-a_1}{p^{1/4}}\right]
\end{align}

\subsection{Exchange r${}^4$SCAN}

\begin{align}
    F_\mathrm{x}^\mathrm{r^4SCAN}(p, \bar\alpha) &= \left\{h_\mathrm{1x}(p) + f_\mr{x}(\bar{\alpha})\left [h_\mathrm{0x} - h_\mathrm{1x}(p)\right ] +\Delta F_4(p,\bar\alpha) \right\} g_\mathrm{x}(p)\\
% 	\Delta f_\mathrm{2} &= \frac{2(1 - \bar\alpha)(c - 1)}{c} \\
% 	\Delta f_\mathrm{4} &= \frac{(1 - \bar{\alpha})^2(c - 1)(4 - c)}{c^2}  \\
	\Delta f_\mathrm{x2} &= -(1 - \bar{\alpha})\sum_{i=1}^7 ic_{\mr{x}i} \\
	\Delta f_\mathrm{x4} &= \frac{(1 - \bar{\alpha})^2}{2} \sum_{i=2}^7i(i - 1)c_{\mr{x}i} \\
	\Delta F_4(p,\bar\alpha) &= \left\{C_2\left[(1-\bar\alpha) - C_\eta p\right] + C_{\bar\alpha\bar\alpha}(1-\bar\alpha)^2 +C_{p\bar\alpha}p(1-\bar\alpha) + C_{pp}p^2 \right\}\frac{2\bar\alpha^2}{1+\bar\alpha^4} \exp\left[-\frac{(1-\bar\alpha)^2 }{d_{\bar\alpha 4}} - \frac{p^2}{d_{p4}^4}\right] \\
	C_{\bar\alpha\bar\alpha} &= \frac{73}{5000} - \frac{1}{2}\sum_{i=2}^7i(i-1)c_i[h_{0\mr{x}}-1] \approx -0.0593531 \\
	C_{p\bar\alpha} &= \frac{511}{13500} - \frac{73}{1500}\eta - \sum_{i=1}^{7}ic_{\mr{x}i}[C_\eta C_2 + \mu] \approx 0.0402684 \\
	C_{pp} &= \frac{146}{2025}\left\{\eta\frac{3}{4} + \frac{2}{3}\right\}^2 - \frac{73}{405}\left\{\eta\frac{3}{4} + \frac{2}{3}\right\} + \frac{\left(C_\eta C_2 + \mu\right)^2}{k_1} \approx -0.0880769
\end{align}

\subsection{Correlation (both r$^2$SCAN and r$^4$SCAN)}

\begin{align}
  E\smc^\mathrm{r^2SCAN}[n_\uparrow,n_\downarrow] &= \int \varepsilon\smc^\mathrm{r^2SCAN} n \, d^3 r \\
	\varepsilon_\mathrm{c}^\mathrm{r^2SCAN} &= \varepsilon_\mathrm{c}^1 + f_{\mr{c}}(\bar\alpha)(\varepsilon_\mathrm{c}^0 - \varepsilon_\mathrm{c}^1) \\
  & \nonumber \\
  \varepsilon_\mathrm{c}^0 &= \left (\varepsilon_\mathrm{c}^\mathrm{LDA0} + H_0\right ) g_\mathrm{c}(\zeta) \label{eq:ec0_scan} \\
  H_0 &= b_\mathrm{1c}\ln\left\{ 1 + w_0[1 - g_\infty(\zeta=0, s)]\right\} \\
  \varepsilon_\mathrm{c}^\mathrm{LDA0} &= -\frac{b_\mathrm{1c}}{1 + b_\mathrm{2c}\sqrt{\rs} + b_\mathrm{3c}\rs} \\
  g_{\infty}(\zeta=0, s) &= \frac{1}{(1 + 4\chi_{\infty}s^2)^{1/4}} \\
  w_0 &= \exp\left [-\frac{\varepsilon_\mathrm{c}^\mathrm{LDA0}}{b_\mathrm{1c}}\right ] - 1 \\
  g_\mr{c}(\zeta) &= \{ 1 - 2.363[d_\mr{x}(\zeta) - 1]\}(1 - \zeta^{12}) \label{eq:gc_zeta} \\
  & \nonumber \\
    f_{\mr{c}}(\bar\alpha) &= \begin{cases}
        \exp\left[-\frac{c_{1\mr{c}}^{\mr{SCAN}}\bar\alpha}{1 - \bar\alpha}\right] & \bar{\alpha} < 0 \\
        \sum_{i=0}^7c_{\mr{c}i}\bar\alpha^i & 0 \le \bar\alpha \leq 2.5 \\
        -c_\mr{dc}\exp\left[\frac{c_{\mr{2c}}}{1-\bar\alpha}\right] & \bar\alpha > 2.5
        \end{cases}\label{eq:fc}\\
  & \nonumber \\
  \varepsilon_\mathrm{c}^1 &= \varepsilon_\mathrm{c}^\mathrm{LSDA} + H_\mathrm{c}^1 \\%
  \Delta f_\mathrm{c2} &=  \sum_{i=1}^7 ic_{\mr{c}i} \\
  C_2 &= -C_\eta \Delta f_\mr{c2} \\
  d_s(\zeta) &= \frac{(1+\zeta)^{5/3} + (1-\zeta)^{5/3}}{2}\\
	H_\mathrm{c}^1 &= \gamma\phi^3\ln\left [1+w_1(1 - g(y, \Delta y))\right ] \\
	w_1 &= \exp\left[-\frac{\varepsilon_\mathrm{c}^\mathrm{LSDA}}{\gamma\phi^3}\right ] - 1 \\
	g(y, \Delta y) &= \frac{1}{\left (1 + 4(y - \Delta y)\right )^{1/4}} \\
% 	SCAN2 g(y, \Delta y) &= \frac{1}{\left (1 + 4(y - \Delta y) + d_{cr}y^2\right )^{1/4}} \\
	y &= \frac{\beta(\rs)}{\gamma w_1}t^2 \\
  \beta(\rs) &= \beta_{\mr{MB}}\frac{1 + 0.1 \rs}{1 + 0.1778 \rs} \\
	\Delta y &= \frac{\Delta f_{c2}}{27 \gamma d_s(\zeta) \phi^3 w_1} \left\{20\rs\left[g_\mr{c}(\zeta)\frac{\partial \varepsilon_c^{\text{LDA0}}}{\partial \rs} - \frac{\partial \varepsilon_c^{\text{LSDA}}}{\partial \rs} \right] - 45\eta[\varepsilon_c^{\text{LDA0}}g_\mr{c}(\zeta) - \varepsilon_c^{\text{LSDA}}] \right\} p \exp[-p^2/d_{p2}^4]
\end{align}

\subsection{Constants needed for \rrscan and \rfscan}

Constants needed for both exchange and correlation in \rrscan and \rfscan:
\begin{align}
  d_{p2} &= 0.361 \\
  \eta &= 0.001
\end{align}

\vspace{1cm}Constants needed for exchange:
\begin{align}
  \bm{c}_{\mr{x}} &= (1, -0.667, -0.4445555, -0.663086601049, 1.451297044490,\nonumber \\
    & -0.887998041597, 0.234528941479, -0.023185843322) \\
    c^{\mr{SCAN}}_{1\mr{x}} &= 0.667 \\
  c^{\mr{SCAN}}_{2\mr{x}} &= 0.8 \\
  c^{\mr{SCAN}}_{\mr{dx}} &= 1.24 \\
  k_0 &= 0.174 \\
  k_1 &= 0.065 \\
  \mu &= 10/81 \\
  a_1 &= 4.9479 \\
\end{align}
Constants for r$^4$SCAN exchange,
\begin{align}
  d_{p4} &= 0.802 \\
  d_{\bar{\alpha}4} &= 0.178
\end{align}
The $\bm{c}_\mr{x}$ are taken from Ref. \cite{Bartok2019}; all other constants in the preceding list are taken from Ref. \cite{Sun2015}.

\vspace{1cm}Constants needed for correlation:
\begin{align}
  \bm{c}_{\mr{c}} &= (1, -0.64, -0.4352, -1.535685604549, 3.061560252175,\nonumber \\
  &  -1.915710236206, 0.516884468372, -0.051848879792) \\
  c^{\mr{SCAN}}_{1\mr{c}} &= 0.64 \\
  c^{\mr{SCAN}}_{2} &= 1.5 \\
  c^{\mr{SCAN}}_{\mr{dc}} &= 0.7 \\
  b_{1\mr{c}} &= 0.0285764 \\
  b_{2\mr{c}} &= 0.0889 \\
  b_{3\mr{c}} &= 0.125541 \\
  \beta_{\mr{MB}} &\approx 0.066725 \\
  \chi_{\infty} &= \left(\frac{3\pi^2}{16}\right)^{2/3} \frac{\beta_{\mr{MB}}}{1.778\{ 0.9-3[3/(16\pi)]^{2/3}\}} \approx 0.128025 \\
  \gamma &= (1 - \ln 2)/\pi^2 \approx 0.031090690869655
\end{align}
The $\bm{c}_{\mr{c}}$ are taken from Ref. \cite{Bartok2019}; all other constants in the preceding list are taken from Ref. \cite{Sun2015}.

\clearpage

\section{Reference Atomic Calculations With Standard Basis Set}

\begin{table}[h]
    \centering
    \begin{tabular}{l|ll}
            & r++SCAN & r$^4$SCAN \\
        \hline
        E & -128.5891915 & -128.5891915 \\
        E\mol{x} & -12.14196798 & -12.14196798 \\
    \end{tabular}
    \caption{Reference self-consistent total energy ($E$) \textbf{using only the exchange part of the functional}, and corresponding exchange energies ($E_{\mr{x}}$) in Hartree for the neon atom. Calculated using the cc-pVTZ basis set \cite{Dunning1989,Woon1995} and the ``reference'' level Turbomole grid.}
    \label{tab:xonly_atoms}
\end{table}

\begin{table}[h]
    \centering
    \begin{tabular}{l|ll}
            & r++SCAN & r$^4$SCAN \\
        \hline
        E & -128.6015519 & -128.5718446 \\
        E\mol{xc} & -12.15409076 & -12.12073578 \\
    \end{tabular}
    \caption{Reference self-consistent total atomic energy ($E$) and exchange-correlation energies ($E_{\mr{xc}}$) in Hartree for the neon atom. Calculated using the cc-pVTZ basis set \cite{Dunning1989,Woon1995} and the ``reference'' level Turbomole grid.}
    \label{tab:xc_atoms}
\end{table}

\section{Individual Xenon Potential Components\label{sec:xenon_components}}

\begin{figure}[h]
    \centering
    \includegraphics[width=0.95\textwidth]{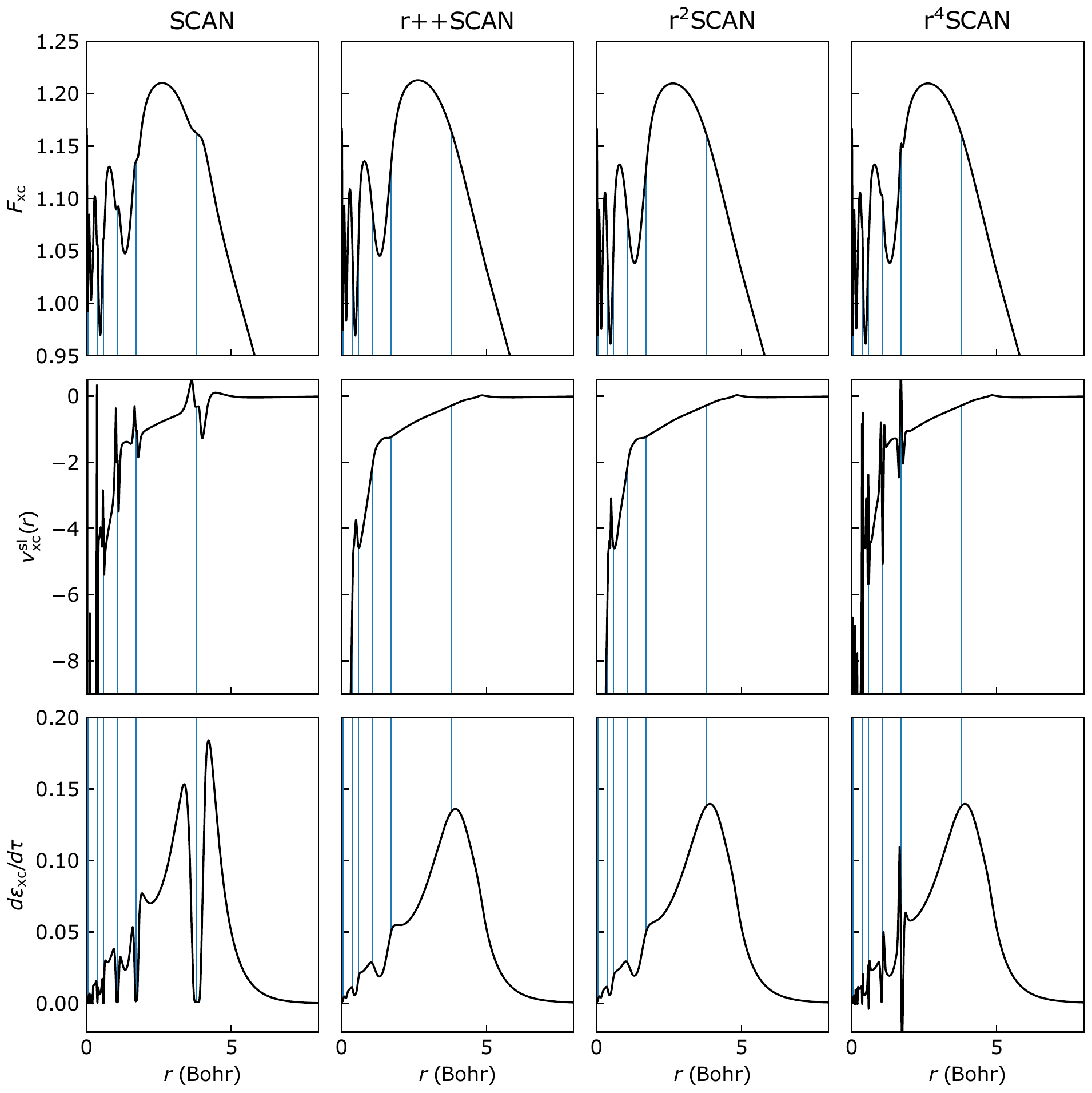}
    \caption{ The XC enhancement factor (top), the multiplicative component of the XC potential (middle), and the derivative of the XC energy density with respect to the orbital dependent kinetic energy density, $\tau$, (bottom). Shown for the SCAN, r++SCAN, \rrscan, and r$^4$SCAN functionals, calculated from reference Hartree--Fock Slater orbitals \cite{Clementi1974, Furness2021a}.}
    \label{fig:xenon_scan_rscan}
\end{figure}

\pagebreak

\section{G3 Atomization Energies}

% Not sure why, but sometimes using \include makes the references fail to link properly
% using \input works more reliably - ADK
\begin{longtable}{l|rrrrrrr}
\caption{Deviation of SCAN atomization energies (kcal/mol) obtained with increasingly dense numerical grids. Mean error (ME) and Mean Absolute Error (MAE) are given at end of table.}\label{tab:G3_SCAN}\\
\toprule
{} & 1 & 2 & 3 & 4 & 5 & 6 & 7 \\
\midrule
\endfirsthead
\caption{Deviation of SCAN atomization energies (kcal/mol) obtained with increasingly dense numerical grids. Mean error (ME) and Mean Absolute Error (MAE) are given at end of table.}\\
\toprule
{} & 1 & 2 & 3 & 4 & 5 & 6 & 7 \\
\midrule
\endhead
\midrule
\multicolumn{8}{r}{{Continued on next page}} \\
\midrule
\endfoot

\bottomrule
\endlastfoot
LiH                                                               &   2.186 &   2.549 &   2.778 &   2.524 &   2.449 &   2.448 &   2.518 \\
BeH                                                               & -10.520 & -10.879 & -11.005 & -10.701 & -10.472 & -10.447 & -10.508 \\
CH                                                                &   1.438 &   1.265 &   0.127 &   1.596 &   2.317 &   2.277 &   1.914 \\
CH\mol{2} (\excite{3}B\mol{1})                                    &  -6.715 &  -6.243 &  -7.581 &  -6.325 &  -5.790 &  -5.911 &  -6.400 \\
CH\mol{2} (\excite{1}A\mol{1})                                    &   4.207 &   4.836 &   2.635 &   4.362 &   5.415 &   5.626 &   5.269 \\
CH\mol{3}                                                         &  -5.871 &  -5.316 &  -7.022 &  -5.373 &  -4.543 &  -4.490 &  -4.867 \\
Methane (CH\mol{4})                                               &   0.207 &  -0.173 &  -1.744 &  -0.707 &   0.398 &   0.746 &   0.570 \\
NH                                                                &  -2.770 &  -3.772 &  -1.302 &  -1.004 &  -1.294 &  -1.994 &  -1.230 \\
\end{longtable}

\begin{longtable}{l|rrrrrrr}
\caption{Deviation of rSCAN atomization energies (kcal/mol) obtained with increasingly dense numerical grids. Mean error (ME) and Mean Absolute Error (MAE) are given at end of table.}\label{tab:G3_rSCAN}\\
\toprule
{} & 1 & 2 & 3 & 4 & 5 & 6 & 7 \\
\midrule
\endfirsthead
\caption{Deviation of rSCAN atomization energies (kcal/mol) obtained with increasingly dense numerical grids. Mean error (ME) and Mean Absolute Error (MAE) are given at end of table.}\\
\toprule
{} & 1 & 2 & 3 & 4 & 5 & 6 & 7 \\
\midrule
\endhead
\midrule
\multicolumn{8}{r}{{Continued on next page}} \\
\midrule
\endfoot

\bottomrule
\endlastfoot
LiH                                                               &    2.022 &    2.093 &    2.081 &    2.144 &    2.139 &    2.125 &    2.130 \\
BeH                                                               &  -11.231 &  -11.181 &  -10.947 &  -11.020 &  -11.013 &  -10.998 &  -11.006 \\
CH                                                                &    1.521 &    1.462 &    1.516 &    1.491 &    1.387 &    1.406 &    1.426 \\
CH\mol{2} (\excite{3}B\mol{1})                                    &   -8.621 &   -8.971 &   -8.715 &   -8.716 &   -8.799 &   -8.790 &   -8.769 \\
CH\mol{2} (\excite{1}A\mol{1})                                    &    3.671 &    3.600 &    3.783 &    3.736 &    3.596 &    3.610 &    3.644 \\
CH\mol{3}                                                         &   -7.135 &   -7.528 &   -7.236 &   -7.324 &   -7.429 &   -7.409 &   -7.387 \\
Methane (CH\mol{4})                                               &   -3.790 &   -3.669 &   -3.563 &   -3.537 &   -3.680 &   -3.668 &   -3.636 \\
NH                                                                &   -1.620 &   -1.740 &   -1.830 &   -2.047 &   -1.987 &   -1.930 &   -1.959 \\
\end{longtable}

\begin{longtable}{l|rrrrrrr}
\caption{Deviation of r++SCAN atomization energies (kcal/mol) obtained with increasingly dense numerical grids. Mean error (ME) and Mean Absolute Error (MAE) are given at end of table.}\label{tab:G3_rppSCAN}\\
\toprule
{} & 1 & 2 & 3 & 4 & 5 & 6 & 7 \\
\midrule
\endfirsthead
\caption{Deviation of r++SCAN atomization energies (kcal/mol) obtained with increasingly dense numerical grids. Mean error (ME) and Mean Absolute Error (MAE) are given at end of table.}\\
\toprule
{} & 1 & 2 & 3 & 4 & 5 & 6 & 7 \\
\midrule
\endhead
\midrule
\multicolumn{8}{r}{{Continued on next page}} \\
\midrule
\endfoot

\bottomrule
\endlastfoot
LiH                                                               &      2.407 &      2.428 &      2.386 &      2.480 &      2.511 &      2.504 &      2.498 \\
BeH                                                               &    -10.768 &    -10.712 &    -10.485 &    -10.553 &    -10.547 &    -10.532 &    -10.540 \\
CH                                                                &      2.197 &      2.136 &      2.188 &      2.160 &      2.058 &      2.079 &      2.098 \\
CH\mol{2} (\excite{3}B\mol{1})                                    &     -7.859 &     -8.212 &     -7.957 &     -7.960 &     -8.042 &     -8.031 &     -8.011 \\
CH\mol{2} (\excite{1}A\mol{1})                                    &      4.728 &      4.662 &      4.840 &      4.788 &      4.650 &      4.667 &      4.699 \\
CH\mol{3}                                                         &     -5.860 &     -6.250 &     -5.963 &     -6.055 &     -6.158 &     -6.136 &     -6.115 \\
Methane (CH\mol{4})                                               &     -2.264 &     -2.142 &     -2.042 &     -2.021 &     -2.163 &     -2.148 &     -2.117 \\
NH                                                                &     -0.853 &     -0.991 &     -1.081 &     -1.297 &     -1.235 &     -1.180 &     -1.208 \\
\end{longtable}

\begin{longtable}{l|rrrrrrr}
\caption{Deviation of r${}^2$SCAN atomization energies (kcal/mol) obtained with increasingly dense numerical grids. Mean error (ME) and Mean Absolute Error (MAE) are given at end of table.}\label{tab:G3_r2SCAN}\\
\toprule
{} & 1 & 2 & 3 & 4 & 5 & 6 & 7 \\
\midrule
\endfirsthead
\caption{Deviation of r${}^2$SCAN atomization energies (kcal/mol) obtained with increasingly dense numerical grids. Mean error (ME) and Mean Absolute Error (MAE) are given at end of table.}\\
\toprule
{} & 1 & 2 & 3 & 4 & 5 & 6 & 7 \\
\midrule
\endhead
\midrule
\multicolumn{8}{r}{{Continued on next page}} \\
\midrule
\endfoot

\bottomrule
\endlastfoot
LiH                                                               &     2.455 &     2.525 &     2.484 &     2.553 &     2.586 &     2.583 &     2.575 \\
BeH                                                               &   -10.418 &   -10.465 &   -10.225 &   -10.267 &   -10.247 &   -10.233 &   -10.242 \\
CH                                                                &     2.617 &     2.454 &     2.557 &     2.508 &     2.417 &     2.440 &     2.455 \\
CH\mol{2} (\excite{3}B\mol{1})                                    &    -6.222 &    -6.642 &    -6.398 &    -6.398 &    -6.469 &    -6.455 &    -6.439 \\
CH\mol{2} (\excite{1}A\mol{1})                                    &     5.934 &     5.708 &     5.938 &     5.884 &     5.748 &     5.769 &     5.797 \\
CH\mol{3}                                                         &    -3.602 &    -4.222 &    -3.879 &    -3.970 &    -4.069 &    -4.042 &    -4.025 \\
Methane (CH\mol{4})                                               &     0.930 &     0.535 &     0.881 &     0.849 &     0.707 &     0.723 &     0.751 \\
NH                                                                &    -0.347 &    -0.728 &    -0.764 &    -0.953 &    -0.901 &    -0.853 &    -0.881 \\
\end{longtable}

\begin{longtable}{l|rrrrrrr}
\caption{Deviation of r${}^4$SCAN atomization energies (kcal/mol) for the G3 test set obtained with increasingly dense numerical grids. Mean error (ME) and Mean Absolute Error (MAE) are given at end of table.}\label{tab:G3_r4scan}\\
\toprule
{} & 1 & 2 & 3 & 4 & 5 & 6 & 7 \\
\midrule
\endfirsthead
\caption{Deviation of r${}^4$SCAN atomization energies (kcal/mol) for the G3 test set obtained with increasingly dense numerical grids. Mean error (ME) and Mean Absolute Error (MAE) are given at end of table.}\\
\toprule
{} & 1 & 2 & 3 & 4 & 5 & 6 & 7 \\
\midrule
\endhead
\midrule
\multicolumn{8}{r}{{Continued on next page}} \\
\midrule
\endfoot

\bottomrule
\endlastfoot
LiH                                                               &     2.403 &     2.537 &     2.562 &     2.503 &     2.547 &     2.553 &     2.583 \\
BeH                                                               &   -10.549 &   -10.555 &   -10.337 &   -10.462 &   -10.363 &   -10.166 &   -10.358 \\
CH                                                                &     2.203 &     2.564 &     2.348 &     2.400 &     2.449 &     2.293 &     2.375 \\
CH\mol{2} (\excite{3}B\mol{1})                                    &    -6.302 &    -6.343 &    -6.300 &    -6.009 &    -6.037 &    -6.256 &    -6.188 \\
CH\mol{2} (\excite{1}A\mol{1})                                    &     5.376 &     5.926 &     5.755 &     5.800 &     5.758 &     5.603 &     5.724 \\
CH\mol{3}                                                         &    -4.008 &    -4.250 &    -4.020 &    -4.022 &    -3.956 &    -4.130 &    -4.078 \\
Methane (CH\mol{4})                                               &     0.527 &     0.574 &     0.493 &     0.530 &     0.544 &     0.419 &     0.530 \\
NH                                                                &     0.413 &    -1.697 &    -0.425 &    -0.814 &    -0.948 &    -0.957 &    -0.987 \\
\end{longtable}

\section{Weakly Bound Complexes}

\begin{longtable}{l|rrrrr}
\caption{Error in dissociation energies (kcal/mol) for the S22 test set. Mean error (ME) and Mean Absolute Error (MAE) are given at end of table.}\label{tab:full_S22}\\
\toprule
{} &  SCAN &  rSCAN &  r++SCAN &  r${}^2$SCAN &  r${}^4$SCAN \\
\midrule
\endfirsthead
\caption{Error in dissociation energies (kcal/mol) for the S22 test set. Mean error (ME) and Mean Absolute Error (MAE) are given at end of table.}\\
\toprule
{} &  SCAN &  rSCAN &  r++SCAN &  r${}^2$SCAN &  r${}^4$SCAN \\
\midrule
\endhead
\midrule
\multicolumn{6}{r}{{Continued on next page}} \\
\midrule
\endfoot

\bottomrule
\endlastfoot
0   &   -0.032 &    -0.259 &      -0.048 &     -0.236 &     -0.245 \\
1   &    0.383 &     0.051 &       0.392 &      0.135 &      0.142 \\
2   &    2.157 &     1.332 &       2.087 &      1.248 &      1.480 \\
3   &    0.478 &    -0.181 &       0.511 &     -0.181 &     -0.131 \\
4   &    0.025 &    -0.573 &       0.105 &     -0.640 &     -0.674 \\
5   &    0.031 &    -0.434 &       0.207 &     -0.547 &     -0.563 \\
6   &   -0.502 &    -1.045 &      -0.343 &     -1.108 &     -1.080 \\
7   &   -0.176 &    -0.298 &      -0.196 &     -0.238 &     -0.233 \\
\end{longtable}

\begin{longtable}{l|rrrrr}
\caption{Error in reaction barrier heights (kcal/mol) for the BH76 test set. Mean error (ME) and Mean Absolute Error (MAE) are given at end of table.}\label{tab:BH76_full}\\
\toprule
{} &  SCAN &  rSCAN &  r++SCAN &  r$^2$SCAN &  r$^4$SCAN \\
\midrule
\endfirsthead
\caption{Error in reaction barrier heights (kcal/mol) for the BH76 test set. Mean error (ME) and Mean Absolute Error (MAE) are given at end of table.}\\
\toprule
{} &  SCAN &  rSCAN &  r++SCAN &  r$^2$SCAN &  r$^4$SCAN \\
\midrule
\endhead
\midrule
\multicolumn{6}{r}{{Continued on next page}} \\
\midrule
\endfoot

\bottomrule
\endlastfoot
0   &   -8.791 &    -8.273 &      -8.096 &     -7.781 &     -8.427 \\
1   &  -18.390 &   -21.305 &     -21.950 &    -20.121 &    -20.565 \\
2   &  -13.497 &   -13.512 &     -13.451 &    -13.044 &    -13.569 \\
3   &  -13.497 &   -13.512 &     -13.451 &    -13.044 &    -13.569 \\
4   &   -8.797 &    -9.233 &      -9.052 &     -8.601 &     -8.918 \\
5   &   -8.797 &    -9.233 &      -9.052 &     -8.601 &     -8.918 \\
6   &  -10.468 &   -11.605 &     -11.294 &    -11.038 &    -10.433 \\
7   &  -10.804 &   -10.046 &     -10.548 &    -10.149 &    -11.007 \\
\end{longtable}

\clearpage
\twocolumngrid

\end{document}